\documentclass[twocolumn]{pasj01}
\usepackage{natbib} 
\usepackage{lscape}

\newcommand{\OII}{[O\,{\sc ii}]}

\newcommand{\CIV}{C{\sc iv}}

\newcommand{\OIII}{[O\,{\sc iii}]}

\newcommand{\luv}{$L_{\rm UV}$}

\newcommand{\lya}{${\rm Ly}\alpha \ $}

\newcommand{\wobs}{$\omega_{\rm obs}(\theta)$}
\newcommand{\Ao}{A_{\omega}}
\newcommand{\Aocor}{A_{\rm \omega,\, corr}}

\Received{$\langle$2017 July 28$\rangle$}
\Accepted{$\langle$2017 November 22$\rangle$}
\Published{$\langle$publication date$\rangle$}

\begin{document}
\title{{\color{black}The Stellar Mass, Star Formation Rate and Dark Matter Halo Properties of LAEs at $z\sim2$}} 

\author{Haruka \textsc{Kusakabe}\altaffilmark{1}%
}
\altaffiltext{1}{Department of Astronomy, Graduate School of Science, The University of Tokyo, 7-3-1 Hongo, Bunkyo-ku, Tokyo 113-0033, Japan}
\email{kusakabe@astron.s.u-tokyo.ac.jp}

\author{Kazuhiro \textsc{Shimasaku}\altaffilmark{1,2}}
\altaffiltext{2}{Research Center for the Early Universe, The University of Tokyo, 7-3-1 Hongo, Bunkyo-ku, Tokyo 113-0033, Japan}

\author{Masami \textsc{Ouchi}\altaffilmark{3,4}}
\altaffiltext{3}{Kavli Institute for the Physics and Mathematics of the Universe (Kavli IPMU, WPI), The University of Tokyo, 5-1-5 Kashiwanoha, Kashiwa, Chiba 277-8583, Japan}
\altaffiltext{4}{Institute for Cosmic Ray Research, The University of Tokyo, 5-1-5 Kashiwanoha, Kashiwa, Chiba 277-8582, Japan}

\author{Kimihiko \textsc{Nakajima}\altaffilmark{5}}
\altaffiltext{5}{European Southern Observatory, Karl-Schwarzschild-Strasse 2, D-85748 Garching, b. M{\"u}nchen, Germany}

\author{Ryosuke \textsc{Goto}\altaffilmark{1}} 

\author{Takuya \textsc{Hashimoto}\altaffilmark{1, 6, 7}}
\altaffiltext{6}{College of General Education, Osaka Sangyo University, 3-1-1 Nakagaito, Daito, Osaka 574-8530, Japan}
\altaffiltext{7}{National Astronomical Observatory of Japan, 2-21-1 Osawa, Mitaka, Tokyo 181-8588, Japan}

\author{Akira \textsc{Konno}\altaffilmark{1,4}}

\author{Yuichi \textsc{Harikane}\altaffilmark{4,8}}
\altaffiltext{8}{Department of Physics, Graduate School of Science, The University of Tokyo, 7-3-1 Hongo, Bunkyo, Tokyo 113-0033, Japan}

\author{John \textsc{D. Silverman}\altaffilmark{3}}

\author{Peter \textsc{L. Capak}\altaffilmark{9, 10}}
\altaffiltext{9}{California Institute of Technology, MC 105-24, 1200 East California Blvd., Pasadena, CA 91125, USA}
\altaffiltext{10}{Infrared Processing and Analysis Center, California Institute of Technology, MC 100-22, 770 South Wilson Ave., Pasadena, CA 91125, USA}

\maketitle

\begin{abstract}
We present average stellar population properties and dark matter halo masses of $z \sim 2$ \lya emitters (LAEs) {\color{black}from SED fitting} and clustering analysis, respectively, {\color{black}using $\simeq$ $1250$ objects ($NB387\le25.5$)} in four separate fields of $\simeq 1$ deg$^2$ in total. With an average stellar mass of {\color{black}$10.2\, \pm\, 1.8\times 10^8\  {\mathrm M_\odot}$} and star formation rate of {\color{black}$3.4\, \pm\, 0.4\ {\mathrm M_\odot}\ {\rm yr^{-1}}$}, the LAEs {\color{black}lie on an extrapolation of the star-formation main sequence (MS) to low stellar mass.} Their effective dark matter halo mass is estimated to be $4.0_{-2.9}^{+5.1} \times 10^{10}\ {\mathrm M_\odot}$ with {\color{black} an effective bias of $1.22^{+0.16}_{-0.18}$} {\color{black}which is} lower than that of $z \sim 2$ LAEs {\color{black}($1.8\, \pm\, 0.3$),} obtained by a previous study based on a three times smaller {\color{black}survey area, with a probability of $96\%$.} {\color{black}However, the difference in the bias values can be explained if cosmic variance is taken into account.  If such a low halo mass implies a low HI gas mass, this result appears to be consistent with the observations of a high \lya escape fraction.}  
{\color{black}With the low halo masses and ongoing star formation, our LAEs have a relatively high stellar-to-halo mass ratio (SHMR) and a high efficiency of converting baryons into stars.} The extended Press-Schechter formalism predicts that at $z=0$ our LAEs are typically embedded in halos with masses similar to that of the Large Magellanic Cloud (LMC); {\color{black}they will also have similar SHMRs to the LMC, if their SFRs are largely suppressed after $z \sim 2$ as some previous studies have reported for the LMC itself.}

\end{abstract}

\setcounter{section}{0}
\section{Introduction}
Galaxies assemble their stellar mass through star formation and {\color{black}galaxy merging under the gravitational influence of their host dark matter halos, which also grow through} 
mass accretion and merging {\color{black} \citep[e.g.,][]{Somerville2015}}. Hence, observations of {\color{black} the intrinsic properties of galaxies and their dependence 
on halo mass in the past are key to tracing the history of the mass growth of galaxies
and constraining the physical processes that control star formation (SF). }

{\color{black}Low-mass galaxies at high redshift are ''building blocks'' of present-day galaxies over a wide mass range.} Nebular emission lines are useful to detect faint (or low-mass) galaxies at high redshift ($z$), among which \lya line has been used most commonly. Tens of thousands of \lya emitters (LAEs) have been selected so far by narrowband (NB) imaging observations \citep[$z\sim2$--$7$: e.g., ][]{Malhotra2002, Taniguchi2005, Shimasaku2006, Gronwall2007, Ota2008,  Ouchi2008, Guaita2010, Hayes2010, Hu2010, Ouchi2010,  Ciardullo2012, Nakajima2012, Yamada2012, Konno2014, Sandberg2015, Ota2017, Shimakawa2017a, Shibuya2017arxiv} and/or spectroscopically identified \citep[$z\sim0$--$7$: e.g., ][]{ Shapley2003, Kashikawa2006,  Reddy2008, Cowie2010, Blanc2011, Dressler2011, Kashikawa2011a, Curtis-Lake2012,  Mallery2012, Nakajima2013, Erb2014, Hayes2014, Hashimoto2013, Hathi2016,  Karman2017, Shibuya2017barXiv} and they are one of the important populations of high-z star forming galaxies.  

Typical LAEs at high redshifts have low stellar masses \citep[$M_{\star} \lesssim 10^9\ {\mathrm M_\odot}$:][]{Ono2010b, Guaita2011, Kusakabe2015, Hagen2016, Shimakawa2017a}. They are also {\color{black} dust poor} \citep{Lai2008, Blanc2011, Kusakabe2015} {\color{black}  and metal poor} \citep{Nakajima2012, Nakajima2013, Nakajima2014, Kojima2017}, and {\color{black} have young stellar populations} \citep{Pirzkal2007, Gawiser2007, Hagen2014}, although a small fraction of them are {\color{black}attributed to} dusty galaxies with high stellar masses \citep{Nilsson2009, Ono2010a, Pentericci2010, Oteo2012}. 

Since their dust emission is typically too faint to be detected by current infrared (IR) telescopes without gravitational lensing, {\color{black}estimates of their star formation rates ($SFR$s) vary greatly depending on the method of measurement,} making it difficult to determine {\color{black} their mode of star formation (i.e., starburst or more typical of main-sequence (MS) galaxies)} \citep{Finkelstein2015, Hagen2016, Hashimoto2017, Shimakawa2017a}. Only at $z\sim2$ has {\color{black} the average $SFR$ of LAEs} been estimated from ultraviolet (UV) and dust emission, by means of stacking, from which they are found to lie on the star formation main sequence \citep[SFMS: e.g.,][]{Daddi2007}, although the analysis is limited to only a single survey field \citep{Kusakabe2015}. Recent observations have revealed that {\color{black} the} stellar properties of LAEs are similar to those of other emission line galaxies {\color{black} at $z\sim2$ }\citep[][]{Hagen2016}. \citet{Shimakawa2017a} have also found that LAEs at $M_{\star}\lesssim10^{10}\ {\rm M_{\odot}}$ obey the same $M_{\star}$-$SFR$ and $M_{\star}$-size relations as H$\alpha$ emitters (HAEs) at $z=2.5$. Thus, there is a possibility that LAEs are normal star-forming galaxies in the low stellar mass regime at high redshift.
  
With regard to {\color{black}their} dark matter halos, LAEs have been found to reside in low-mass halos from clustering analysis \citep[$M_{\rm h}\sim10^{10}$--$10^{12}\ {\rm M_{\odot}}$ over $z\sim2$--$7$: e.g., ][]{Ouchi2005b, Kovac2007, Gawiser2007, Shioya2009, Guaita2010, Ouchi2010, Bielby2016, Diener2017arXiv, Ouchi2017arXiv}. These results imply that LAEs at $z\sim 4$--$7$ and $z\sim2$--$3$ evolve into massive elliptical galaxies and $L_{\star}$ galaxies at $z=0$, respectively. {\color{black} For both cases,} high-$z$ LAEs are {\color{black} likely candidates of the ``building blocks''} of mature galaxies in the local Universe \citep[see also][]{Rauch2008, Dressler2011} because they are embedded in the lowest-mass halos among all the high-$z$ galaxy populations. 

With stellar masses, $SFR$s, and halo masses in hand, one can obtain {\color{black}the} stellar to halo mass ratios ($\equiv M_{\star}/M_{\rm h}$: $SHMR$) and baryon conversion efficiencies ($\equiv SFR/ {\rm baryon\ accretion\ rate}$: $BCE$) to quantify the star formation efficiency in dark matter halos. The $SHMR$ measures the time-integrated (time-averaged) efficiency of star formation 
{\color{black}up to the observed epoch, }
while the $BCE$ measures the efficiency {\color{black}at the observed epoch.  Previous studies show tight relations of the $SHMR$ and $BCE$ of galaxies as a function of $M_{\rm h}$ over a wide redshift range \citep[e.g.,][]{Behroozi2013,Moster2013,Rodriguez-Puebla2017a}. These relation are usually given as the average relations in the literature thus presented here as such.} The SF mode also tells us the nature of star formation in terms of stellar mass growth.
 
For LAEs, {\color{black} these parameters are most reliably measured at $z \sim 2$, 
because this redshift is high enough that the} \lya line is redshifted into the optical regime where a wide-field ground-based \lya survey, critical for clustering analysis, is possible,  and low enough that deep rest-frame near-infrared (NIR) photometry, critical for SED fitting of faint galaxies like LAEs, is still possible with Spitzer/IRAC. This redshift is also scientifically interesting {\color{black}because star-formation activity in the universe is at a global maximum \citep{Madau2014}.}

{\color{black} To date, there is only one clustering study carried out at $z \sim 2$, by \citet{Guaita2010}, for which they obtain }a relatively high halo mass of ${\rm log} (M_{\rm h}/ {\rm M_{\odot}}) \sim11.5^{+0.4}_{-0.5}$, which implies {\color{black} an $SHMR$ comparable to or lower than the average relations by \citet{Behroozi2013} and \citet{Moster2013} at the same {\color{black} dark halo} mass. Their LAEs are estimated to have a comparable $BCE$ with the average relation by \citet{Behroozi2013} but its uncertainty is as large as $\sim1$ dex.} However, this {\color{black}halo} mass estimate may suffer from statistical uncertainties due to a small sample size ($N\sim 250$ objects) and systematic uncertainties from cosmic variance due to a small survey area ($\sim0.3\ {\rm deg}^{2}$). A larger number of sources from a larger survey area with deep multi-wavelength data {\color{black} is needed to obtain SHMRs and BCEs accurately and to overcome these uncertainties.}

In this paper, we study star forming activity and its dependence on halo mass for $z \sim 2$ LAEs using $\sim$ $1250$ NB-selected LAEs from four deep survey fields with a total area of $\simeq1 \ {\rm deg}^2$. Section 2 summarizes the data and sample used in this study. In section 3 we estimate halo masses from clustering analysis. In section 4 we perform SED fitting to stacked imaging data to measure stellar population parameters. The $SHMR$ and $BCE$ are calculated and compared with literature results in section 5.  Section 6 is devoted to discuss the results obtained in the previous sections. Conclusions are given in Section 7.

Throughout this paper, we adopt a flat cosmological model with the matter density $\Omega_{\rm m} = 0.3$, the cosmological constant $\Omega_{\Lambda} = 0.7$, the baryon density
$\Omega_{b} = 0.045$, the Hubble constant $H_{0} = 70~{\rm km\,s^{-1}Mpc^{-1}}\>(h_{100}=0.7)$, the power-law index of the primordial power spectrum $n_{\rm s} = 1$, and the linear amplitude of mass fluctuations in the universe $\sigma_{8}=0.8$, which are consistent with the latest Planck results \citep{Plankcollaboration2016}. We assume a Salpeter initial mass function \citep[IMF: ][]{Salpeter1955}\footnote{\label{ft:imf}To rescale stellar masses in previous studies assuming a Chabrier or Kroupa IMF \citep{Kroupa2001, Chabrier2003}, we divide them by a constant factor of 0.61 or 0.66, respectively. Similarly, to convert SFRs in the literature with a Chabrier or Kroupa IMF, we divide them by a constant factor of 0.63 or 0.67, respectively.}. Magnitudes are given in the AB system \citep{oke1983} and coordinates are given in J2000. Distances are expressed in comoving units. We use ``log'' to denote a logarithm with a base $10$ (${\rm log}_{10}$).

%
%
%
\section{Data and Sample}
\subsection{Sample Selection}\label{subsec:selection}
Our LAE samples are constructed in four deep survey fields, the
Subaru/XMM-Newton Deep Survey (SXDS) field \citep{Furusawa2008}, the Cosmic Evolution Survey (COSMOS) field \citep{Scoville2007}, the Hubble Deep Field North \citep[HDFN:][]{ Capak2004}, and the Chandra Deep Field South \citep[CDFS:][]{Giacconi2001}. We select LAEs at $z=2.14$--$2.22$ using the narrow band $NB387$ \citep{Nakajima2012} as described in selection papers \citep[][]{Nakajima2012, Nakajima2013, Kusakabe2015, Konno2016}. The threshold of rest-frame equivalent width, $EW_0$, of \lya emission is $EW_0({\rm Ly}\alpha) \geq 20$--$30$\AA$\,$ \citep{Konno2016}\footnote{\color{black}The threshold varies from 20 to 30 \AA$\,$ because the response curves of the selection bands $U$ (or $u^*$) and $B$ are slightly different among the four fields. Two-color diagrams of U (or $u^\star$)--$NB387$ and $B$--$NB387$ for selection in each of the four fields are shown in figure 1 in \citet{Konno2016}}. {\color{black} While the SXDS field consists of five sub-fields, we use the three regions (SXDS-C, N and S) with deeper $NB387$ images.} {\color{black}The 5$\,\sigma$ depths in a $2''$ diameter aperture are $\simeq$ 25.7 (SXDS-C,N,S), 26.1 (COSMOS), 26.4 (HDFN), and 26.6 (CDFS).} 
{\color{black}{For accurate clustering analysis,} we remove LAEs in regions with short net exposure times, resulting from the dither pattern.} 
In the SXDS field (SXDS--C, N, and S), we use {\color{black}the overlapping regions to examine} if there exists an offset in the $NB387$ zero point. A non-negligible offset of $0.06$ mag is found in SXDS-N {\color{black}and appropriately corrected.} In the other three fields, we examine the $NB387$ zero point using {\color{black}the colors of the Galactic stars from} \citet{Gunn1983} and apply {\color{black}a $0.1$ mag correction to LAEs in} CDFS. Note that such {\color{black}a} correction values change {\color{black}the} \lya luminosities only slightly. Our entire sample consists of $2441$ LAEs from $\simeq1$ square degree (each survey area size is shown in table \ref{tbl:data}). {\color{black} Of these, we use {\color{black}$1937$ LAEs with $NB387_{\rm tot} \le 26.3$}, where $NB387_{\rm tot}$ is the NB387 total magnitude, {\color{black} for the} clustering analysis to examine {\color{black}the} halo mass dependence on $NB387_{\rm tot}$} (see figure \ref{fig:NBBNB_NB}, table \ref{tbl:NBcr} and section \ref{subsec:subsample}). Note that $1248$ LAEs with $NB387_{\rm tot} \le 25.5$ are used to calculate a four-field average effective bias (see section \ref{subsec:bias}) and derive the SHMR and BCE of our LAEs.

\subsection{Contamination Fraction}\label{subsec:fc} 
Possible interlopers in our LAE samples are categorized into {\color{black}
(i) spurious sources without continuum, 
(ii) active galactic nuclei (AGNs), 
(iii) low-$z$ line emitters whose line emission (not Ly$\alpha$) is strong enough to meet our color selection, (iv) low-$z$ line emitters with weaker emission lines which happen to meet the color selection owing to photometric errors in the selection bands, 
(v) low-EW ($\lesssim 20-30\,$\AA) LAEs at our target redshift selected owing to photometric errors in the selection bands, and (vi) continuum sources at any redshifts selected as LAEs owing to photometric errors in the selection bands. We describe each in further detail here.}

\begin{description}
\item[(i)]
{\color{black}Spurious sources without continuum are possibly included in our LAE sample even after visual inspection was performed as described in the original papers based on selection. About 1.6\% of all 2441 LAEs have neither $U$ (or $u^*$) nor $B$ band detection at more than 2$\,\sigma$, and this fraction reduces to 0.2\% for the 1248 objects with $NB387 \le 25.5$.}

\item[(ii)]
{\color{black} All sources detected in either X-ray, UV, or radio are regarded as AGNs and have been removed as described in the selection papers. Their fraction of  
the entire sample is about 2\%. Obscured faint AGNs at these wavelengths may contaminate our sample, although heavily obscured  AGNs are unlikely to have emission lines strong enough to pass our color selection.  Following \citep{Guaita2010}, we estimate the possible fraction of obscured AGNs in our LAE sample to be $\sim2$\%, i.e. similar to that of X- ray, UV, or radio detected AGNs \citep[i.e.,][]{Xue2010, Assef2012, Heckman2014, Aird2017,Ricci2017}. }

\item[(iii)]
Candidate emitters are \OII\ $\lambda\,3727$ emitters at $z\simeq0.04$, Mg\,{\sc ii} $\lambda\,2798$ emitters at $z\simeq0.4$, and \CIV $\lambda\,1550$ and C{\sc iii}] $\lambda\,1909$ emitters at $z\simeq1.5$. However, the survey volume of \OII\ emitters at $z\simeq0.04$ is three orders of magnitude smaller than that of LAEs at $z=2.2$. {\color{black}Moreover, the $EW_{0}$(\OII) of the vast majority of \OII\ emitters is too small ($\sim8\,$\AA) to meet our color selection of $EW_{0}$(\OII) $\geq70\ $\AA\ \citep[see][]{Konno2016, Ciardullo2013}. \OII\ emitters with such a large $EW_{0}$(\OII) should be AGNs. Mg\,{\sc ii}, \CIV\ and C{\sc iii}] emitters which satisfy our selection criteria are also likely to be AGNs. X-ray, UV, or radio detected AGNs have been removed. Therefore, the fraction of contaminants (iii) is expected to be negligibly small and is included in the possible fraction of obscured AGNs as described in category (ii).}

\item[(iv), (v), (vi)]
We evaluate the contamination fraction {\color{black}contributed by (iv), (v) and (vi) sources that} do not satisfy the selection criteria {\color{black} if they have no photometric error (hereafter, intrinsically unselected sources)}, using Monte Carlo simulations. {\color{black}We use bright sources with $NB387{\le}24.0$ mag where photometric errors are negligible in all three selection bands of  $U$ (or $u^*$), $B$, and $NB387$ in the four fields. Assuming that the relative distribution of $NB387$-detected objects in the two-color selection plane, $U$ (or $u^*$) --$NB387$ vs. $B$ -- $NB387$, is unchanged with $NB387$ magnitude intrinsically, we create a mock catalog by adding photometric errors to the three selection bands. Here, the distribution of NB387 magnitudes of simulated sources is set equal to that of real $NB387$-detected objects down to the 5$\sigma$ limiting magnitude of $NB387$ in each of the four fields as described in section \ref{subsec:selection}. }

{\color{black} We then apply the same selection as for the real catalog to obtain the number of objects passing the selection. The contamination fraction is calculated by dividing the number of intrinsically unselected sources passing the selection by the number of all sources passing the selection. The latter are a mixture of real LAEs with $EW_{0}({\rm Ly}\alpha) \ge 20$--$30$ \AA$\ $ and intrinsically unselected sources passing the selection (i.e., (iv), (v) and (vi)). We find that the contamination fraction at $NB387 \le 25.5$ is $10$--$20$\% for all four fields. This contamination fraction is conservative in the sense that (v) real LAEs with $EW_{0}({\rm Ly}\alpha)\le 20$--$30$ \AA $\ $ are categorized as intrinsically unselected sources, whose fraction is expected to be significantly higher than that of (iv). }
\end{description}

{\color{black}To summarize, the fractions of possible interlopers (i), (ii), and (iii) are negligibly small and those of (iv), (v), and (vi) are estimated to be $10$--$20$\% in total for all four fields.} 

Spectroscopic follow-up observations of \lya emission of bright LAEs in our sample ($NB387{\le}24.5\ {\rm mag}$) have also been carried out with Magellan/IMACS, MagE, and Keck/LRIS by \citet{Nakajima2012}, \citet{Hashimoto2013}, \citet{Shibuya2014a}, \citet{Hashimoto2015}, \citet{Hashimoto2017}, and M. Rauch et al. (2017, in preparation). In total, more than $40$ LAEs are spectroscopically confirmed and no foreground interlopers such as [OII] emitters at $z = 0.04$ are found \citep{Nakajima2012}. Although faint LAEs cannot be confirmed spectroscopically, the contamination fraction is probably not high. Indeed, \citet{Konno2016} have not applied contamination correction in deriving luminosity functions.
On the basis of the results of the Monte Carlo simulations and the spectroscopic follow-up observations, $0$--$20$\%, we conservatively adopt $10\, \pm\, 10\%$ for the contamination fraction. {\color{black} This value is similar to a previous result for NB-selected LAEs at $z\sim2$, $7\,\pm\,7$\%, which is a sum of (i), (ii), (iii) and (vi) \citep[][]{Guaita2010}.} The effect of contamination sources is taken into account in clustering analysis (see section \ref{subsec:acf}). On the other hand, it is negligible in SED fitting for median-stacked subsamples in section \ref{sec:sed}.

\subsection{Imaging Data for SED Fitting}\label{subsec:data}
We use ten broadband images for SED fitting: five optical bands -- $B, V, R$ (or $r$), $i$ (or $i'$) and $z$ (or $z'$); three NIR bands -- $J$, $H$ and $K$ (or $Ks$); and two mid-infrared (MIR) bands -- IRAC ch1 and ch2. The PSFs of the images are matched in each field (not in each sub-field). The aperture corrections {\color{black} for converting $3''$ MIR aperture magnitudes} to total magnitudes are taken from \citet[][see table\ref{tbl:data}]{Ono2010b}. For each field, a K-band or NIR detected catalog is used to {\color{black} obtain secure IRAC photometry} in section \ref{subsec:irac}. Here we summarize the data used in SED fitting and IRAC cleaning in the four fields. 

\begin{landscape}
\begin{table}
\tbl{Details of the data. }{
\begin{tabular}{l|ccc|ccc|ccc|ccc}
\hline
&\multicolumn{3}{|l|}{SXDS ($\sim1240\,{\rm arcmin^2}$)}  &\multicolumn{3}{|l|}{COSMOS ($\sim740\,{\rm arcmin^2}$)}&\multicolumn{3}{|l|}{HDFN ($\sim780\,{\rm arcmin^2}$)}&\multicolumn{3}{|l}{CDFS ($\sim580\,{\rm arcmin^2}$)} \\ 
band & PSF & aperture  & aperture  & PSF & aperture  & aperture& PSF & aperture  & aperture& PSF & aperture  & aperture\\ 
 & ($''$)& diameter ($''$)&correction (mag)& ($''$)& diameter ($''$)&correction (mag)& ($''$)& diameter ($''$)&correction (mag)& ($''$)& diameter ($''$)&correction (mag)\\ \hline
$NB387$ & 0.88 & 2.0 & 0.17 & 0.95 & 2.0 & 0.25& 0.89 & 2.0 & 0.14& 0.85 & 2.0 & 0.13\\ 
$B$ & 0.84 & 2.0 & 0.17 & 0.95 & 2.0 & 0.12& 0.77 & 2.0 & 0.15& 1.0 & 2.0 & 0.20\\ 
$V$ & 0.8 & 2.0 & 0.15 &1.32 & 2.0 & 0.33& 1.24 & 2.0 & 0.20& 0.94 & 2.0 & 0.18\\ 
$R$($r'$) & 0.82 & 2.0 & 0.16 & 1.04 & 2.0 & 0.19& 1.18 & 2.0 & 0.22& 0.83 & 2.0 & 0.16\\ 
$i'$($I$) & 0.8 & 2.0 & 0.16 & 0.95 & 2.0 & 0.12& 0.80 & 2.0 & 0.13& 0.95 & 2.0 & 0.22\\ 
$z'$ & 0.81 & 2.0 & 0.16 & 1.14 & 2.0 & 0.25& 0.81 & 2.0 & 0.15& 1.1 & 2.0 & 0.24\\ 
$J$ & 0.85 & 2.0 & 0.15& 0.79 & 2.0 & 0.3& 0.84 & 2.0 & 0.17& 0.80 & 2.0 & 0.22\\ 
$H$ & 0.85 & 2.0 & 0.15& 0.76 & 2.0 & 0.2& 0.84 & 2.0 & 0.17& 1.5 & 2.0 & 0.55\\ 
$K$($Ks$) & 0.85 & 2.0 & 0.16& 0.75 & 2.0 & 0.2& 0.84 & 2.0 & 0.18& 0.70 & 2.0 & 0.18  \\ 
IRAC ch1 & 1.7 & 3.0 & 0.52  & 1.7 & 3.0 & 0.52& 1.7 & 3.0 & 0.52& 1.7 & 3.0 & 0.52 \\ 
IRAC ch2 & 1.7 & 3.0 & 0.55 & 1.7 & 3.0 & 0.55& 1.7 & 3.0 & 0.55& 1.7 & 3.0 & 0.55\\ \hline 
\end{tabular}
}\label{tbl:data}
\tabnote{Note. The FWHM of PSF, aperture diameter, and aperture correction are shown. The value in parentheses shows the area used in clustering analysis.} 
\end{table}
\end{landscape}

\begin{description}
\item[SXDS fields]
The images used for SED fitting are {\color{black} as follows}: $B, V, R, i'$, and $z'$ images with Subaru/Suprime-Cam from the Subaru/XMM-Newton Deep Survey project \citep[SXDS]{Furusawa2008}; $J, H$, and $K$ images from the data release $8$ of the UKIRT/WFCAM UKIDSS/UDS project \citep[][Almaini et al. in prep.]{Lawrence2007}; Spitzer/IRAC $3.6$~$\mu$m (ch1) and $4.5$~$\mu$m (ch2) images from the Spitzer Large Area Survey with Hyper-Suprime-Cam (SPLASH) project \citep[SPLASH: PI: P. Capak;][]{Laigle2016}. All images are publicly available except the SPLASH data. The aperture corrections for optical and NIR {\color{black}images} are given in \citet{Nakajima2013}. The catalog used to clean IRAC photometry is constructed from the $K$-band image of the UKIDSS/UDS data release 11 (Almaini et al. in prep).
\item[COSMOS field]
We use the publicly available $B, V, r', i'$, and $z'$ images with Subaru/Suprime-Cam by the Cosmic Evolution Survey \citep[COSMOS:][]{Capak2007, Taniguchi2007} and $J, H$, and $Ks$ images with the VISTA/VIRCAM from the first data release of the UltraVISTA survey \citep{McCracken2012}. We also use Spitzer/IRAC ch1 and ch2 images from the SPLASH project. The aperture corrections for the optical images are derived in \citet{Nakajima2013} and those for the NIR images follow \citet{McCracken2012}. The catalog used to clean IRAC photometry is from \citet{Laigle2016}, for which sources have been detected in the z'YJHKs images.

\item[HDFN field]
The images used for SED fitting are: $B, V, R, I$, and $z'$ images with Subaru/Suprime-Cam from the Hubble Deep Field North Survey \citep[HDFN:][]{ Capak2004}; $J$ \citep{Lin2012}, $H$ (Hsu et al. 2017 in prep.), and $Ks$ \citep{Wang2010} images with CFHT/WIRCAm {\color{black}(PI of the $J$ \& $H$ imaging observations: L. Lin)}; Spitzer/IRAC ch1 and ch2 images from the Spitzer Extended Deep Survey \citep[SEDS:][]{Ashby2013}. We use reduced $J$-band and $Ks$-band images given in \citet{Lin2012}. All images are publicly available. The aperture corrections for the optical images are given in \citet{Nakajima2013}. Those of the NIR images with a $2''$ radius aperture are evaluated using bright and isolated point sources in each band. We measure fluxes for $20$ bright point sources in a series of apertures from $2''$ with an interval of $0.''1$ and find that the fluxes level off for $>7.''8$ apertures. We measure the difference in magnitude between the $2''$ and $7.''8$ apertures of $100$ bright and isolated sources and perform Gaussian fitting to the histogram of differences. We adopt the best-fit mean as the aperture correction term. The catalog used to clean IRAC photometry is constructed from the $K$-band image \citep{Wang2010}.

\item[CDFS fields] 
We use the publicly available $B, V, R$, and $I$ images with the MPG $2.2$m telescope/WFI by the Garching-Bonn Deep Survey \citep[GaBoDS:][]{ Hildebrandt2006, Cardamone2010}, the $z'$ image with the CTIO $4$m Blanco telescope/Mosaic-II camera from the MUltiwavelength Survey by Yale-Chile \citep[MUSYC:][]{Taylor2009, Cardamone2010}, the $H$ image with the ESO-NTT telescope/SofI camera by the MUSYC \citep{Moy2003, Cardamone2010}, and the $J$ and $Ks$ images by the Taiwan ECDFS Near-Infrared Survey \citep[TENIS:][]{Hsieh2012}. We also use the Spitzer/IRAC ch1 and ch2 images from the Spitzer IRAC/MUSYC Public Legacy Survey in the Extended CDF-South \citep[SIMPLE:][]{Damen2011}. The aperture corrections for optical and NIR photometry are derived in a similar manner to those in HDFN. The catalog used to clean IRAC photometry is from \citet{Hsieh2012}, for which sources have been detected in the $J$ image.
\end{description}
The FWHM of the PSF, aperture diameters, and aperture corrections are summarized in table \ref{tbl:data}.

%
\section{Clustering Analysis}\label{sec:cl}

\subsection{Subsamples Divided by $NB387$ Magnitude}\label{subsec:subsample}

\begin{figure}[ht]
\includegraphics[width=1.0\linewidth]{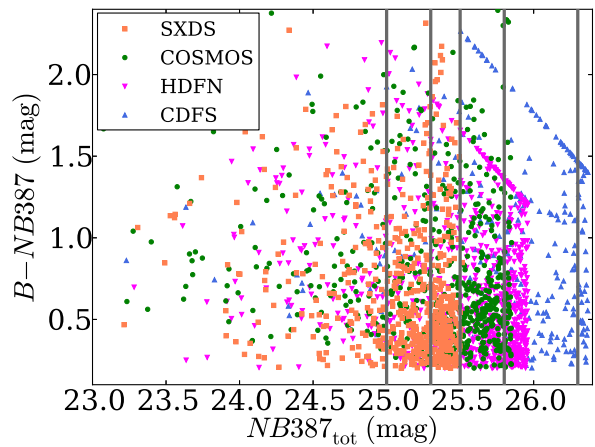}
\caption{
$B-NB387$ ($NB387$ excess) plotted against $NB387$ total magnitude. Orange, green, magenta, and blue points show LAEs in SXDS, COSMOS, HDFN, and CDFS, respectively. LAEs are divided into cumulative subsamples with different limiting magnitudes shown by gray solid lines: $NB387_{\rm tot}{\le}$ $25.0$ mag, $25.3$ mag, $25.5$ mag, $25.8$ mag, and $26.3$ mag. (Color online)
}
\label{fig:NBBNB_NB}
\end{figure}

\begin{table}
\tbl{{\color{black}Number of objects in each subsample. }
}{
\begin{tabular}{lccccc}
\hline
&\multicolumn{5}{c}{$NB387_{\rm tot}$ magnitude limit (mag)} \\
{\color{black}Field} 
& 25.0 &25.3 & 25.5 & 25.8  &26.3 \\
\hline 
SXDS&161 &368 & 601 (93)& - & -\\ 
COSMOS&119 &205 & 297 (21)&526  &- \\ 
HDFN&119 &200 & 299 (56)&588 &-\\
CDFS&27 &41 & 51 (4)&92  &222   \\
\hline
\end{tabular}
}\label{tbl:NBcr}
 \tabnote{Note. The value in parentheses shows the number of objects used for SED fitting. }
 \end{table}
 
The distribution of $B-NB387$ as a function of total $NB387$ magnitude, $NB387_{\rm tot}$, is shown in figure \ref{fig:NBBNB_NB}. {\color{black}To examine} the dependence of halo mass on the total $NB387$ magnitude, we divide our LAE sample of each field {\color{black}in} up to five cumulative subsamples with different limiting magnitudes, as shown in table \ref{tbl:NBcr} and figure \ref{fig:NBBNB_NB}. {\color{black}There are} $1937$ LAEs with {\color{black}$NB387_{\rm tot} \le 26.3$} used in the clustering analysis.

\subsection{Angular Correlation Function}\label{subsec:acf}
\begin{figure*}[ht]
  \begin{center}
    \begin{tabular}{c}
      \begin{minipage}{0.5\hsize}
        \begin{center}
          \includegraphics[width=1.0\linewidth]{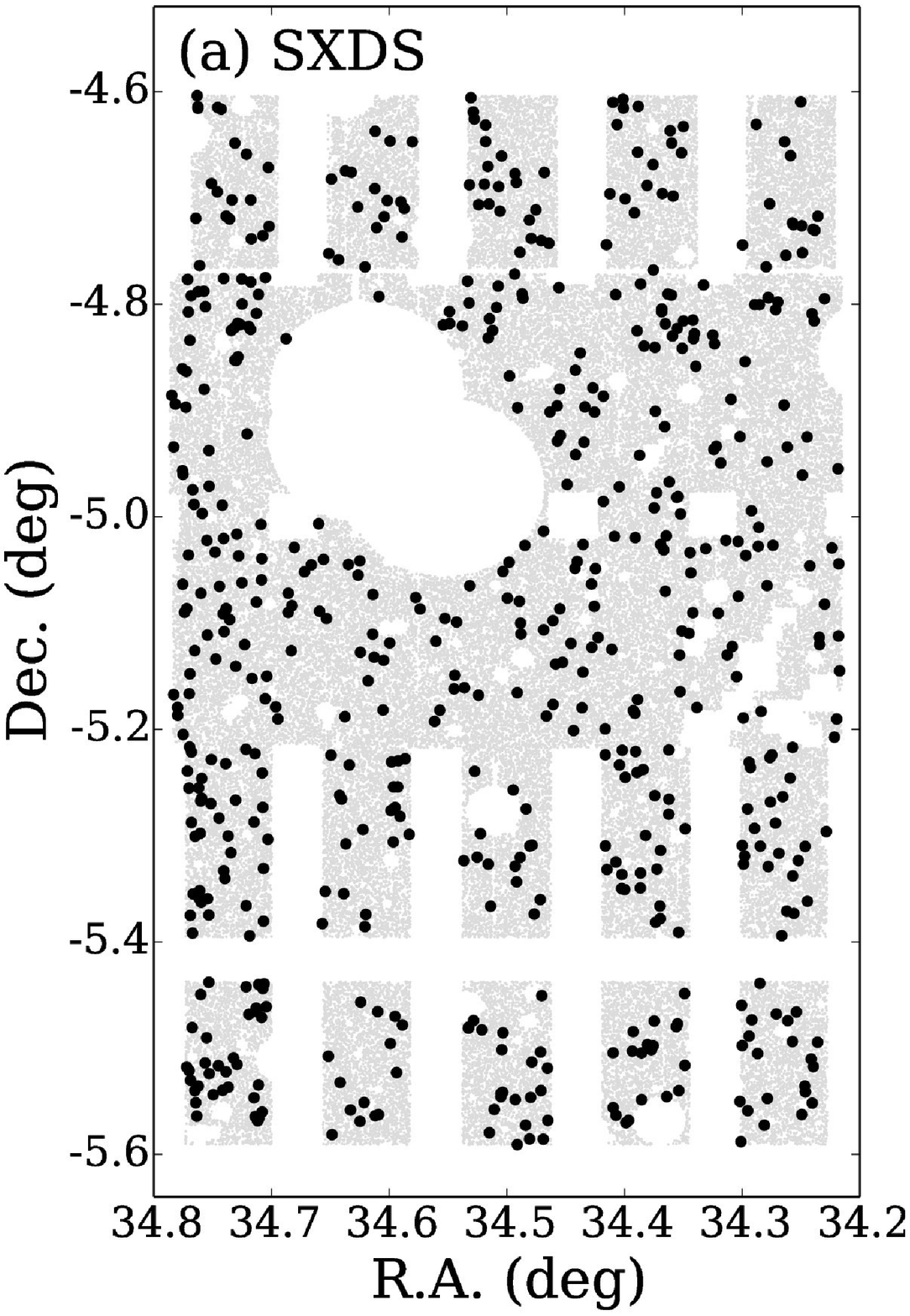}
        \end{center}
      \end{minipage}
      \hfill
      \begin{minipage}{0.5\hsize}
        \begin{center}
           \vspace{2cm}         
     \includegraphics[width=1.0\linewidth]{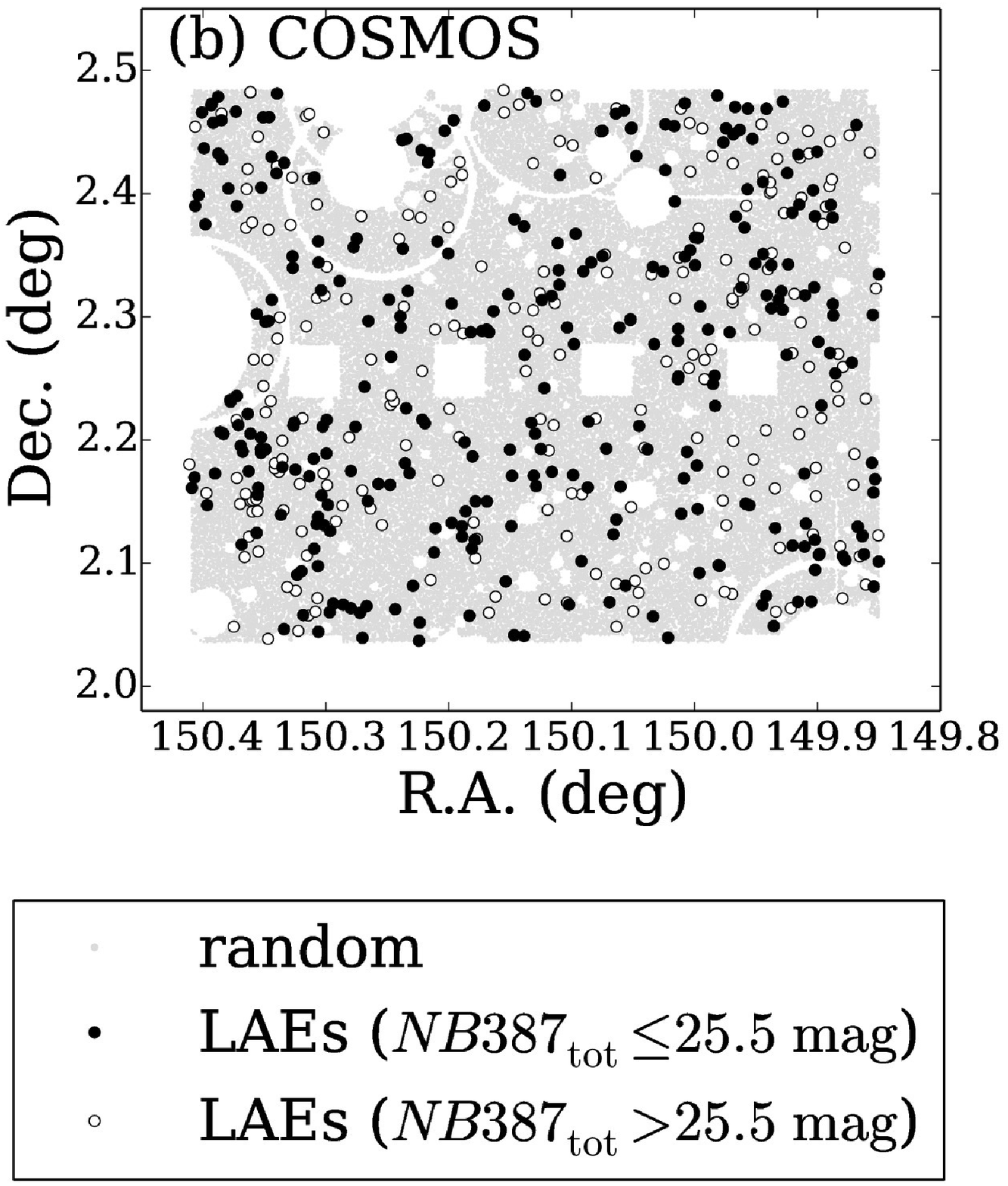}     
        \end{center}
      \end{minipage}\\
      \begin{minipage}{0.55\hsize}
        \begin{center}
               \includegraphics[width=1.0\linewidth]{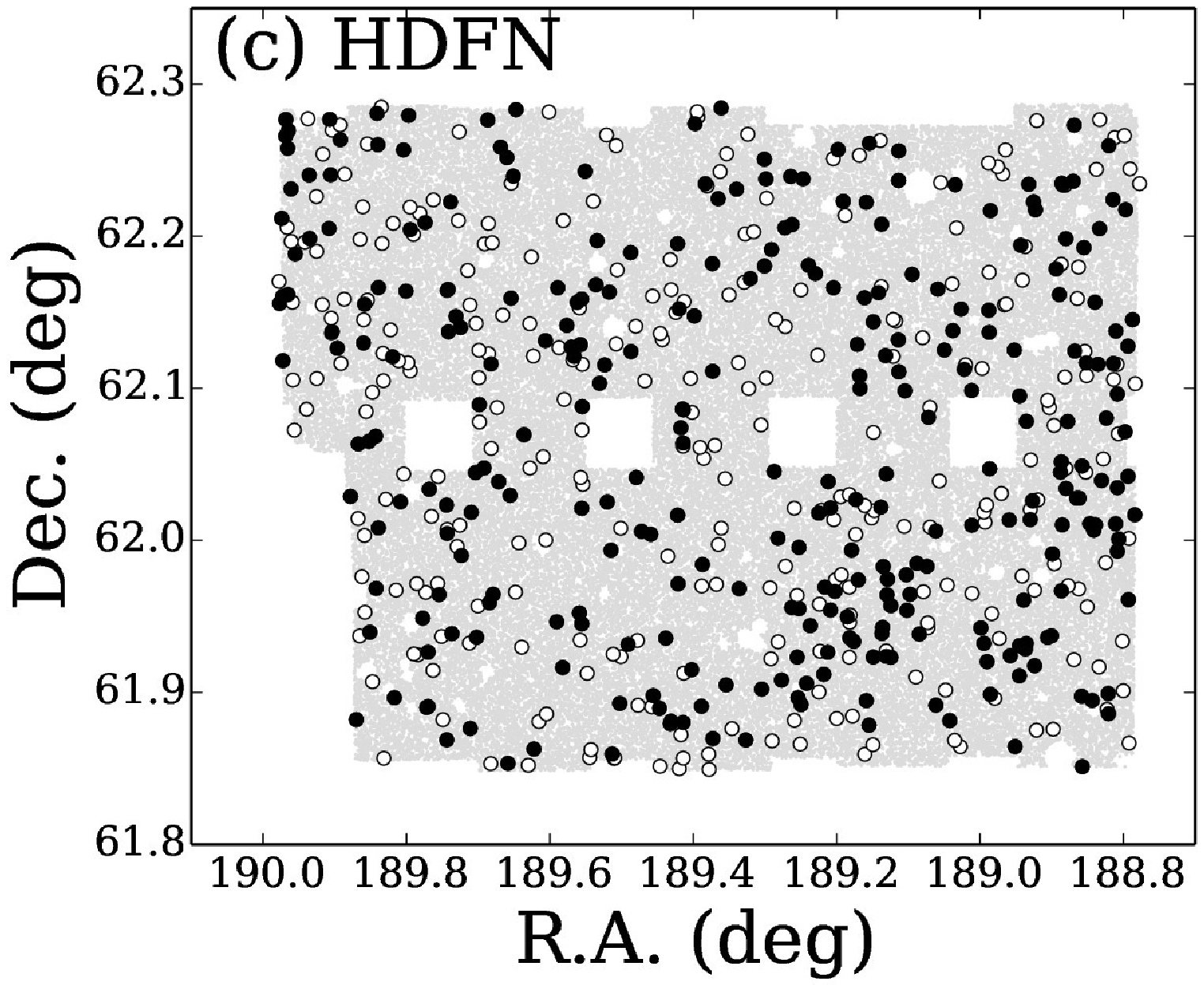}
        \end{center}
      \end{minipage}     
       \hfill
      \begin{minipage}{0.45\hsize}
        \begin{center}
         \includegraphics[width=1.0\linewidth]{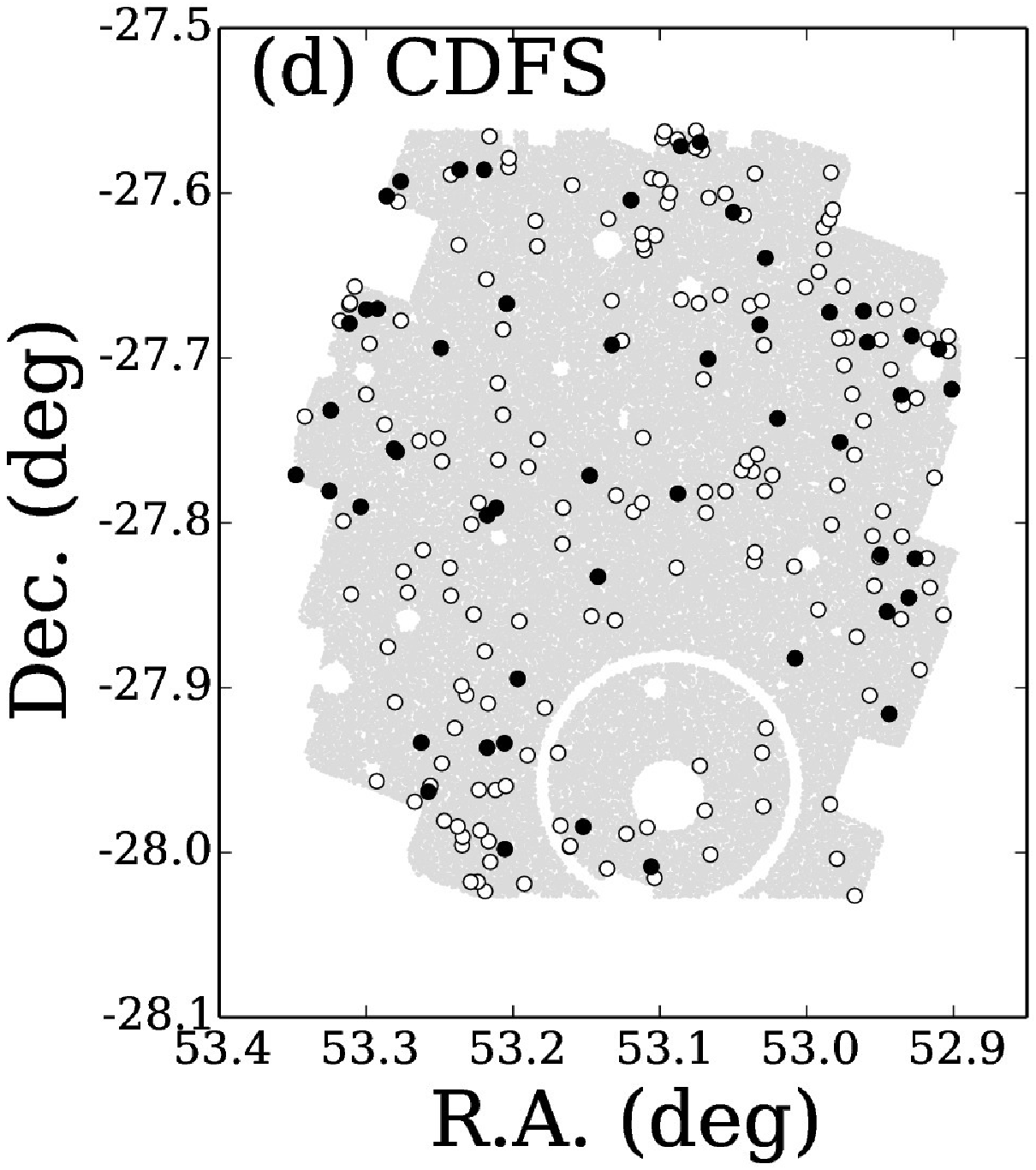}
        \end{center}
      \end{minipage}
  \end{tabular} 
 \end{center}  
  \caption{
Sky distribution of LAEs in SXDS (panel [a]), COSMOS ([b]), HDFN ([c]), and CDFS ([d]). Filled and open black circles represent objects with $NB_{\rm tot}\le25.5\ {\rm mag}$ and $NB_{\rm tot} > 25.5\ {\rm mag}$, respectively. {\color{black} Gray points indicate $100,000$ random sources used in the clustering analysis. Masked regions are shown in white.}  
}
  \label{fig:radec}
\end{figure*}

{\color{black} Angular correlation functions} of our LAEs are derived from clustering analysis. 
The sky distributions of the LAEs in the four fields are shown in figure \ref{fig:radec}
\footnote{In the COSMOS field, \citet[][hereafter M16]{Matthee2016} find an overdense region in their HAE sample at $z=2.231\pm0.016$ {\color{black}{(see their figure 2)}} and a part of their survey region overlaps with that of our LAEs at $z=2.14$--$2.22$. In their overdense region, two X-ray sources at $z=2.219$ and $z=2.232$ have bright \lya emission. The first one is roughly at the center of the overdense region but just outside of our $NB387$ image coverage (ID:$1139$: see figure 2 and table 2 in M16). The second one is included in our coverage but not selected by our color-color criteria probably because its redshift is too large (ID:$1037$). Indeed, {\color{black}{we do not find, by eye inspection, any overdense region in figure \ref{fig:radec}(d) as significant as the one discovered by M16.}}}.
We measure the angular two-point correlation function (ACF), {\wobs}, for a given (sub) sample
using the calculator given in \citet{Landy1993}: 
\begin{equation}
\omega_{\rm obs}(\theta)=\frac{DD(\theta)-2DR(\theta)+RR(\theta)}{RR(\theta)},
\end{equation}
where DD($\theta$), RR($\theta$), and DR($\theta$) are the normalized numbers of galaxy-galaxy, galaxy-random, and random-random pairs, respectively:  
 \begin{equation}
DD(\theta) = \frac{DD_0(\theta)\times2}{N_{\rm D}(N_{\rm D}-1)},
\end{equation}
\begin{equation}
RR(\theta) = \frac{RR_0(\theta)\times2}{N_{\rm R}(N_{\rm R}-1)},
\end{equation}
\begin{equation}
DR(\theta) = \frac{DR_0(\theta)}{N_{\rm D}{\times}N_{\rm R}},
\end{equation}
Here, $N$ is the total number of pairs with subscripts ``D'' and ``R'' indicating galaxies and random points, respectively, and subscript ``0'' indicates the raw number of pairs. We use a random sample composed of $100,000$ sources with the same geometrical constraints as the data sample (see figure \ref{fig:radec}). The $1\,\sigma$ uncertainties in ACF measurements are estimated as: 
\begin{equation}
\color{black}
\Delta\omega_{\rm obs}(\theta)=\frac{1+\omega(\theta)}{\sqrt{DD_0(\theta)}}
\label{eq:e_LS1993}
\end{equation}
following \citet{Guaita2010}. While \citet{Norberg2009} find that Poisson errors underestimate the $1\,\sigma$ uncertainties in ACF measurements and that bootstrapping errors overestimate them $40$\% using a large number of sources ($\sim10^5$--$10^6$), \citet{Khostovan2017arxiv} show that Poisson errors and bootstrapping errors are comparable in the case of a small sample size using $\sim200$ H$\beta$ + \OIII $\ $ emitters at $z\sim3.2$ (see also our footnote \ref{ft:cohn} and figure \ref{fig:cv}(b)). 

We approximate the spatial correlation function of LAEs by a power law: 
\begin{equation}
\xi(r)=\left(\frac{r}{r_0}\right)^{-\gamma},
\end{equation}
where $r$, $r_0$, and $\gamma$ are the spatial separation between two objects in comoving scale, the correlation length, and the slope of the power law, respectively \citep{Totsuji1969, Zehavi2004}. 
We then convert $\xi(r)$ into the ACF, 
$\omega_{\rm model}(\theta)$, following \citet{Simon2007},
and describe it as:
\begin{equation}
\omega_{\rm model}(\theta) 
= C\, \omega_{\rm model,\, 0}(\theta),
\end{equation}
where $\omega_{\rm model,\, 0}(\theta)$ is the ACF in the case of $r_0= 1 \ h^{-1}_{100}{\rm Mpc}$ and $C$ is a normalization constant: 
\begin{equation}
C = \left(\frac{r_0 \ h^{-1}_{100}{\rm Mpc}}{1\ h^{-1}_{100}{\rm Mpc}}\right)^{\gamma}.
\label{eq:C}
\end{equation}

The correlation amplitude of the ACF at $\theta=1''$, $A_{\omega}$, is
\begin{eqnarray}
A_{\omega} &=& C\,\omega_{\rm model,\, 0}(\theta=1'')
\label{eq:A-C}
\end{eqnarray}

An observationally obtained ACF, $\omega_{\rm obs}(\theta)$, includes an offset due to the fact that the measurements are made over a limited area.  
This offset is given by the integral constraint (IC), 
\begin{equation}
\omega(\theta)=\omega_{\rm obs}(\theta)+IC,
\end{equation}
\begin{equation}
IC= \frac{\Sigma_{\theta} RR(\theta)\,C\,\omega_{\rm model,\, 0}(\theta)}{\Sigma_{\theta} RR(\theta)},
\end{equation}
where $\omega(\theta)$ is the true ACF. We fit the $\omega_{\rm model}(\theta)$ to this $\omega(\theta)$ over $\sim40''$ $-1000''$ by minimizing $\chi^2$:
\begin{eqnarray}
\chi^2&=&\Sigma_{\theta}\left( \frac{\omega_{\rm obs}(\theta)+IC-\omega_{\rm model}(\theta)}{\Delta\omega_{\rm obs}(\theta)}  \right)^2\\
&=&\Sigma_{\theta}\left( \frac{\omega_{\rm obs}(\theta)+C\,(IC_0-\omega_{\rm model,0}(\theta))}{\Delta\omega_{\rm obs}(\theta)}  \right)^2,
\end{eqnarray} 
where $IC_0 =IC/C$. 
This $\theta$ range is determined conservatively avoiding the one-halo term at small scales and large sampling noise at large scales.
We fix $\gamma$ to the fiducial value $1.8$ following previous clustering analyses \citep[e.g., ][]{Ouchi2003}. 
The analytic solution of the best-fit correlation amplitude is
\begin{equation}
\Ao =\frac{ \Sigma_{\theta}\left( \frac{\omega_{\rm obs}(\theta)(\omega_{\rm model,0}(\theta)-IC_0)}{\Delta\omega_{\rm obs}(\theta)^2}  \right) }{\Sigma_{\theta}\left( \frac{IC_0-\omega_{\rm model,0}(\theta)}{\Delta\omega_{\rm obs}(\theta)}   \right)^2}\,\omega_{\rm model,\, 0}(\theta=1'').
\end{equation}

\clearpage

\begin{figure*}[ht]
	      \begin{tabular}{l}
        		   \begin{minipage}{0.5\hsize}
      				\includegraphics[width=1.0\linewidth]{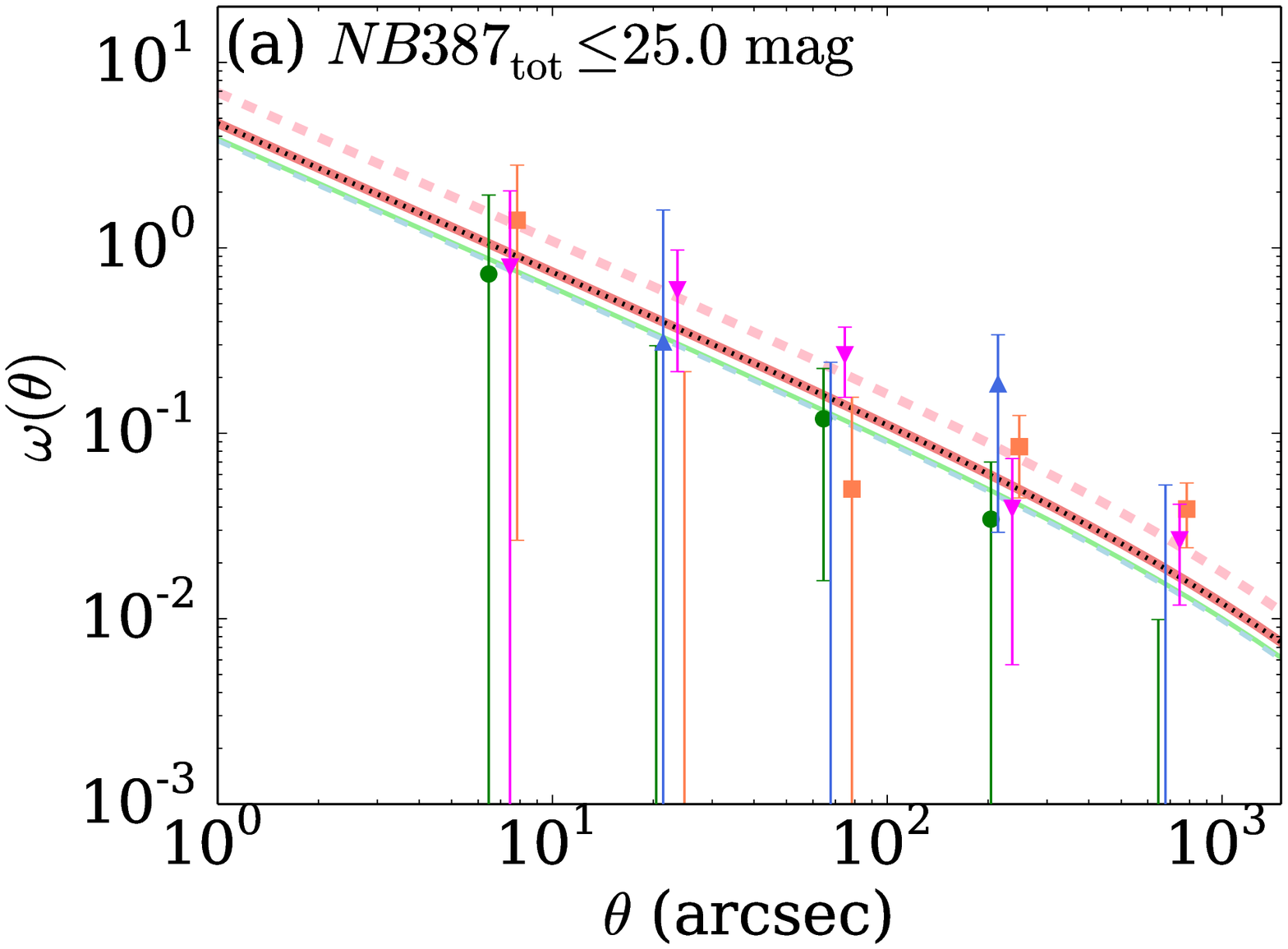}
        		   \end{minipage}
        		   \begin{minipage}{0.5\hsize}
      				\includegraphics[width=1.0\linewidth]{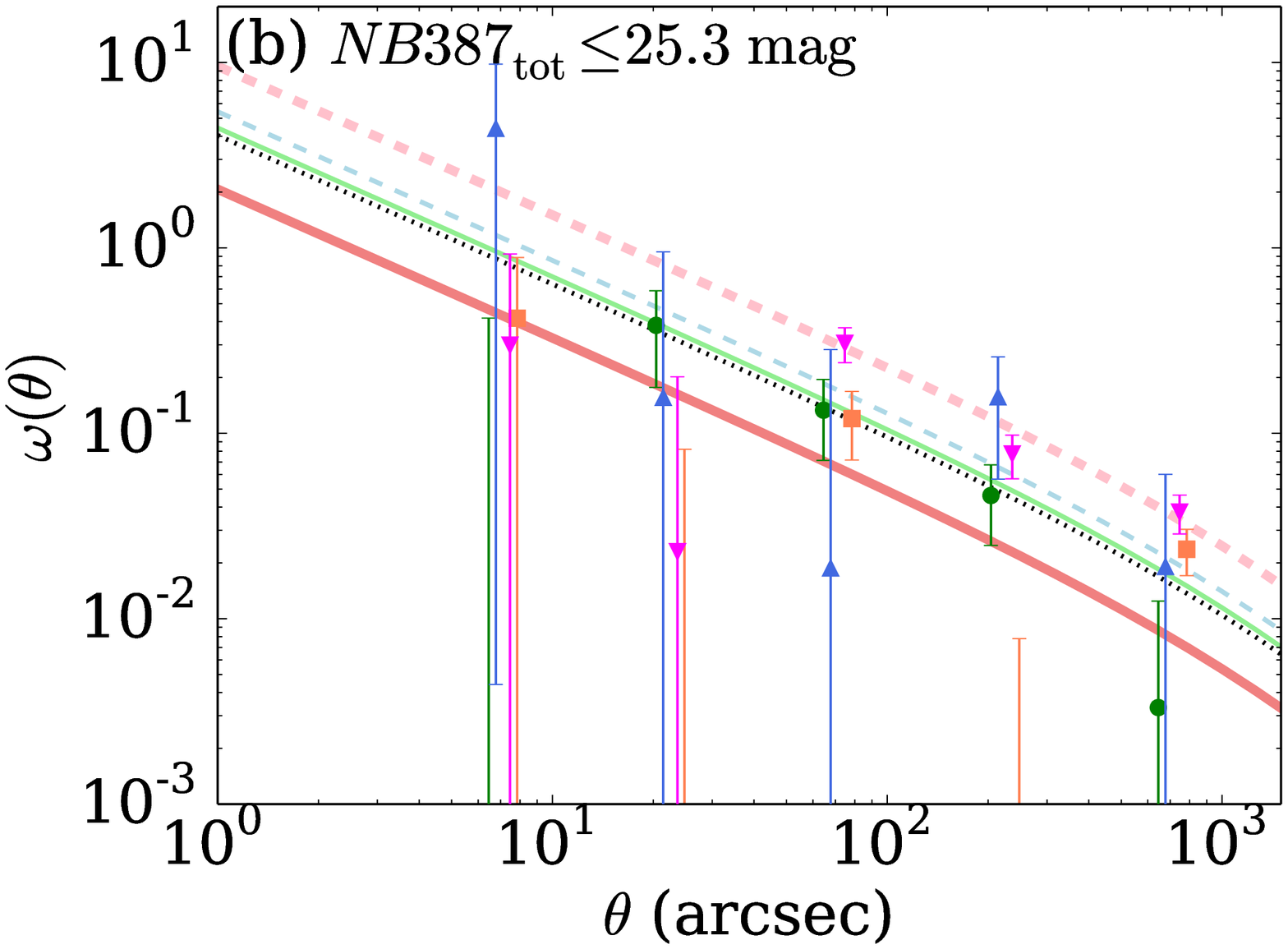}
        		   \end{minipage} \\
        		   \begin{minipage}{0.50\hsize}
                    \includegraphics[width=1.0\linewidth]{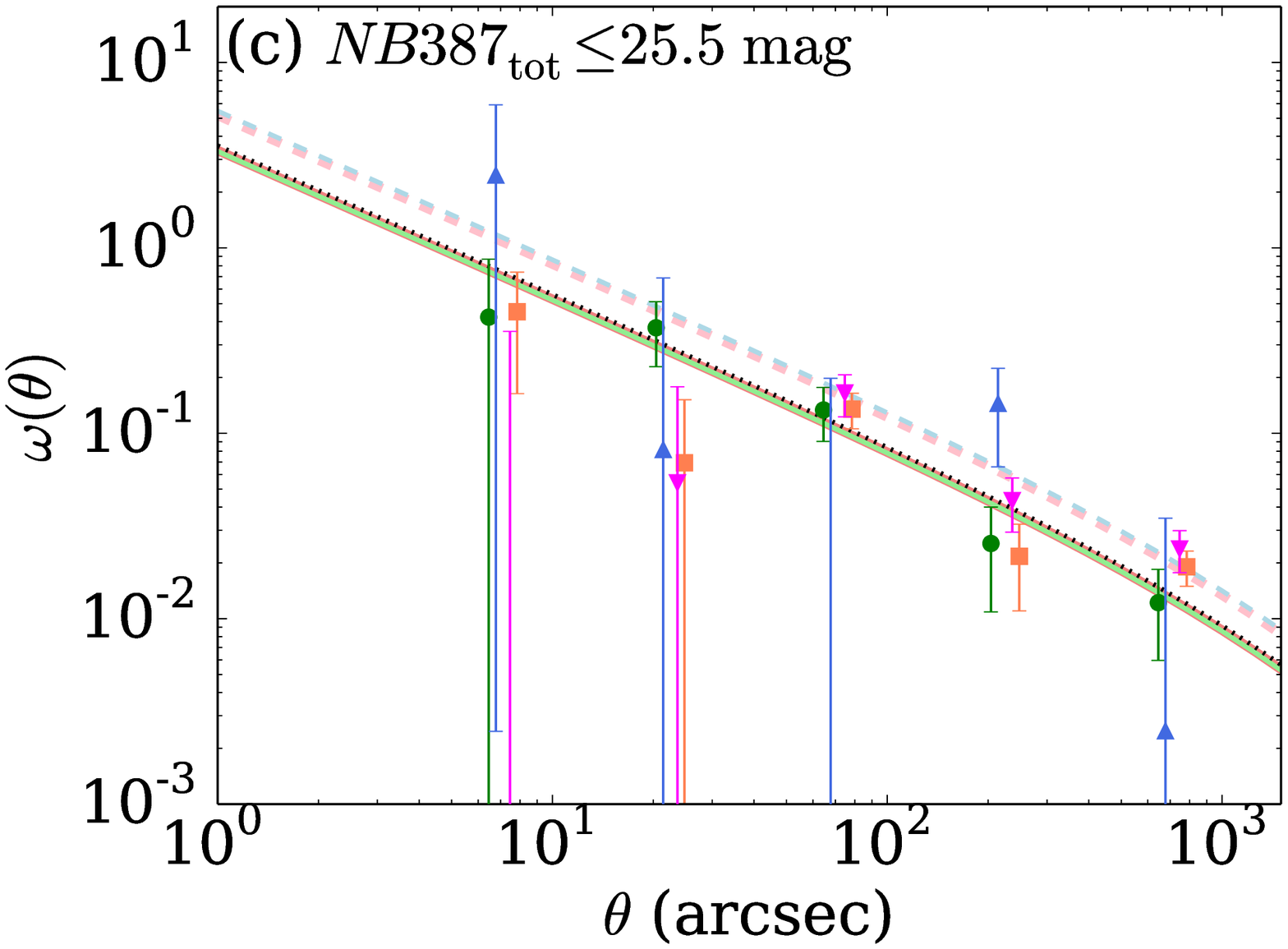}
        		   \end{minipage}
        		   \begin{minipage}{0.50\hsize}
                \includegraphics[width=1.0\linewidth]{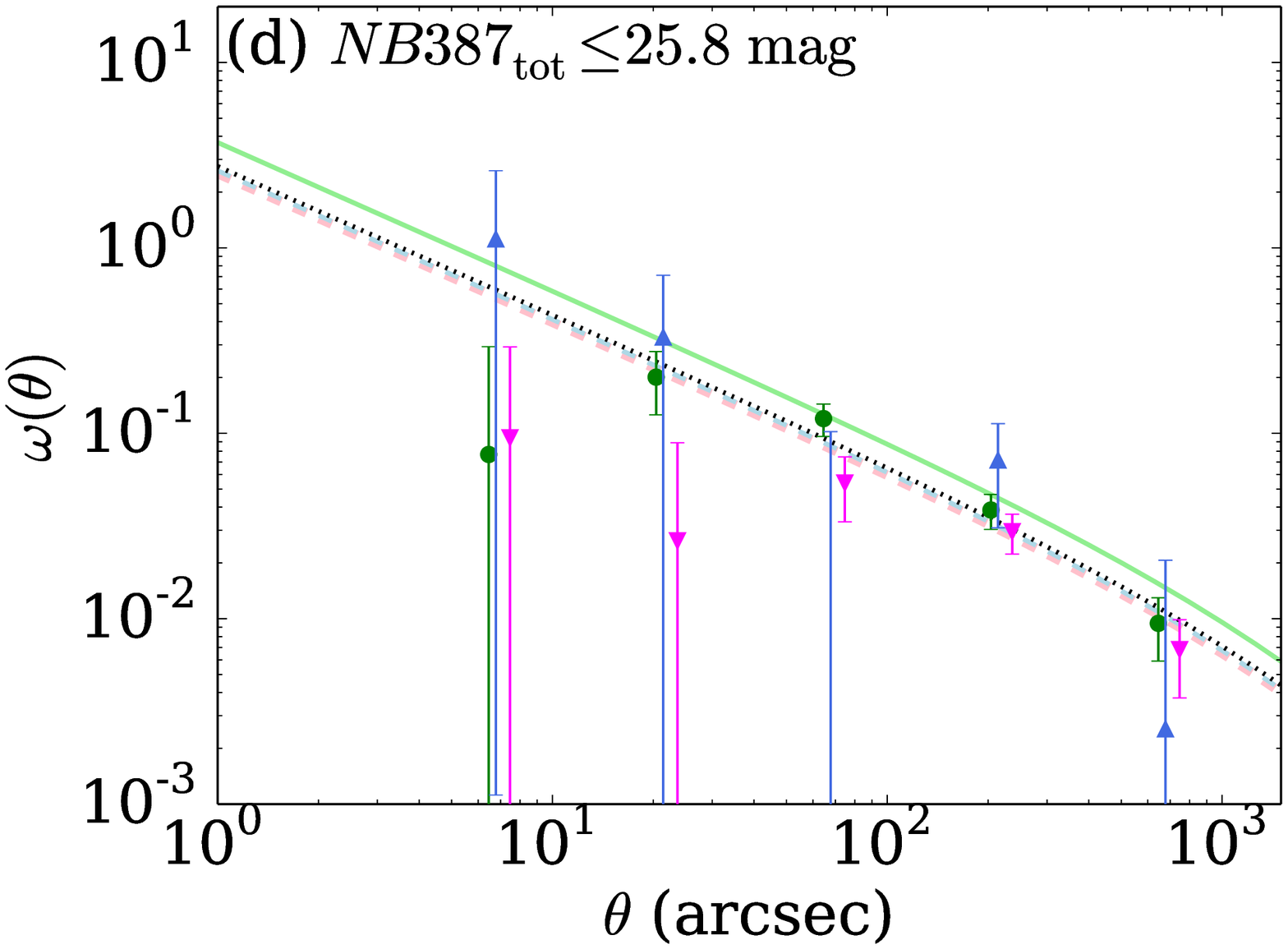}
        		   \end{minipage} \\
        		   \begin{minipage}{0.9\hsize} 
             \begin{flushright}
                \includegraphics[width=1.1\linewidth]{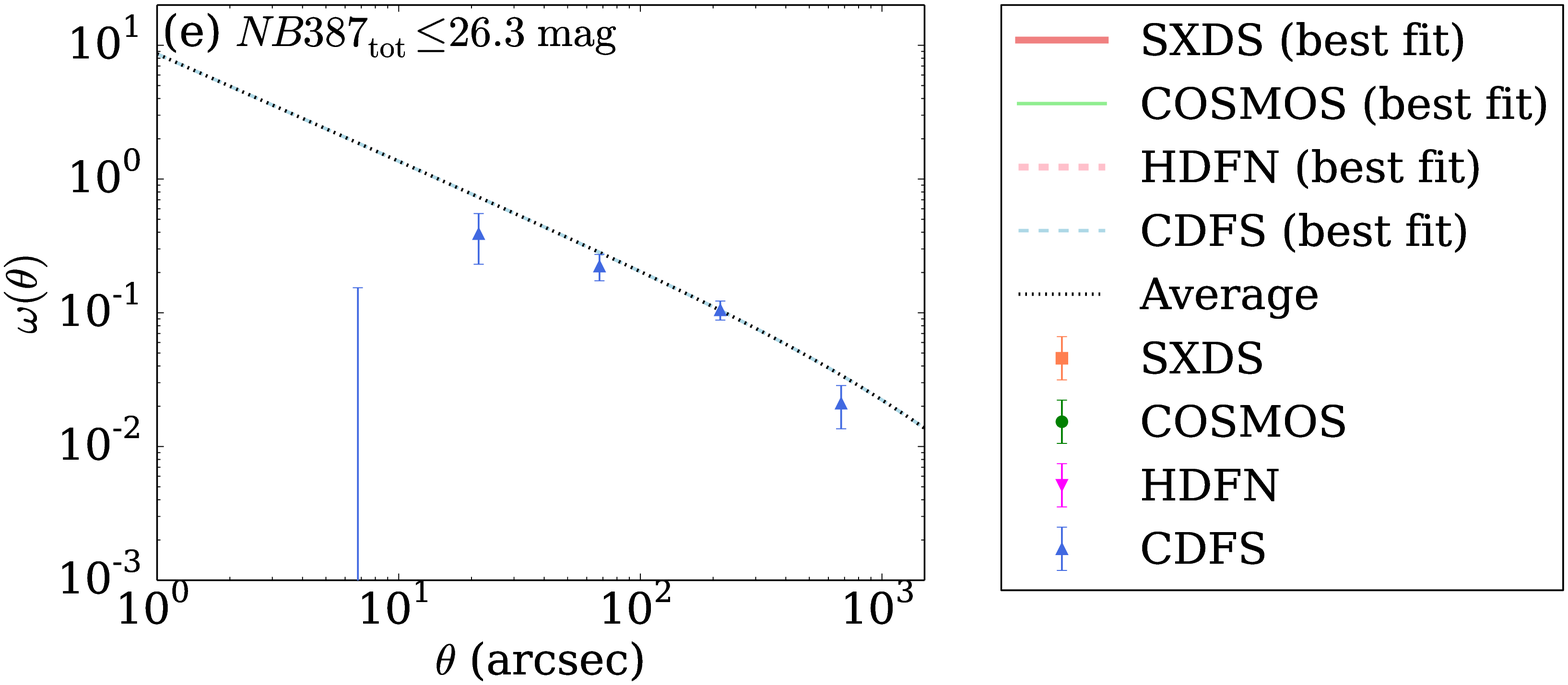}
              				   
              \end{flushright}               
        		   \end{minipage}
               \begin{minipage}{0.1\hsize}  
                 		\end{minipage}
                   
    \end{tabular}
    \caption{
  ACF measurements for LAEs with $NB387_{\rm tot}{\le}25.0$ (panel [a]), $NB387_{\rm tot}{\le}25.3$ ([b]), $NB387_{\rm tot}{\le}25.5$ ([c]), $NB387_{\rm tot}{\le}25.8$ ([d]), and $NB387_{\rm tot}{\le}26.3$ ([e]). For each panel, {\color{black} colored symbols (orange squares, green circles, magenta inverted triangles, and blue triangles)} represent measurements in SXDS, COSMOS, HDFN, and CDFS, respectively. {\color{black}Colored lines, as labeled in the lower right panel,}
indicate the best-fit ACFs with fixed $\beta=0.8$ in SXDS, COSMOS, HDFN, and CDFS, respectively. A {\color{black} dotted} black line shows the average of the best-fit ACFs over the four fields. {\color{black} In panels (a)-(d), we slightly shift all data points along the abscissa by a value depending on the field for presentation purposes.} (Color online)} 
\label{fig:ACF}
\end{figure*}
\clearpage 

The $1\,\sigma$ fitting error in $A_{\omega}$, $\Delta \Ao$, is estimated from $\chi^2_{\rm min} +1$, where $\chi^2_{\rm min}$ is the minimum $\chi^2$ value. 
We also derive, for each limiting magnitude, the field-average correlation amplitude over the four survey fields by minimizing the summation of $\chi^2$ over the four fields:
\begin{eqnarray}
A_{\omega, ave} &=&\frac{ \Sigma_{\theta, i=field}\left( \frac{\omega_{{\rm obs},i}(\theta)(\omega_{\rm model,0}(\theta)-IC_{0,i})}{\Delta\omega_{{\rm obs}, i}(\theta)^2}  \right) }{\Sigma_{\theta, i=field}\left( \frac{IC_{0,i}-\omega_{\rm model,0}(\theta)}{\Delta\omega_{{\rm obs},i}(\theta)}   \right)^2} \omega_{\rm model,\, 0}(\theta=1'').
\label{eq:ave_Aomega}
\end{eqnarray}
The best-fit ACFs are shown in figure \ref{fig:ACF}. 

Contaminations by randomly-distributed foreground and background interlopers dilute the apparent clustering amplitude. The correlation amplitude corrected for randomly distributed interlopers, $\Aocor$, is given by
\begin{equation}
\Aocor = \frac{A_{\omega}}{(1-f_{\rm c})^2},
\end{equation}
where  $f_{\rm c}$ is the contamination fraction. The contamination fraction of our LAEs is estimated to be $10\,\pm\,10$\% ($0$--$20$\%) conservatively from the Monte Carlo simulations and the spectroscopic follow-up observations (see section \ref{subsec:fc}). This $\Aocor$ is the maximum permitted value because interlopers themselves are also clustered in reality. Indeed, some previous clustering studies \citep[e.g.,][]{Khostovan2017arxiv} have not applied any contamination correction. In this study, we apply this equation assuming $f_{\rm c} = 10\, \pm\, 10 \%$ so that the error range in $\Aocor$ include both the no correction case and the maximum correction case. The $1\,\sigma$ error in the contamination-corrected correlation amplitude, $\Delta \Aocor$, is derived by summing the $1\,\sigma$ error in the ACF fitting, $\Delta \Ao$, and the uncertainty in the contamination estimate, $\Delta f_{\rm c}= 0.1$, in quadrature (error propagation): 
\begin{equation}
\frac{\Delta\Aocor }{\Aocor} \simeq \sqrt{\left(\frac{\Delta\Ao}{\Ao}\right)^2 + \left(\frac{2\Delta f_{\rm c}}{f_{\rm c}}\right) ^2  }.
\label{eq:delta_a}
\end{equation}

The value of {\color{black} the contamination-corrected correlation length, $r_{\rm 0,\,corr}$} and its $1\,\sigma$ error are calculated from $\Aocor$ and $\Delta \Aocor$. Table \ref{tbl:acf} summarizes the results of the clustering analysis.

\begin{table*}
\tbl{Clustering Measurements of our LAEs. 
}{
\begin{tabular}{lccccccc}
\hline
Field & $A_{\omega}$ &   $A_{\rm \omega,\, corr}$  &  ${\color{black}r_{\rm 0,\, corr}}$ &  $b_{\rm g,\,eff}$  &$M_{\rm h}$ & reduced $\chi^2_{\nu}$ & $IC$\\
$NB387_{\rm tot}$ (mag) &  &     &  $(\mathrm{h^{-1}_{100}Mpc})$ &   &$(\mathrm{\times10^{10}\ {\rm M_{\odot}} })$ &  & \\
& (1) & (2) & (3) & (4) & (5) & (6) & (7)\\
\hline 
 SXDS\\
${\le}25.0$ &  4.70 $\pm$  2.86 & 5.80 $\pm$  3.75 &$ 2.78^{+ 0.89}_{- 1.22}$& $ 1.40^{+ 0.40}_{- 0.57}$& $10.1^{+28.8}_{-10.1}$& 1.74& 0.0137\\
${\le}25.3$ & 2.07 $\pm$  1.27 & 2.56 $\pm$  1.67 &$ 1.77^{+ 0.57}_{- 0.78}$& $ 0.93^{+ 0.27}_{- 0.38}$& $ 0.4^{+ 3.2}_{- 0.4}$& 5.40& 0.0060\\
${\le}25.5$ & 3.35 $\pm$  0.78 & 4.14 $\pm$  1.33 &$ 2.31^{+ 0.39}_{- 0.45}$& $ 1.18^{+ 0.18}_{- 0.21}$& $ 3.3^{+ 5.2}_{- 2.7}$& 3.02& 0.0097\\
\hline
COSMOS\\
${\le}25.0$ &  3.88 $\pm$  3.03 & 4.79 $\pm$  3.88 &$ 2.50^{+ 0.98}_{- 1.51}$& $ 1.27^{+ 0.44}_{- 0.72}$& $ 5.5^{+25.3}_{- 5.5}$& 0.89& 0.0176\\
${\le}25.3$ &4.44 $\pm$  1.81 & 5.48 $\pm$  2.54 &$ 2.70^{+ 0.64}_{- 0.79}$& $ 1.36^{+ 0.29}_{- 0.36}$& $ 8.5^{+16.6}_{- 7.7}$& 1.11& 0.0201\\
${\le}25.5$ & 3.32 $\pm$  1.25 & 4.10 $\pm$  1.79 &$ 2.29^{+ 0.51}_{- 0.63}$& $ 1.18^{+ 0.23}_{- 0.29}$& $ 3.1^{+ 7.5}_{- 2.9}$& 0.62& 0.0150\\
${\le}25.8$ & 3.70 $\pm$  0.70 & 4.57 $\pm$  1.33 &$ 2.44^{+ 0.37}_{- 0.42}$& $ 1.24^{+ 0.17}_{- 0.20}$& $ 4.7^{+ 6.0}_{- 3.5}$& 0.95& 0.0168\\
\hline
HDFN\\
${\le}25.0$ & 6.89 $\pm$  3.77 & 8.51 $\pm$  5.03 &$ 3.44^{+ 1.01}_{- 1.35}$& $ 1.70^{+ 0.44}_{- 0.61}$& $29.3^{+55.5}_{-27.6}$& 0.81& 0.0319\\
${\le}25.3$ & 9.55 $\pm$  2.28 &11.79 $\pm$  3.84 &$ 4.13^{+ 0.70}_{- 0.81}$& $ 2.00^{+ 0.30}_{- 0.36}$& $62.9^{+52.0}_{-38.3}$& 1.33& 0.0441\\
${\le}25.5$ &5.18 $\pm$  1.51 & 6.40 $\pm$  2.34 &$ 2.94^{+ 0.56}_{- 0.66}$& $ 1.47^{+ 0.25}_{- 0.30}$& $13.6^{+17.7}_{-10.5}$& 0.95& 0.0240\\
${\le}25.8$ &2.52 $\pm$  0.75 & 3.11 $\pm$  1.15 &$ 1.97^{+ 0.38}_{- 0.45}$& $ 1.03^{+ 0.18}_{- 0.21}$& $ 1.0^{+ 2.6}_{- 0.9}$& 1.12& 0.0116\\
\hline
CDFS\\
${\le}25.0$ & 3.78 $\pm$ 11.89 & 4.67 $\pm$ 14.72 &$ 2.47^{+ 2.97}_{- 2.47}$& $ 1.26^{+ 1.30}_{- 1.26}$& $ 5.0^{+170.0}_{- 5.0}$& 0.71& 0.0215\\
${\le}25.3$ &  5.43 $\pm$  8.12 & 6.70 $\pm$ 10.14 &$ 3.02^{+ 2.02}_{- 3.02}$& $ 1.51^{+ 0.88}_{- 1.51}$& $15.5^{+117.8}_{-15.5}$& 0.61& 0.0309\\
${\le}25.5$ & 5.47 $\pm$  6.34 & 6.75 $\pm$  7.97 &$ 3.03^{+ 1.64}_{- 3.03}$& $ 1.51^{+ 0.72}_{- 1.51}$& $15.8^{+85.5}_{-15.8}$& 1.07& 0.0311\\
${\le}25.8$ &2.61 $\pm$  3.43 & 3.22 $\pm$  4.29 &$ 2.01^{+ 1.21}_{- 2.01}$& $ 1.04^{+ 0.55}_{- 1.04}$& $ 1.2^{+20.0}_{- 1.2}$& 0.94& 0.0148\\
${\le}26.3$ & 8.62 $\pm$  1.49 &10.64 $\pm$  2.99 &$ 3.90^{+ 0.58}_{- 0.65}$& $ 1.90^{+ 0.25}_{- 0.29}$& $50.2^{+35.9}_{-28.0}$& 1.66& 0.0490\\
\hline
field average (number of fields)\\
${\le}25.0\ (4)$ &4.69 $\pm$  1.70 & 5.80 $\pm$  2.46 &$ 2.78^{+ 0.60}_{- 0.74}$& $ 1.40^{+ 0.27}_{- 0.34}$& $10.1^{+17.0}_{- 8.8}$& 0.75&   \\
${\le}25.3\ (4)$ & 4.04 $\pm$  0.90 & 4.99 $\pm$  1.57 &$ 2.56^{+ 0.42}_{- 0.48}$& $ 1.30^{+ 0.19}_{- 0.22}$& $ 6.3^{+ 8.3}_{- 4.8}$& 2.04&  \\
${\le}25.5\ (4)$ & 3.55 $\pm$  0.58 & 4.39 $\pm$  1.21 &$ 2.38^{+ 0.34}_{- 0.39}$& $ 1.22^{+ 0.16}_{- 0.18}$& $ 4.0^{+ 5.1}_{- 2.9}$& 1.01& \\
${\le}25.8\ (3)$ & 2.75 $\pm$  0.45 & 3.40 $\pm$  0.94 &$ 2.07^{+ 0.30}_{- 0.34}$& $ 1.07^{+ 0.14}_{- 0.16}$& $ 1.5^{+ 2.4}_{- 1.2}$& 1.08&  \\
${\le}26.3\ (1)$ &8.62 $\pm$  1.49 &10.64 $\pm$  2.99 &$ 3.90^{+ 0.58}_{- 0.65}$& $ 1.90^{+ 0.25}_{- 0.29}$& $50.2^{+35.9}_{-28.0}$& 1.66&  \\
\hline
\end{tabular}
}
\label{tbl:acf}
\tabnote{Note. (1) The best fit correlation amplitude without $f_{\rm c}$ correction; (2) the best fit correlation amplitude with $f_{\rm c}$ correction used to derive (3)--(5); (3) the best fit {\color{black}(contamination-corrected)} correlation length; (4) the best fit effective bias factor {\color{black}(contamination-corrected)}; (5) the best fit effective dark matter halo mass {\color{black}(contamination-corrected)}; (6) reduced chi-squared value; (7) the best fit integral constant; The value in parentheses shows the number of fields used to calculate the field-average correlation amplitude using equation \ref{eq:ave_Aomega}.
}
\end{table*}

\subsection{Bias Factor}\label{subsec:bias}
 \begin{figure*}[t]
	      \begin{tabular}{c}
        		   \begin{minipage}{0.5\hsize}
        			\centering
      \includegraphics[width=1\linewidth]{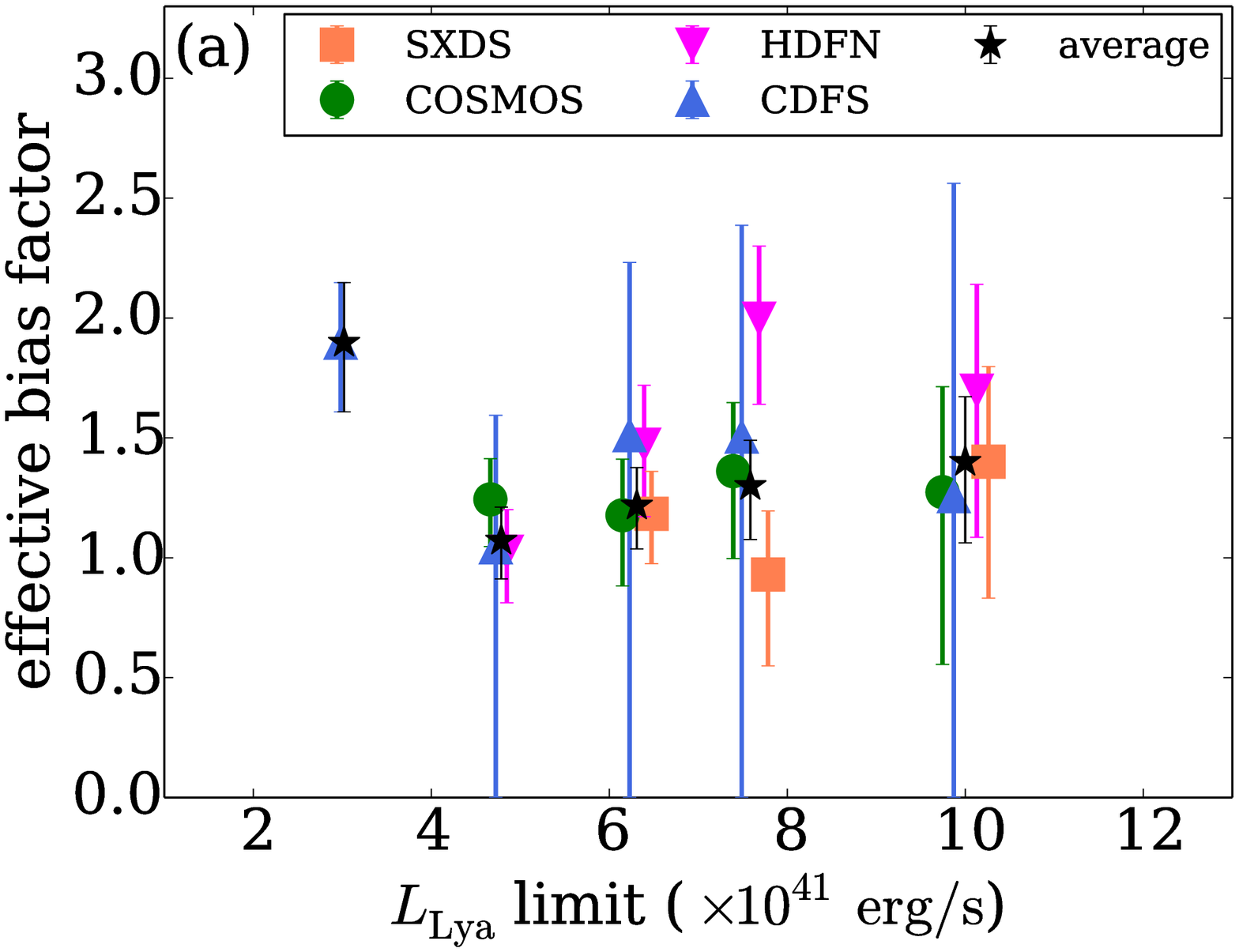}
 				        		   \end{minipage}
        		   \begin{minipage}{0.5\hsize}
        			\centering
      				\includegraphics[width=1\linewidth]{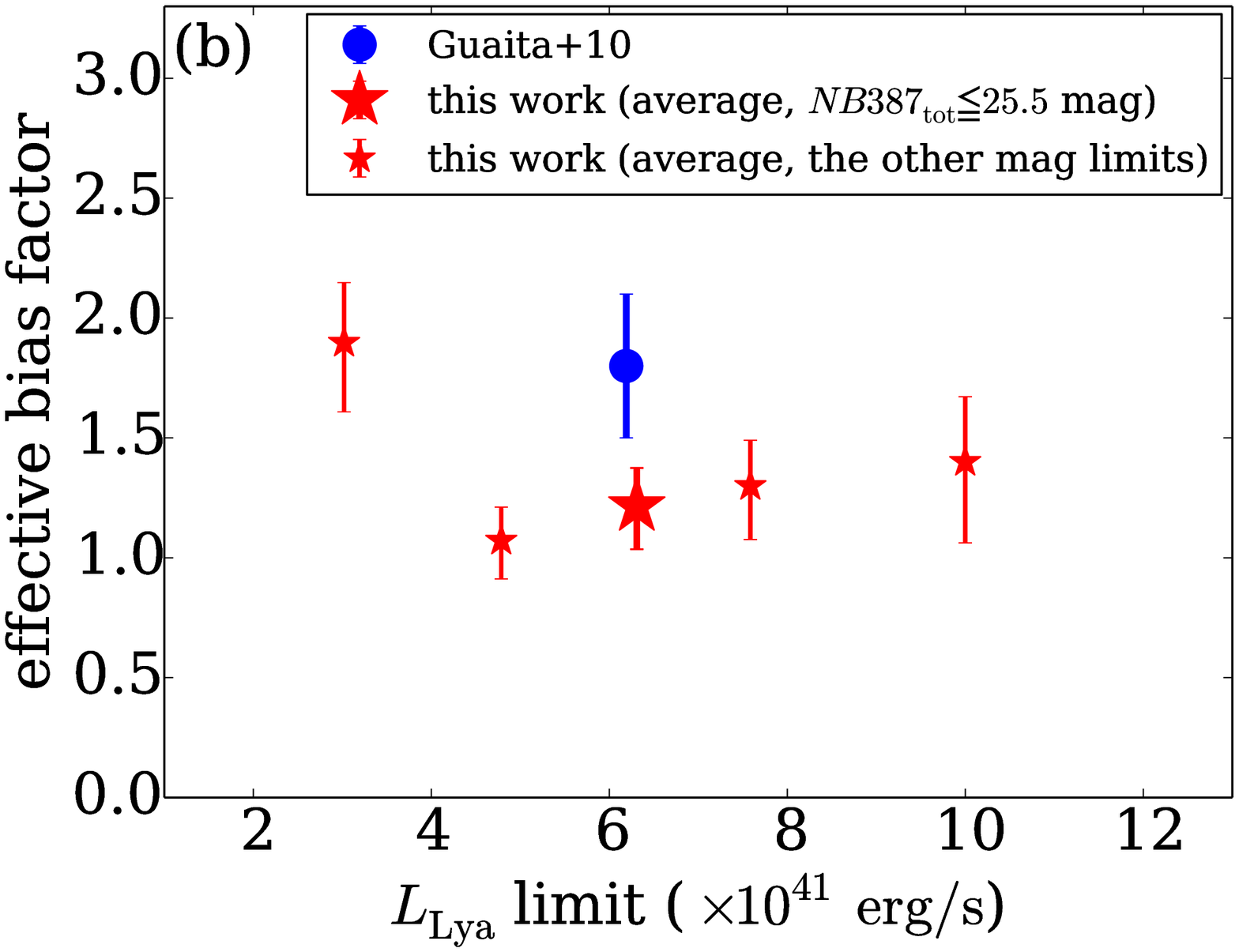}
 				        		   \end{minipage} \\
            \end{tabular}

  \caption{
Bias value plotted against \lya limiting luminosity for the four fields. Panel (a). Orange squares, green circles, magenta inverted triangles, and blue triangles represent the SXDS, COSMOS, HDFN, and CDFS fields, respectively. Black stars indicate the average (weighted mean) over available fields at each limiting luminosity (also shown by red stars in panel (b)). For presentation purposes, we slightly shift all of the points except for black stars along the abscissa.　Panel (b). The measurements shown by small black stars in panel (a) are plotted by small red stars except for the value at \lya$_{\rm limit} \simeq 6\times10^{41}\ {\rm erg\ s^{-1}}$ (or $NB387_{\rm tot} \le 25.5$ mag) shown by a large red star.  \citet{Guaita2010}'s measurement is also plotted by a blue circle. (Color online)
}
  \label{fig:lyalim_bias}
\end{figure*}

The galaxy-matter bias, $b_{\rm g}$, is defined as 
 \begin{equation}
 b_{\rm g}(r)=\sqrt{\frac{\xi(r)}{\xi_{\rm DM}(r,z)}}, 
 \end{equation}
 where $\xi_{\rm DM}(r,z)$ is the spatial correlation function of underlying dark matter, 
\begin{equation}
\xi_{\rm DM}(r,z)= \int \frac{k^2 {dk}}{2\pi^2} \frac{\sin(k{r})}{k{r}}P_{m}(k,z),
\end{equation}
where $P_{m}(k,z)$ is the linear dark matter power spectrum as a function of wave number, $k$, at redshift $z$ \citep{ Eisenstein1999} with the \citet{Eisenstein1998} transfer function. 
We estimate the effective galaxy-matter bias, $b_{\rm g,\, eff}$, at $r=8\, h^{-1}_{100}{\rm Mpc}$ following previous clustering analyses \citep[e.g., ][]{Ouchi2003} using a suite of cosmological codes called Colossus \citep{Diemer2015}.

Figure \ref{fig:lyalim_bias}(a) shows $b_{\rm g,\, eff}$ for the cumulative subsamples in the four fields, where \lya luminosity limits are calculated from the limiting $NB387$ magnitudes of the subsamples. We find that the average bias value of our LAEs (represented by black stars in panel (a) and also by red stars in panel (b)) does not significantly change with the \lya luminosity limit. A possible change in {\color{black}$b_{\rm g,\, eff}$} over $L_{{\rm Ly}\alpha} \simeq 3$--$10 \times 10^{41}\, {\rm erg\ s^{-1}}$ is less than $20$\% since the uncertainties in the average biases are $\sim10$--$20\%$. 

This weak dependence may be partly due to radiative transfer effects on \lya photons. Star forming galaxies in more massive (i.e., larger bias) halos are thought to have higher $SFRs$ and thus brighter nebular emission lines. Indeed, \citet{Cochrane2017} have found a significant positive correlation between H$\alpha$ luminosity and bias for bright $z=2.23$ HAEs, indicating a similarly strong correlation between intrinsic \lya luminosity and bias for bright galaxies. However, such a strong correlation, if any, weakens when observed \lya luminosity is used in place, because brighter (i.e., more massive) galaxies have lower \lya escape fractions, $f_{\rm esc}^{\rm Ly\alpha}$ \citep[e.g.,][]{Vanzella2009, Matthee2016}. 
Indeed, our cumulative subsamples do not show a significant correlation between the observed \lya luminosity and the total $SFR$ (derived from SED fitting in the same manner as described in section \ref{sec:sed}) but rather show a positive correlation between the observed \lya luminosity and the \lya escape fraction, where the intrinsic \lya luminosity is calculated from the total $SFR$ \citep{Brocklehurst1971, Kennicutt1998}.

Moreover, some previous studies have found that high-redshift UV-selected galaxies with comparably faint UV luminosities (\luv) to our LAEs (the average absolute magnitude of our LAEs is $M_{\rm UV}\sim-19$ mag) have weak dependence of $b_{\rm g}$ on UV luminosity \citep[$z\sim3$--$4$ Lyman break galaxies (LBGs):][see however, \citet{Lee2006} who find significant dependence for $z\sim4$--$5$ LBGs]{Ouchi2004b, Ouchi2005b, Harikane2016, Bielby2016}, suggesting that the correlation between intrinsic \lya luminosity and bias is not so strong for typical LAEs with modest \lya luminosities.

The faintest limiting \lya luminosity at which $b_{\rm g,\, eff}$ measurements are available for all four fields is $L_{{\rm Ly}\alpha} = 6.2 \times 10^{41}\, {\rm erg\ s^{-1}}$ (corresponding to $25.5$ mag in $NB387$). 
In order to reduce the uncertainty due to cosmic variance as much as possible, we adopt the average $b_{\rm g,\, eff}$ at this limiting luminosity, $b_{\rm g,\, eff}^{\rm ave}=1.22^{+0.16}_{-0.18}$, as the average $b_{\rm g,\, eff}$ of our entire sample. 

This average bias is lower than that of the previous work on narrow-band-selected LAEs at $z\sim2.1$, $b_{\rm g,\,  eff}=1.8\,\pm\,0.3$ \citep[][see the blue point in panel (b) of figure \ref{fig:lyalim_bias}]{Guaita2010}, with a probability of $96\%$. The median \lya luminosity of their sample is $L_{\rm Ly\alpha}=1.3\times10^{42}\, {\rm erg\  s^{-1}}$ and their $5\sigma$ detection limit in \lya luminosity is $L_{\rm Ly\alpha}=6.3\times10^{41}\,{\rm erg\ s^{-1}}$, which is similar to the luminosity limit of our $NB387\le25.5$ samples. Our clustering method is essentially the same as of \citet{Guaita2010} and in both studies the bias value is calculated at $r=8\, h^{-1}_{100}{\rm Mpc}$. Although we use a slightly different cosmological parameter set, ($\Omega_{\rm m}$, $\Omega_{\Lambda}$, $h$, $\sigma_{8}$)$=$($0.3$, $0.7$, $0.7$, $0.8$), from theirs, ($\Omega_{\rm m}$, $\Omega_{\Lambda}$, $h$, $\sigma_{8}$)$=$($0.26$, $0.74$, $0.7$, $0.8$), using \citet{Guaita2010}'s set changes $b_{\rm g,\, eff}$ only negligibly. Our contamination fraction, $f_{\rm c}=10\,\pm\,10\%$, is comparable to or slightly conservative than theirs, $f_{\rm c}=7\,\pm\,7\%$. The error in \citet{Guaita2010}'s $b_{\rm g,\, eff}$ is a quadrature sum of the uncertainty in $f_{\rm c}$ and the fitting error (statistical error), with the latter dominating because of the small sample size ($250$ objects). As discussed in section \ref{subsec:cvbias}, their high $b_{\rm g, eff}$ value is attributable to cosmic variance since their survey area is approximately one third of ours (see figure \ref{fig:cv}(b)).　Indeed, the sky distribution of their LAEs has a large scale excess at the north-west part and the ACF measurements seem to deviate to higher values from the best-fit power law at large scales because of it
\footnote{We do not include the result of \citet{Guaita2010} when calculating the average bias.}.

\subsection{Cosmic Variance on Bias Factor}\label{subsec:cvbias}

 \begin{figure*}[ht]
	      \begin{tabular}{c}
        		   \begin{minipage}{0.5\hsize}
        			\centering
      \includegraphics[width=1\linewidth]{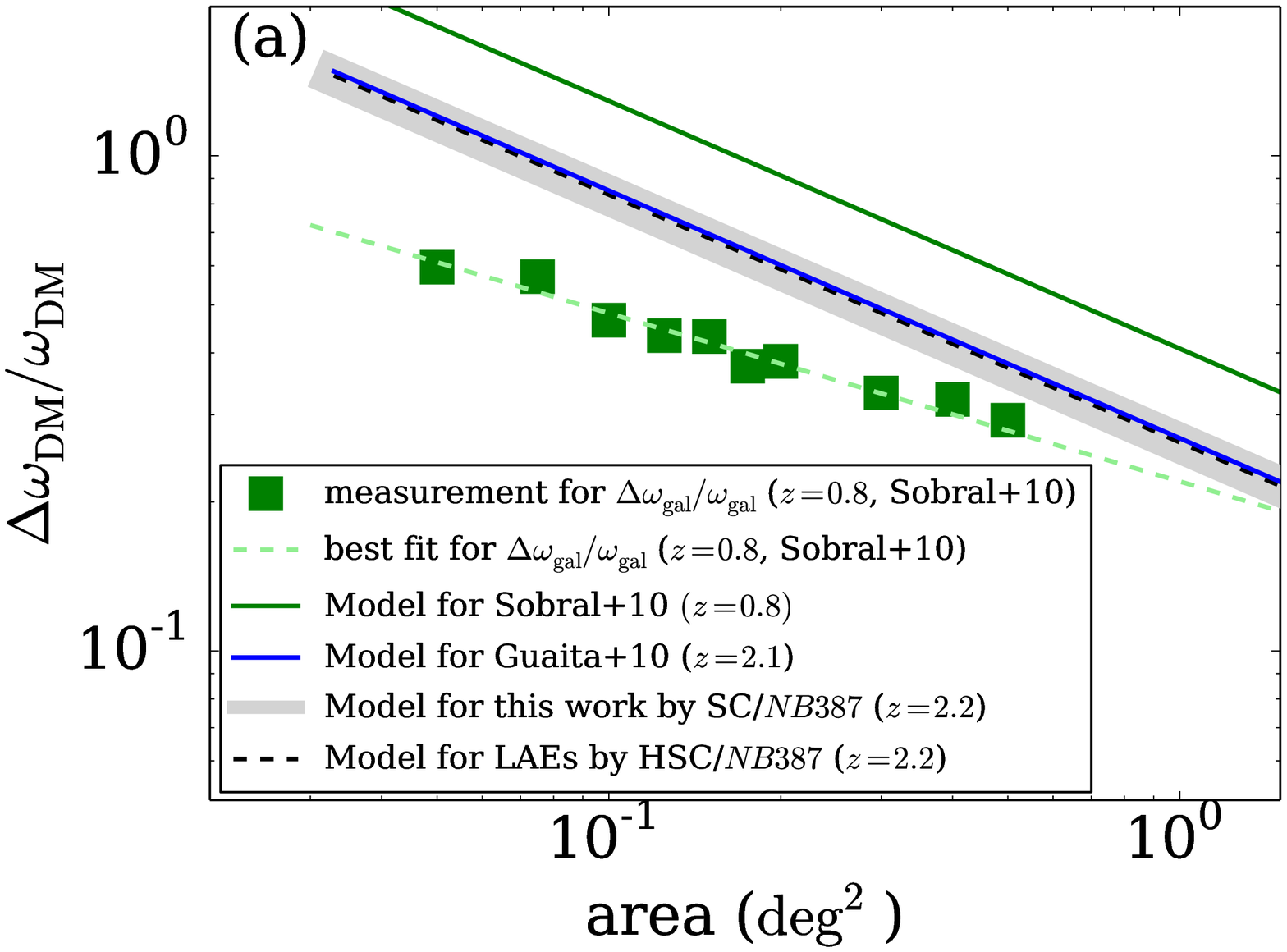}
 				        		   \end{minipage}
        		   \begin{minipage}{0.5\hsize}
        			\centering
      				\includegraphics[width=1\linewidth]{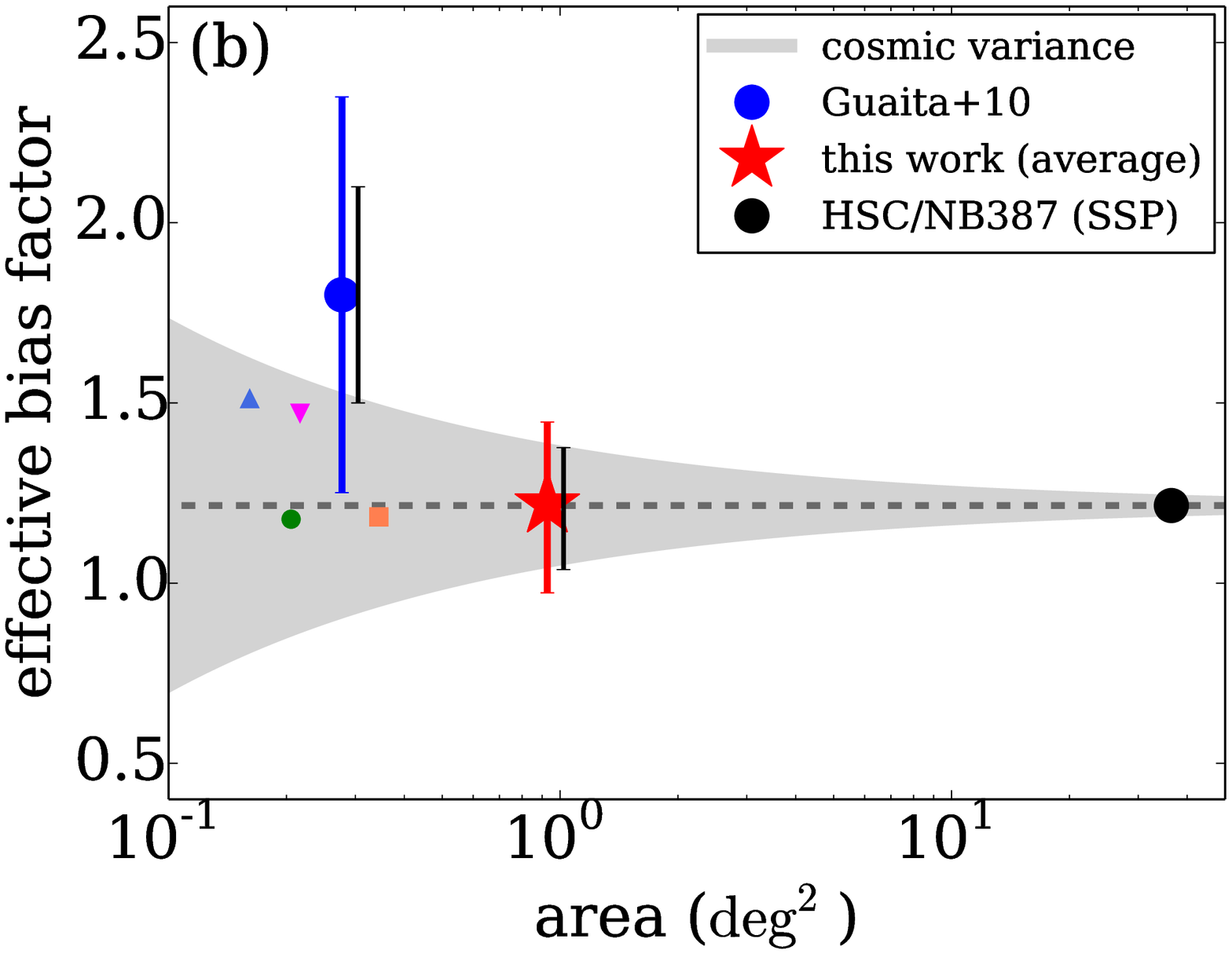}
 				        		   \end{minipage} \\
            \end{tabular}
  \caption{
 {\color{black} Effect of} cosmic variance on clustering analysis. Panel (a). Uncertainties in the amplitude of the dark matter ACF as a function of survey area. Green squares and a light green dashed line denote the empirical measurements at $z\sim0.8$ and the best-fit power law to them, respectively, by \citet[][: $\Delta\omega_{\rm gal}/\omega_{\rm gal}$]{Sobral2010}. Other lines show our analytic calculations for four NB surveys: green solid line for \citet[][]{Sobral2010}, lightgray thick solid line for this study (Suprime-Cam/$NB387$), blue solid line for \citet{Guaita2010}, and black dashed line for an on-going Hyper Suprime-Cam/$NB387$ survey (see section \ref{subsec:HSC}). 
Panel (b). Effective bias factor as a function of survey area. 
The cosmic variance on $b_{\rm g,\, eff}^{\rm ave}$, which is indicated by a light gray thick solid line in panel (a), is shown by a light gray filled region around $b_{\rm g,\, eff}^{\rm ave}$ (fixed) shown by a dim gray dashed line. A red star and a blue circle indicate the $b_{\rm g,\, eff}^{\rm ave}$ in this work and the $b_{\rm g,\, eff}$ in \citet{Guaita2010}, respectively, where colored error bars include the uncertainty due to cosmic variance while black bars next to them do not. {\color{black}A black circle corresponds to the expected HSC/$NB387$ survey area when completed}. A small orange square, green circle, magenta inverted triangle, and blue triangle represent $b_{\rm g,\, eff}$ with $NB387\le25.5$ mag from SXDS, COMOS, HDFN, and CDFS, respectively. (Color online)}
\label{fig:cv}
\end{figure*}

Our average effective bias value and that of \citet{Guaita2010} are not consistent within the $1\,\sigma$ uncertainties in spite of similar limiting Ly$\alpha$ luminosities. Biases derived from limited survey areas possibly suffer from cosmic variance due to spatial variations in the ACF of dark matter. We analytically estimate cosmic variance in the bias value derived from clustering analysis for the first time. With the ACF the galaxy-matter bias can be expressed as
$
b(\theta) = \sqrt{\omega_{\rm gal}(\theta) / \omega_{\rm DM}(\theta)}.
$
Assuming that the cosmic variance in $b$ originates solely from the spatial variation of the dark matter ACF, we can express the $b$ of a given galaxy sample in a given survey field as:
\begin{equation}
b({\rm field}) = \sqrt{ \frac{\omega_{\rm DM}({\rm field})}{\langle\omega_{\rm DM}\rangle} \frac{\omega_{\rm gal}({\rm field})}{\omega_{\rm DM}({\rm field)}}} = \sqrt{ \frac{\omega_{\rm DM}({\rm field})}{\langle\omega_{\rm DM}\rangle}} b_{\rm int},
\end{equation}
where $\langle\omega_{\rm DM}\rangle$ is the cosmic average of the dark matter ACF, $\omega_{\rm DM}({\rm field})$ is the dark matter ACF in the field,  $\omega_{\rm gal}({\rm field})$ is the observed galaxy ACF in the field, and 
\begin{equation}
b_{\rm int} \equiv \sqrt{\frac{\omega_{\rm gal}({\rm field})}{\omega_{\rm DM}({\rm field)}}}
\label{eq:bint}
\end{equation}
is the intrinsic bias of this galaxy population which we assume to be unchanged from field to field (parameter $\theta$ is omitted for clarity). 
This assumption is the same as the one assumed to predict cosmic variance in number density \citep[e.g.,][]{Moster2011}, as explained below. 
Field to field fluctuations of number density, $\sigma_{\rm ND,\, g'}$, are assumed to come from field to field fluctuations of dark matter distribution (i.e., cosmic variance in the density of dark matter), $\sigma_{\rm ND,\, DM}$, as 
\begin{equation}
\sigma_{\rm ND,\, g'}=b_{\rm g'}\ \sigma_{\rm ND,\, DM},
\end{equation}
where the intrinsic galaxy bias, $b_{\rm g'}$, is uniform and independent of fields by definition. We also assume that $\omega_{\rm gal}^2({\rm field})$ is proportional to $\omega_{\rm DM}^2({\rm field)}$ by a factor of $b_{\rm int}$. 

The covariance in $\omega_{\rm DM}$ between two angular separations for area $\Omega_{\rm s}$ is given by the first term of equation 19 of \citet{Cohn2006}\footnote{\citet{Cohn2006}'s equation (19) corresponds to the full covariance including those due to a discrete sampling with a finite number of objects; the second term is proportional to $P_2(K)/N\Omega_{\rm S}$, where $N$ is the number density of objects, and the subsequent terms correspond to the uncertainty shown in our equation \ref{eq:e_LS1993}. Inclusion of the second term in our equation \ref{eq:cv} increases $\Delta \omega_{\rm DM}$ by $\sim 30\%$ for our LAE survey, although in this study we neglect this term and only consider cosmic variance not dependent on $N$.\label{ft:cohn}}:
\begin{equation}
{\rm Cov}(\omega_{\rm DM}(\theta),\ \omega_{\rm DM}(\theta')) = \frac{1}{\pi\Omega_{\rm s}}\int\ {\rm K}\ {\rm dK} J_0({\rm K}\theta)\ _0({\rm K}\theta')\ {\rm P^2_2({\rm K})},
\label{eq:cv}
\end{equation}
where K, $P_2({\rm K})$ and $J_0({\rm K}\theta)$ are the Fourier transform of $\theta$, the projected power spectrum calculated using the redshift distribution defined by the filter, and the zeroth-order Bessel function of the first kind, respectively. With this equation we calculate $\omega_{\rm DM}$ and
its standard deviation, $\sigma_{\rm DM}$, for the three angular bins used to determine the $A_\omega$ of our LAEs. We then fit a power-law correlation function to those values in the same manner as for observed data but also considering the intrinsic covariance given in equation (\ref{eq:cv}),  
and obtain the relative uncertainty in $A_\omega$ due to the variation in $\omega_{\rm DM}$, $\frac{\Delta {\omega_{\rm DM}}}{{\omega_{\rm DM}}}$. According to equation \ref{eq:cv}, the relative uncertainty in $A_\omega$ depends on $\Omega_{\rm s}$ as:
\begin{equation}
\frac{\Delta {\omega_{\rm DM}}}{{\omega_{\rm DM}}} \propto \Omega_{\rm s}^{-0.5}, 
\label{eq:delta_cv}
\end{equation}
as shown by a light gray solid line in figure \ref{fig:cv} (a).

We find $\frac{\Delta {\omega_{\rm DM}}}{{\omega_{\rm DM}}} \simeq 53 \%$ for $\Omega_{\rm s}=0.25$ deg$^2$, a typical area of the four survey fields, and $\simeq 26 \%$ for the entire survey area ($\simeq1$ deg$^2$). 

\citet{Sobral2010} have empirically estimated relative uncertainties in ACF measurements for NB-selected $z=0.85$ HAEs as a function of area by dividing their survey regions, $\simeq1.3$ deg$^2$ in total, into sub regions with different sizes (green squares in figure \ref{fig:cv}(a)). This empirical relation has been used to estimate cosmic variance in ACF measurements in a $\simeq2\,{\rm deg^2}$ survey area of emission line galaxies at $z\sim0.8$--$4.7$ in \citet{Khostovan2017arxiv}. Our analytic method applied to the \citet{Sobral2010} survey with their own NB filter (over the same fitting range of $\theta$ as that for our LAEs for simplicity), however, gives larger uncertainties as shown by a green solid line in figure \ref{fig:cv}(a). This may be partly because the area of \citet{Sobral2010}'s survey is not large enough to catch the total variance. Our analytic estimation seems to be more conservative than theirs. 

We expect that \citet{Guaita2010}'s {\color{black}$b_{\rm g,\ eff}$} obtained from $\sim0.28$ deg$^2$ area has also a $\simeq 51 \%$ uncertainty using their $NB3727$ filter (solid blue line {\color{black} in figure \ref{fig:cv}(a)}). The $1\,\sigma$ uncertainty in an observed bias including cosmic variance, $\Delta b_{\rm g,\ eff,\, CV}$, is given by:
\begin{eqnarray}
\frac{\Delta b_{\rm g,\ eff,\, CV}}{ b_{\rm g,\ eff}} &\simeq& \frac{1}{2}\sqrt{\left(\frac{\Delta\Ao}{\Ao}\right)^2 + \left(\frac{2\Delta{f_c}}{f_c}\right)^2 + \left(\frac{\Delta {\omega_{\rm DM}}}{{\omega_{\rm DM}}}\right)^2  }\\
&\simeq&\frac{1}{2}\sqrt{ \left(\frac{2\Delta{b_{\rm g,\,eff}}}{b_{\rm g,\, eff}}\right)^2+ \left(\frac{\Delta {\omega_{\rm DM}}}{{\omega_{\rm DM}}}\right)^2   }, 
\label{equ:err_b}
\end{eqnarray}
where $\Delta{b_{\rm g,\,eff}}$ is the $1\,\sigma$ error in $b_{\rm g,\,eff}$. 

By updating the errors using this equation (where for our $b_{\rm g,\,eff}$ the plus and minus errors are treated separately), our average effective bias and that of \citet{Guaita2010} are written as $b_{\rm g,\, eff}^{\rm ave}=1.22^{+0.23}_{-0.26}$ and $b_{\rm g,\,  eff}=1.8\,\pm\,0.55$, respectively, thus becoming consistent with each other {\color{black}within the errors} (see figure \ref{fig:cv} (b)). We also note that the relatively large scatter of {\color{black}$b_{\rm g,\ eff}$} among the four fields at each limiting Ly$\alpha$ luminosity seen in figure \ref{fig:lyalim_bias}(a) may be partly due to cosmic variance although the observational errors are too large to confirm it (see figure \ref{fig:cv} (b)). All the best-fit {\color{black}$b_{\rm g,\ eff}$} values for the four fields fall within the $1\sigma$ uncertainty range from cosmic variance shown by a shaded light gray region in figure\ref{fig:cv} (b).

\subsection{Dark Matter Halo Mass}\label{subsec:Mh}
 We estimate {\color{black}the} effective dark matter halo masses from $b_{\rm g,\, eff}$ directly assuming that each halo hosts only one galaxy and that our sample has a narrow range of dark matter halo mass. We use the formula of bias and peak height in the linear density field, $\nu$, given in \citet{Tinker2010}, which is based on a large set of collisionless cosmological simulations in flat $\Lambda$CDM cosmology. The obtained $\nu$ is converted to the effective dark matter halo mass with the top-hat window function and the linear dark matter power spectrum \citep{Eisenstein1998, Eisenstein1999} using a cosmological package for Python called CosmoloPy\footnote{http://roban.github.com/CosmoloPy/}. 

The effective halo mass of each sub-sample is listed in table \ref{tbl:acf}. The field average of effective halo masses corresponding to the field average of effective biases of our LAEs with $NB387_{\rm tot}\leq25.5\ {\rm mag}$, $b_{\rm g,\,  eff}^{\rm ave}=1.22^{+0.16}_{-0.18}$, is $4.0_{-2.9}^{+5.1}\times10^{10}\ {\rm M_{\odot}}$. This value is roughly comparable to previous measurements for $z \sim 3$--$7$ LAEs with similar Ly$\alpha$ luminosities, $M_{\rm h}\simeq10^{10}$--$10^{12}\ {\rm M_{\odot}}$ \citep[e.g.,][]{Ouchi2005b, Ouchi2010, Kovac2007, Gawiser2007, Shioya2009, Bielby2016, Diener2017arXiv, Ouchi2017arXiv}, suggesting that the mass of dark haloes which can host typical LAEs is roughly unchanged with time. 

The average $M_{\rm h}$ of our LAEs is smaller than those of HAEs at $z\sim1.6$ \citep{Kashino2017arxiv}, $M_{\rm h}\sim7\times10^{12}\ {\rm M_{\odot}}$, and at $z\sim2.2$, a few times $10^{12}\ {\rm M_{\odot}}$ \citep{Cochrane2017}. The typical dust-corrected H$\alpha$ luminosity, $L_{\rm H\alpha,\, corr}$, of our LAEs is estimated to be $4.3\pm0.9\times10^{41}\ {\rm erg\ s^{-1}}$ from the $SFR$ obtained by SED fitting in section \ref{sec:sed} using the conversion formula given in \citet{Kennicutt1998} on the assumption of case B recombination. This H$\alpha$ luminosity corresponds to an effective halo mass of $M_{\rm h, eff}=5.2^{+4.8}_{-2.7}\times10^{10}\ {\rm M_{\odot}}$ according to the redshift independent relation between the normalized luminosity $L_{\rm H\alpha,\, corr}/L_{\rm H\alpha}^{\star}(z)$ and $M_{\rm h, eff}$ found by \citet{Cochrane2017}. The estimated halo mass of our LAEs, $M_{\rm h}=4.0_{-2.9}^{+5.1} \times 10^{10}\ {\mathrm M_\odot}$, is thus consistent with this relation. This result supports the result by \citet{Shimakawa2017a} and \citet{Hagen2016} that the stellar properties of LAEs at $z\sim2-3$ do not significantly differ from those of other emission galaxies such as HAEs and \OIII\ emitters. However, \citet{Cochrane2017} assume a constant dust attenuation against H$\alpha$ luminosity, $A_{H\alpha}=1.0\ {\rm mag}$, for all HAEs, which is larger than that of our LAEs, $A_{H\alpha}\sim0.13\pm0.04\ {\rm mag}$, derived from the average $E(B-V)$ in section \ref{sec:sed}. If the (extrapolated) relation overestimates $L_{\rm H\alpha,\, corr}$ at low halo masses owing to overestimation of $A_{\rm H\alpha}$, then the true log-log slope of $L_{\rm H\alpha,\, corr}$ as a function of $M_{\rm h}$ would be steeper, implying that our LAEs would lie above the relation (see also section \ref{subsec:sh_bce} and figure \ref{fig:bce}). 

%

\section{SED fitting}\label{sec:sed}
\begin{table*}[t]
\tbl{Results of SED fitting.  }{
\begin{tabular}{lccccc}
\hline
field & $M_{\star}$ &$E(B-V)_{\star}\ [A_{1600}]$ &Age &$SFR$ &$\chi^2_r$   \\
 &($10^8 {\rm M_{\odot}}$) &(mag) &($10^8$ yr) &(M$_{\odot}$yr$^{-1}$) &    \\
 & (1) & (2) & (3) & (4) & (5)\\
\hline 
SXDS 
& $9.7^{+3.6}_{-1.7}$
& $0.05^{+0.01}_{-0.02}\ [0.6^{+0.1}_{-0.2}]$ 
& $3.6^{+2.8}_{-1.1}$
& $ 3.3^{+ 0.5}_{- 0.7}$
& $0.604$ \\ 
COSMOS 
& $14.0^{+3.4}_{-3.6}$
& $0.07^{+0.02}_{-0.02}\ [0.8^{+0.2}_{-0.2}]$ 
& $4.1^{+2.4}_{-1.8}$
& $ 4.2^{+ 1.2}_{- 0.8}$
& $0.473$ \\ 
HDFN
& $7.6^{+4.0}_{-1.9}$
& $0.06^{+0.02}_{-0.03}\ [0.7^{+0.2}_{-0.4}]$ 
& $3.2^{+4.0}_{-1.4}$
& $ 2.9^{+ 0.8}_{- 0.8}$
& $1.298$ \\ 
CDFS 
& $10.3^{+11.1}_{-9.7}$
& $0.02^{+0.07}_{-0.01}\ [0.2^{+0.8}_{-0.1}]$ 
& $5.7^{+8.6}_{-5.7}$
& $ 2.2^{+534}_{- 0.4}$
& $0.120$ \\ 
\hline Average
& $10.2\pm{\color{black}1.8}$ 
& $0.06\pm{\color{black}0.01}\ [0.6\pm{\color{black}0.1}]$ 
& $3.8\pm{\color{black}0.3}$
& $ 3.4\pm{\color{black}0.4}$ 
& \\ 
\hline
\end{tabular}}\label{tbl:sed_para}
\tabnote{Note. (1) The best fit stellar mass; (2) the best-fit color excess [UV attenuation]; (3) the best fit age;  (4) the best fit SFR; (5) reduced chi-squared value. The UV attenuation is derived from a SMC-like attenuation curve. Metallicity, redshift, and $f_{\rm esc}^{\rm ion}$ are fixed to $0.2Z_{\odot}$, 2.18, and 0.2, respectively.   }
 \end{table*}
 
\begin{figure*}[t]
  \begin{center}  
     \includegraphics[width=1.0\linewidth]{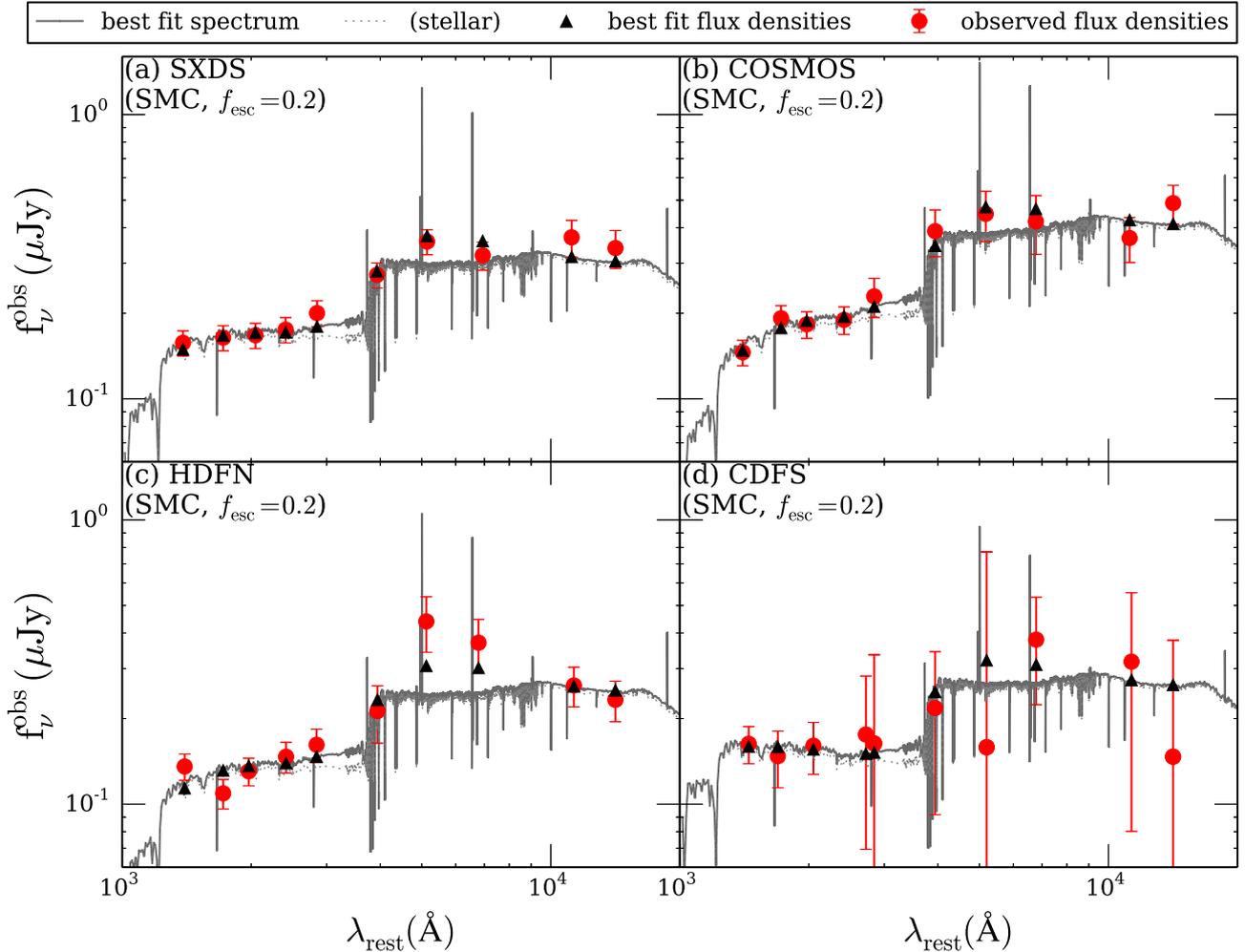}
   \end{center}
  \caption{ 
  Results of SED fitting to stacked LAEs with $NB387_{\rm tot} \le 25.5$ mag in the SXDS, COSMOS, HDFN, and CDFS fields from panels (a) to (d). {\color{black}For each panel, a gray solid line and a light gray dotted line show the best-fit model spectrum and its stellar continuum component, respectively. The difference of these two lines shows a contribution of its nebular continuum component.} Red filled circles and black filled triangles represent the observed flux densities and the flux densities calculated from the best-fit spectrum, respectively. (Color online)
}
  \label{fig:sedfit_smc}
\end{figure*}

We derive {\color{black}parameters that characterize the stellar populations} of LAEs with $NB_{tot}\le 25.5$ mag in each of the four fields by {\color{black}fitting SEDs based on} stacked multiband images. This threshold magnitude is the same as that adopted in the clustering analysis to {\color{black}determine the} average halo masses. We only use $170$ objects ($\sim14$\% of the entire sample, $1248$) {\color{black}that} have data in ten {\color{black}broadband filters} ($B, V, R, i, z, J, H, K, {\rm ch1}$, and ${\rm ch2}$) and {\color{black}are} not contaminated by other objects in the IRAC images (sec. \ref{subsec:selection} and {\color{black}table} \ref{tbl:NBcr}). The procedure to select {\lq}IRAC-clean{\rq} objects is described in the next subsection.

\subsection{Selection of IRAC-clean Objects}\label{subsec:irac}

The IRAC images have {\color{black}lower spatial resolution} (i.e., larger FWHMs of the PSF) compared with {\color{black}images in other bands.} Moreover, they have large-scale residual backgrounds (contaminated {\color{black}sky regions}) around bright objects and in crowded regions due to the extended {\color{black}profile of the IRAC PSF}. {\color{black}Contamination by nearby objects and large-scale sky residuals} can give significant systematic errors in the photometry of stacked images because our LAEs are expected to have very low stellar masses, or very faint IRAC magnitudes. To minimize {\color{black}such contamination}, we select clean LAEs {\color{black}through a two-step process.}

First, we exclude all LAEs which have one or more neighbors. {\color{black}Assuming} that objects bright in IRAC are similarly bright in {\color{black}the $K$ band}, we exclude all LAEs which have one or more $K$-detected {\color{black}objects} with a separation between $0.''85$ and $4.''5$; an object within $0.''85$ separation is considered to be the counterpart to the LAE conservatively (the typical separation is $\sim0.''2${\color{black}; see section \ref{subsec:data}  for the $K$-detected catalogs})\footnote{\color{black}$0.''85$ is the largest PSF FWHM among the $K$ (or $K_s$) bands shown in table \ref{tbl:data}. }. $4''.5$ is $2.5$ times larger than the PSF size of {\color{black}IRAC} ch1.  

Second, we exclude all LAEs with a high sky background {\color{black}as determined} in the following manner. For each field, we randomly select $5,000$ positions with no $K$-band objects within $4.''5$ (i.e., passing the first step) and measure the sky background in an annular region of $3.''5$ radius centered at these positions. We then make a histogram of {\color{black}the} sky background values, which is skewed toward higher values because of {\color{black}contamination} by bright or crowded objects outside of {\color{black}the} $4.''5$ radius.  We fit a Gaussian to the low-flux side (including the peak) of the histogram and obtain its average, $\mu_{\mathrm rand}$, which we consider to be the true sky background. If cutout images at all the random positions are median-stacked, its annular-region sky background will be brighter than $\mu_{\mathrm rand}$. A similar systematic sky-background difference will also be seen when all LAEs are stacked, possibly introducing some systematic errors in photometry. The sky background of the median-stacked random image becomes equal to $\mu_{\mathrm rand}$ if positions whose sky background is higher than a certain threshold, ${\rm sky}_{\mathrm thres}$, are removed, {\color{black}where ${\rm sky}_{\mathrm thres}$ can be determined so that the total number of the remaining positions (i.e., positions with faint sky background below ${\rm sky}_{\mathrm thres}$) is twice as large as the number of positions below $\mu_{\mathrm rand}$.}  Thus, we conservatively remove LAEs with a higher annular-region sky background than ${\rm sky}_{\mathrm thres}$, and are left with $93, 21, 56$, and $4$ IRAC-clean LAEs in SXDS, COSMOS, HSFN and CDFS, respectively. The stacked flux densities of the IRAC-clean LAEs in the $B$ to $K$ bands are mostly consistent with those of the all LAEs before cleaning.

\subsection{Stacking Analysis and Photometry}\label{subsec:stack_photo}
We perform a stacking analysis for each subsample in almost the same manner as \citet{Nakajima2012} and \citet{Kusakabe2015}. {\color{black}Images of size 50$''$ $\times$ 50 $''$} are cut out at the position of LAEs in the $NB387$ image with IRAF/imcopy task.  {\color{black}For each of the $B$ to $K$ bands of the SXDS field}, PSFs are matched to the {\color{black}largest} among the SXDS-Center, North, and South sub-fields using IRAF/gauss task (see {\color{black}table }\ref{tbl:data}). {\color{black}We use the task IRAF/imcombine to create a $NB387$--centered median image.} {\color{black} While} a stacked SED is not necessarily a good representation of individual objects \citep{Vargas2014}, stacking is still useful for our faint objects to obtain {\color{black}a} SED covering rest-frame $\sim1000$--$10000\ $ \AA. 

An aperture flux is measured for each stacked image using the {\color{black}task} PyRAF/phot. Following \citet{Ono2010b}, we use an aperture diameter of $2''$ for the $NB387$, optical, and NIR band images and $3''$ for the MIR (IRAC) images. 
For the $NB387$- to $K$-band images, the inner radius of the annulus to measure the sky flux is set to twice the FWHM of the largest PSF among these images\footnote{The PSF size of the CDFS $H$-band image is exceptionally large and we determine the radius of the annulus for this image independently.}, 
and the area of the annulus is set to five times larger than that of the aperture.  For each of the ch1 and ch2 images, we obtain the net $3''$-aperture flux density of LAEs by subtracting the offset{\color{black},} between the annular-region and {\color{black}the} $3''$-aperture flux densities of the stacked image of IRAC-clean random positions generated in the previous subsection, from the $3''$-aperture flux density of the LAE image (output of the PyRAF/phot task)\footnote{The sky background value on a $3.''5$-radius annulus placed at the image center is consistent between the stacked LAE images and the stacked images of IRAC-clean random positions. For stacked images of random positions, annular-region sky flux densities are brighter than aperture-region sky flux densities with differences corresponding to $\sim7$--$28$\% of the aperture fluxes of median-stacked LAEs.}.

We use the original zero-point magnitudes (ZP) {\color{black}from references given in Section \ref{subsec:data},
} although some previous work argues that some ZPs need to be corrected \citep[e.g.,][]{Yagi2013, Skelton2014}{\color{black}, especially since} the direction of the correction given by \citet{Yagi2013} is opposite to that by \citet{Skelton2014} {\color{black}for} optical bands {\color{black}of} the SXDS field. All aperture magnitudes are corrected for Galactic extinction, $\rm{E(B-V)_b}$, of $0.020$, $0.018$, $0.012$, and $0.008$ for the SXDS, COSMOS, HDFN, and CDFS fields, respectively \citep{Schlegel1998}. 

The aperture magnitudes are then converted into total magnitudes using the aperture correction values summarized in table \ref{tbl:data} (see also section \ref{subsec:data}). The stacked SEDs thus obtained for individual subsamples are shown in figure \ref{fig:sedfit_smc}. The errors include photometric errors and errors in aperture correction and the ZP. For the ch1 and ch2 data, errors in sky subtraction, $\sim0.02$--$0.17$ mag, are also included. The photometric errors are determined following the procedure of \citet{Kusakabe2015}. The aperture correction errors in the $NB387$, optical, and NIR bands are estimated to be less than $0.03$ mag, and those in the ch1 and ch2 bands are set to $0.05$ mag. {\color{black} We adopt $0.1$ mag as the ZP error for all bands, which is the typical value of the offsets of the images used in this paper \citep[e.g.,][]{Yagi2013, Skelton2014} and is twice as large as those adopted in previous studies \citep[e.g.,][]{Nakajima2012}.}

\subsection{SED Models}\label{subsec:sed_model}
We perform SED fitting on the stacked SEDs to derive stellar population parameters in a similar manner to \citet{Kusakabe2015}. Nebular emission (lines and continuum) is added to the stellar population synthesis model of GALAXEV with constant star formation history and 0.2$Z_{\odot}$ stellar metallicity{\color{black},} following previous SED studies of LAEs \citep{Bruzual2003, Ono2010b, Vargas2014}. We assume a SMC{\color{black}-like dust extinction model for} the attenuation curve \citep[hereafter a SMC-like attenuation curve;][]{Gordon2003}, 
which is suggested to be more appropriate for LAEs at $z\sim2$ than the Calzetti curve \citep{Calzetti2000} {\color{black} used by} \citet{Kusakabe2015} and at $z\geq2$ by \citet[][]{Reddy2017arxiv}\footnote{ While \citet{HagenL2017} have found that the SMC indeed has a flatter extinction curve in average than the classical \citep{Pei1992, Gordon2003} curve, we adopt the classical curve which is consistent with recent observations of high-$z$ galaxies including LAEs.\label{ft:smc}
\citet{Reddy2017arxiv} find that galaxies at $z=1.5$--$2.5$ prefer a SMC-like attenuation curve combined with sub-solar metallicity stellar population models. } {\color{black}for star forming galaxies}.
We also examine the {\color{black}case} of the Calzetti {\color{black} attenuation} curve for comparison (see appendix \ref{sec:appendix_cal}). We also assume $\rm{E(B-V)}_{\rm gas}=\rm{E(B-V)}_{\star}$ \citep{Erb2006b}.  
The Lyman continuum escape fraction, $f^{\rm ion}_{\rm esc}$, is fixed to $0.2$ considering recent observations of $f^{\rm ion}_{\rm esc} \sim 0.1$--$0.3$ for $z\sim{3}$ LAEs by \citet{Nestor2013}\footnote{We also perform SED fitting with models without nebular emission, $f^{\rm ion}_{\rm esc}=1$, to examine to what extent $SFRs$ and $M_\star$ change in appendix \ref{sec:appendix_fesc}.}. This means that $80\%$ of ionizing photons produced are converted into nebular emission \citep[see][]{Ono2010b}. 

For each field's stacked SED we search for the best-fitting model SED that minimizes $\chi^2$ and derive the following stellar parameters: stellar mass ($M_{\star}$), color excess ($\rm{E(B-V)}_{\star}$ or UV attenuation of $A_{1600}$), age, and $SFR$. Stellar masses are calculated by solving $\frac{\partial \chi^2}{\partial M_{\star}} = 0$ since it is the amplitude of the model SED. $SFR$ is not a free parameter in the fit but determined from {$M_{\star}$} and age and thus the degree of freedom is $7$. The 1$\sigma$ confidence interval in these stellar parameters is estimated from $\chi^2_{\rm min} +1$, where $\chi^2_{\rm min}$ is the minimum $\chi^2$ value. 

\subsection{Results of SED Fitting}\label{subsec:sed_result}
Table \ref{tbl:sed_para} summarizes the best-fit parameters and figure \ref{fig:sedfit_smc} compares the best-fit SEDs with the observed {\color{black}SEDs} \footnote{\color{black} The uncertainties in the best fit parameters in the CDFS are large since the number of LAEs used in stacking analysis is smaller than those in the other fields as shown in table \ref{tbl:NBcr}. Moreover, the $i$, $z$ and $H$ band images in this field are $\sim0.5$--$2$ mag shallower than those in the other fields.\label{ft:cdfser}}. 
The mean {\color{black}value for each parameter} over the four fields is: {\color{black} $M_{\star}\,=\,10.2\,\pm\,1.8\times10^{8}\ {\rm M_{\odot}}$, $A_{1600}\,=\,0.6\pm0.1\ {\rm mag}$, age$\,=\,3.8\,\pm\,0.3\, \times10^8{\rm\, yr}$, and $SFR\,=\,3.4\pm0.4\ {\rm M_{\odot}}\ {\rm yr^{-1}}$}.  
We discuss the infrared excess and the star formation mode in the following subsections using the results with a SMC-like curve. 

{\color{black}While the SMC-like and Calzetti attenuation curves fit the data equally well,} the resulting parameter values are different (see Appendix \ref{sec:appendix_cal} and figure \ref{fig:sedfit_cal}). 
The Calzetti curve tends to give a smaller stellar mass, a higher attenuation, a younger age, and a higher SFR as the best fit value compared with a SMC-like curve. The difference in the average stellar mass is a factor of $\sim3$ but that in the average SFR reaches a factor of $\sim4$.

\subsubsection{$M_{\star}$--$IRX$ relation}\label{subsubsec:sed_Ms-IRX}
\begin{figure*}[ht]
      \includegraphics[width=1.0\linewidth]{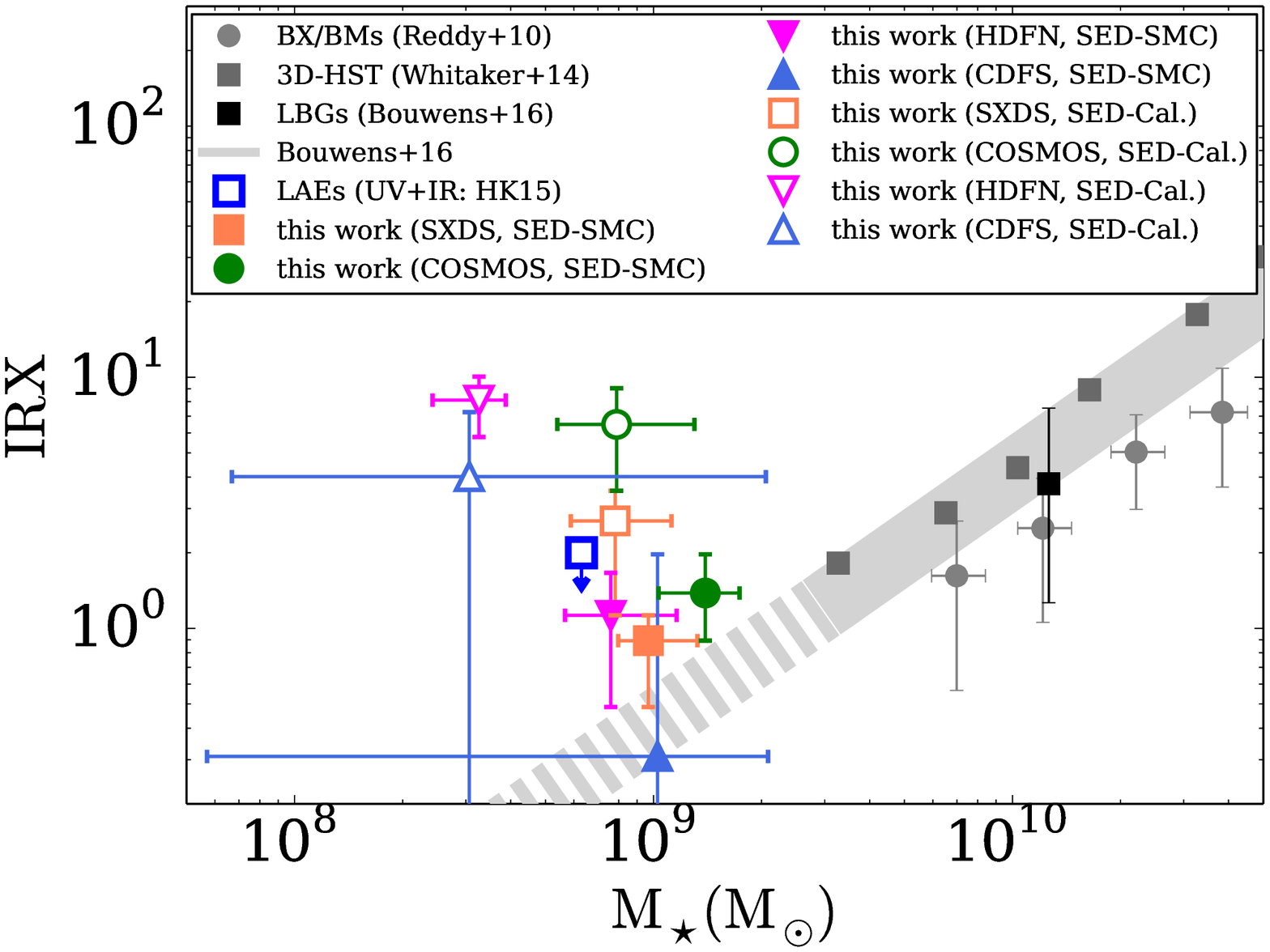}
  \caption{
$IRX$ vs $M_{\star}$. Dim gray squares, dim gray circles, a black square, and a light gray solid band represent, respectively, 3D-HST galaxies at $z\sim2$ in \citet{Whitaker2014}, UV selected galaxies at $z\sim2$ in \citet{Reddy2010}, LBGs at $z\sim2-3$ in \citet{Bouwens2016}, and the consensus relation of them determined by \citet{Bouwens2016}, with its extrapolation indicated by a gray striped band (see also footnote \ref{ft:imf}). A filled {\color{black} (open)} orange square, green circle, magenta inverted triangle, and blue triangle indicate the SXDS, COSMOS, HDFN, and CDFS fields, respectively, on the assumption of a SMC-like attenuation curve {\color{black}(the Calzetti curve)}. {\color{black}An open blue square} represents the 3$\sigma$ upper limit of stacked LAEs at $z\sim2$ with IR observations in \citet[][{\color{black}hereafter HK15}]{Kusakabe2015}. {\color{black}All data are rescaled to a Salpeter IMF according to footnote \ref{ft:imf}}. (Color online)
}
  \label{fig:ms_irx}
\end{figure*}

As shown in figure \ref{fig:ms_irx}, galaxies with higher stellar masses tend to have higher infrared excesses, $IRX \equiv L_{\rm IR}/L_{\rm UV}$, where $L_{\rm IR}$ is the IR luminosity (see also footnote \ref{ft:IRX}), which is an indicator of dustiness \citep[the consensus relation:][]{Reddy2010, Whitaker2014, Bouwens2016}. The dust emission of typical LAEs with $M_{\star} \sim 10^{9}\ {\rm M_{\odot}}$ is too faint to be detected, although a few LAEs at $z\sim2$--$3$ are detected by {\color{black} Herschel/PACS} and {\color{black}Spitzer}/MIPS \citep[e.g.,][]{Pentericci2010, Oteo2012}. 
In order to compare $IRXs$ and stellar masses of LAEs with the consensus relation, we convert the $A_{1600}$ of our LAEs obtained above to $IRXs$ using equation (1) in \citet{Overzier2011a}\footnote{We shift the derived $IRXs$ downward by $10$\% because the $L_{\rm IR}$ of the consensus relation is defined as $L_{\rm IR}\equiv L_{8-1000\mu{m}}$ instead of $L_{\rm IR}\equiv L_{3-1000\mu{m}}$ \label{ft:IRX}}. We find that our LAEs are located near an extrapolation of the consensus relation {\color{black}(see filled color symbols in figure \ref{fig:ms_irx})}. Their $IRX$ values are also consistent with that ($\lesssim2.0$ ($3\,\sigma$)) of typical LAEs obtained by \citet{Kusakabe2015} who constrain the upper limit of the IR luminosity from stacked Spitzer/MIPS $24$ $\mu$m images\footnote{This $IRX$ has also been $10$\% corrected from the original value in \citet[][see our footnote \ref{ft:IRX}]{Kusakabe2015}.}. While unlikely, {\color{black}for our LAEs to require a} Calzetti attenuation curve, they would be dusty galaxies whose {\color{black}values of $IRX$} are more than $10$ times higher than expected from the extrapolated consensus relation {\color{black}(see open colored symbols in figure \ref{fig:ms_irx})} and comparable to those of $10$ times more massive average galaxies.

\clearpage
\begin{figure*}[ht]
     \includegraphics[width=0.93\linewidth]{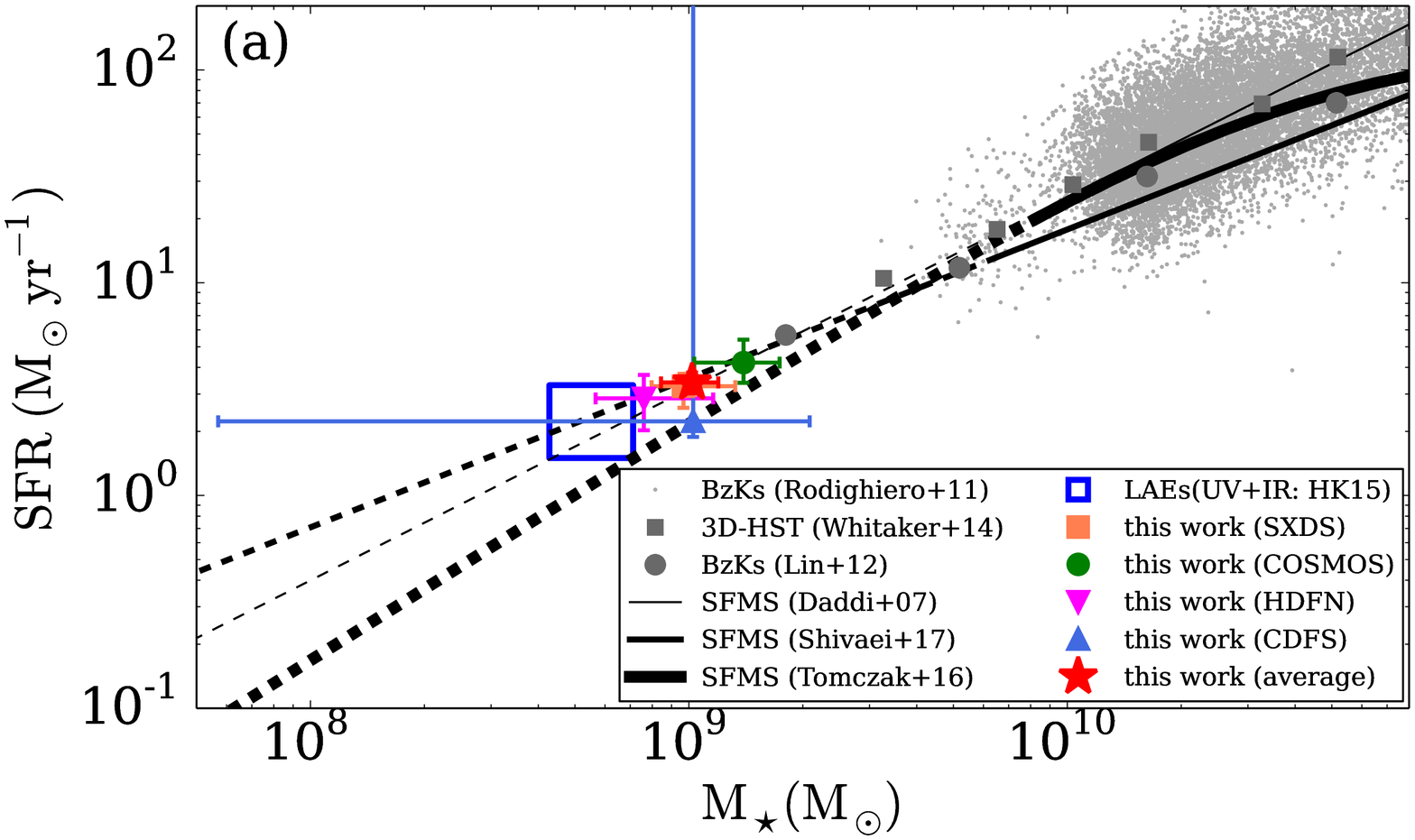}
      \includegraphics[width=0.93\linewidth]{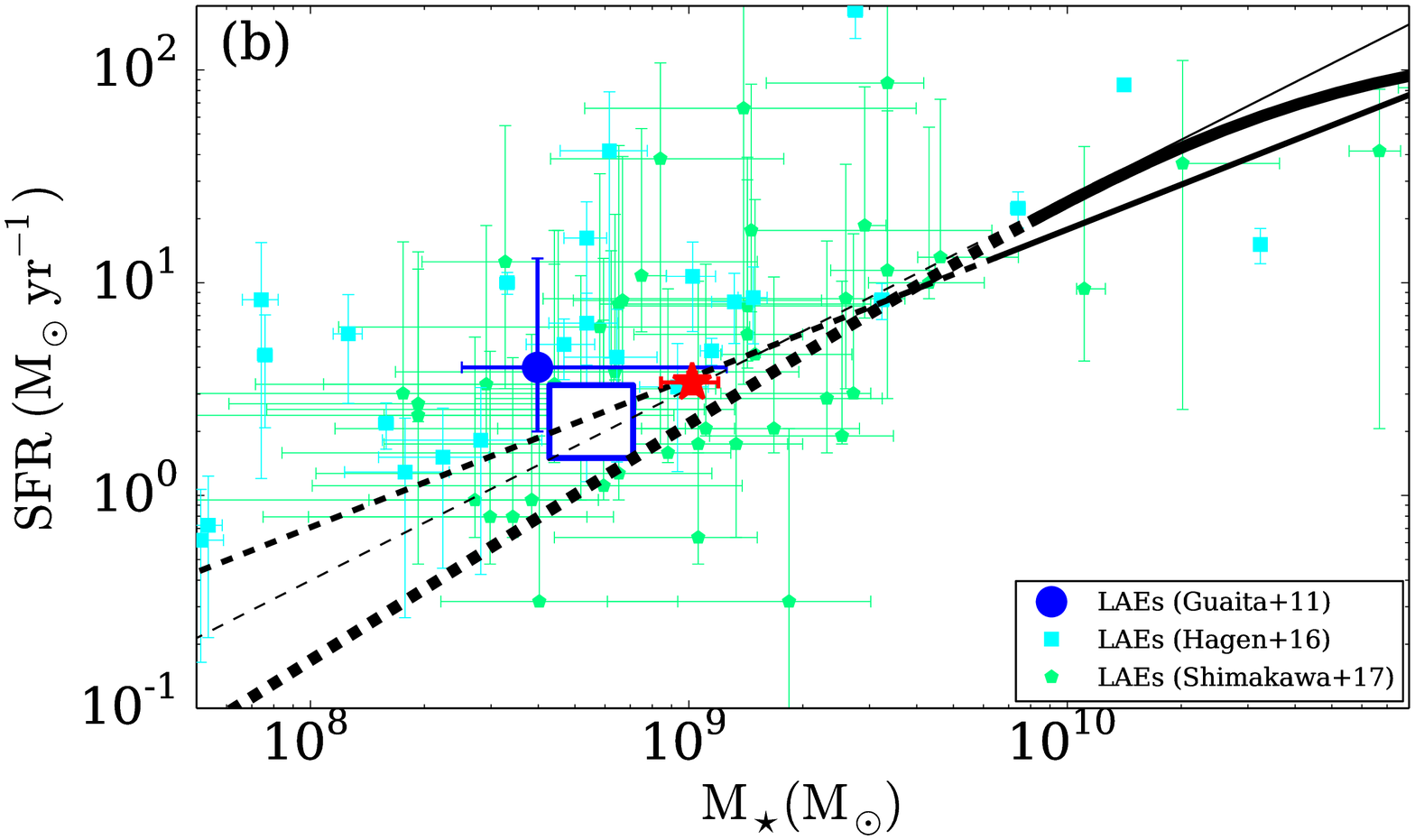} 
  \caption{$SFR$ plotted against $M_\star$. Panel (a). An orange square, green circle, magenta inverted triangle, and blue triangle represent stacked LAEs with $NB387_{\rm tot} \le 25.5$ mag in the SXDS, COSMOS, HDFN, and CDFS fields, respectively, and a red star shows the average over the four fields. The orange square and the red star over lap with each other. A blue open rectangle denotes the permitted range for stacked LAEs from $L_{\rm UV}$ and $L_{\rm IR}$ in \citet{Kusakabe2015}. Light gray dots, dim gray squares, and dim gray circles indicate BzKs from \citet{Rodighiero2011}, BzKs from \citet{Lin2012}, and 3D-HST galaxies from \citet{Whitaker2014}, respectively. Black thin middle-width, and thick solid lines represent the star formation main sequence at $z \sim 2$ in \citet[][hererafter T16]{Tomczak2016}, \citet[][hereafter S17]{Shivaei2017}, and \citet{Daddi2007}, respectively (determined well using $L_{\rm UV}$ and $L_{\rm IR}$), with extrapolated parts shown by dashed lines. (b) Same as panel (a) but LAEs taken from the literature are also plotted. Cyan squares and light green pentagons show individual LAEs at $z \sim 2$ in \citet{Hagen2016} and \citet{Shimakawa2017a}, respectively. A blue circle indicates stacked LAEs at $z \sim 2$ in \citet{Guaita2011}. $SFRs$ in \citet{Hagen2016} and \citet{Shimakawa2017a} are derived from the $IRX-\beta$ relation with the Calzetti curve \citep{Meurer1999} and $SFRs$ in \citet{Guaita2011} are derived from SED fitting with the Calzetti curve, while $SFRs$ in this work are derived from SED fitting with a SMC-like curve. We also show our results with the $IRX-\beta$ and SED fitting with the Calzetti curve in figure \ref{fig:ms_sfr_cal}. {\color{black}All data are rescaled to a Salpeter IMF according to footnote \ref{ft:imf}}. (Color online)}
  \label{fig:ms_sfr_smc}
\end{figure*}
\clearpage

\subsubsection{$M_{\star}$--$SFR$ Relation}\label{subsubsec:sed_Ms-SFR}
The mode of star formation in star-forming galaxies can be divided into two {\color{black}categories}: the main-sequence (MS) mode where galaxies form stars at moderate rates, making a well-defined sequence in the $SFR$-$M_\star$ plane \citep[SFMS; e.g., ][]{Elbaz2007, Speagle2014a}, and the burst mode where galaxies have much higher {\color{black} specific star formation rates, $sSFRs(=SFR/M_\star)$,} than MS galaxies with similar masses \citep[e.g., ][]{Rodighiero2011}. While it is well established that LAEs are mostly low-mass galaxies, which mode they typically have is still under some debate because of differences in $SFR$ estimates.

The SFMS itself at $z \sim 2$ has been determined well using rest UV to far-infrared (FIR) data at $M_{\star}\gtrsim10^{10}\ {\rm M_{\odot}}$ \citep[e.g.,][]{Whitaker2014, Tomczak2016}. Below this stellar mass, the SFMS is suggested to continue at least down to $M_{\star}\sim 10^{8}$--$10^{9}\ {\rm M_{\odot}}$ keeping its power-law slope unchanged \citep[e.g., by][using gravitationally-lensed galaxies in the HST Frontier Fields]{Saintini2017arxiv}, {\color{black}although SFRs have large uncertainties since without FIR data}. In this paper{\color{black}, we simply extrapolate the SFMS, given in the literatures \citep{Daddi2007,Tomczak2016,Shivaei2017} towards lower masses without changing the power-law slope.}

Figure \ref{fig:ms_sfr_smc}(b) show previous results for LAEs at $z\sim2-2.5$. \citet{Hagen2016} have found that bright individually detected LAEs lie along or above the SFMS, while \citet{Shimakawa2017a} have found that fainter, individually detected LAEs lie on the SFMS. \citet{Guaita2010}'s estimates based on stacking analysis have too large errors to distinguish the star formation mode although they are consistent with the MS mode. \citet{Kusakabe2015} have stacked IR and UV images of $z\sim2$ LAEs to show that they are MS galaxies in average.

The $M_\star$ and $SFR$ of our LAEs averaged over the four fields are {\color{black}$M_{\star}=10.2\,\pm\,1.8\times10^{8}\ {\rm M_{\odot}}$ and SFR$=3.4\,\pm\,0.4\ {\rm M_{\odot}}\ {\rm yr^{-1}}$,} respectively. Thus, our LAEs are on average placed near a lower-mass extrapolation of the SFMS as shown by a red star in figure \ref{fig:ms_sfr_smc}(b), confirming the result obtained by \citet{Kusakabe2015} with a $6$ times larger survey area using deep IRAC data. We also find in figure \ref{fig:ms_sfr_smc}(a) that the LAEs in individual fields also lie on the extrapolated SFMS, {\color{black} although that in the CDFS has large uncertainties (blue triangle in figure \ref{fig:ms_sfr_smc}(a))}. {\color{black} This result is unchanged even when we stack all objects including those with $NB387_{tot}\geq25.5$ mag.}

\citet{Hagen2016}'s sample is a mixture of two samples: bright spectroscopically-selected LAEs at $z = 1.90-2.35$ from the HETDEX survey \citep[$L_{\rm Ly\alpha}>10^{43}\ {\rm erg\ s^{-1}}$: ][]{Hagen2014} and bright NB-selected LAEs at $z\simeq 2.1$ from \citet{Guaita2010} and \citet{Vargas2014} with a counterpart in the 3D-HST catalog. They derive $SFR$s from the IRX-$\beta$ relation with the Calzetti curve. Note that we also find our LAEs to have higher $sSFR$s similar to theirs if we use the Calzetti curve as shown in figures \ref{fig:ms_sfr_cal} (a) -- (c)\footnote{\citet{Hagen2016} suggest either that their LAEs are undergoing starbursts, that the SFMS becomes shallower at low stellar masses and their LAEs are distributed around it, or that their LAEs are biased towards high \lya luminosities, not representing typical LAEs. }. {\color{black} They also expect that their objects would move downward toward the SFMS in the $M_\star$--$SFR$ plane if they adopt a SMC-like curve.} \citet{Shimakawa2017a} select LAEs using a narrow-band ($NB\le 26.55\ {\rm mag}\ (5\sigma)$) and only include those with a counterpart in the 3D-HST catalogue \citep{Skelton2014}. They also derive $SFR$s from the $IRX-\beta$ with the Calzetti curve, while stellar masses are derived from SED fitting  without IRAC photometry. Since their LAEs have blue $\beta$ ($\sim-1.9$ in average), their $SFR$s and stellar masses do not change so much if a SMC-like curve is used instead. \citet{Hashimoto2017} have also examined six LAEs with $EW_{0}(\rm Ly\alpha)\simeq200$--$400$\AA$\ $ selected from the same sample as ours and found that they are star-burst galaxies with $M_{\star}\sim10^{7}$--$10^{8}\ {\rm M_{\odot}}$. However, as suggested in \citet{Hashimoto2017}, their high $sSFR$s are probably a consequence of high $EW_{0}(\rm Ly\alpha)$s (because younger galaxies have a larger $EW_{0}(\rm Ly\alpha)$) and the stellar population properties of these six LAEs do not represent those of our LAE sample.

We infer that our sample better represents the majority of $z \sim2$ LAEs because of a wide luminosity coverage \citep[$\sim 0.1$--$2\times L_{{\rm Ly}\alpha}^\star$: see][]{Konno2016} and a simple selection based only on $EW_{0}(\rm Ly\alpha)\geq20$--$30$\AA, being less biased toward/against other quantities such as UV luminosity. The majority of $z \sim 2$ LAEs are probably normal star-forming galaxies with low stellar masses in terms of star formation mode.

\section{Stellar and Halo Properties}\label{sec:sh}
In this section, we combine the stellar masses, $SFR$s, and  halo masses derived in the previous sections (summarized in tables \ref{tbl:acf} and \ref{tbl:sed_para}) to evaluate {\color{black}the} star formation efficiency in {\color{black} dark matter halos}. 

\subsection{Relation between $M_{\star}$ and $M_{\rm h}$}\label{subsec:sh_shmr}

The stellar to halo mass ratio ($=M_{\star}/M_{\rm h}$: $SHMR$) indicates the efficiency of star formation in dark {\color{black}matter} halos integrated over time from the onset of star formation {\color{black}to the observed epoch}, which we {\color{black}refer to as the} integrated SF efficiency.
{\color{black}The $SHMR$ as a function of halo mass is known to have a peak and the halo mass at the peak (pivot mass) is $\simeq2-3\times10^{12}\ {\mathrm M_\odot}$ at $z \sim 2 $ \citep[e.g.,][]{Behroozi2013, Moster2013}. The shape of the average relation show almost no evolution at $z\sim0$--$5$, although the behavior of the $z \sim 2$ $SHMR$ below $M_{\rm h} \sim 10^{11}\ {\mathrm M_\odot}$ has not been constrained well. We plot the $SHMR$s of LAEs at $z\sim2$ comparing them with the average relations for the first time and discuss the typical $SHMR$ of our LAEs with largest survey area so far.}

\begin{figure*}[ht]
      \includegraphics[width=0.9\linewidth]{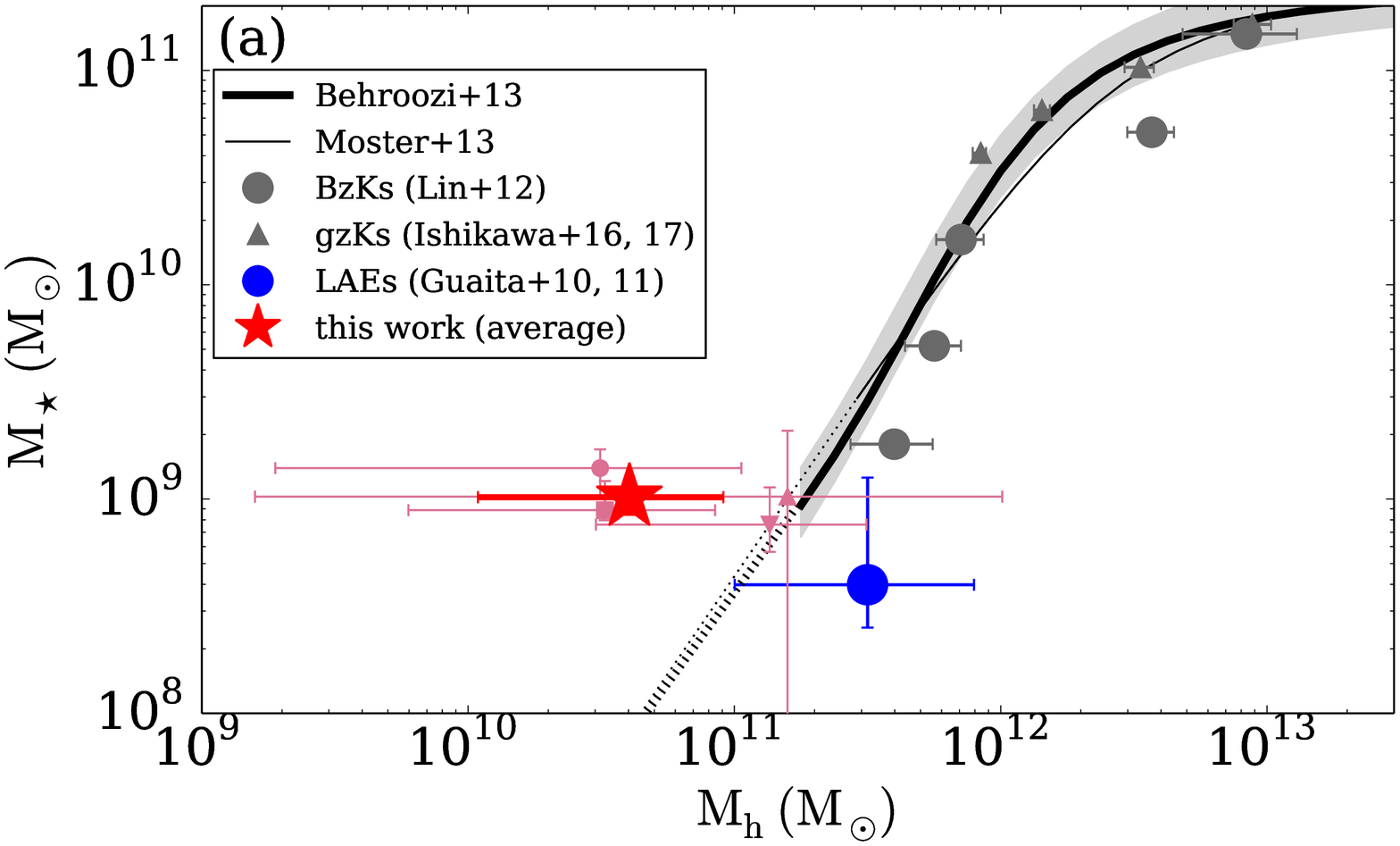}
      	\includegraphics[width=0.9\linewidth]{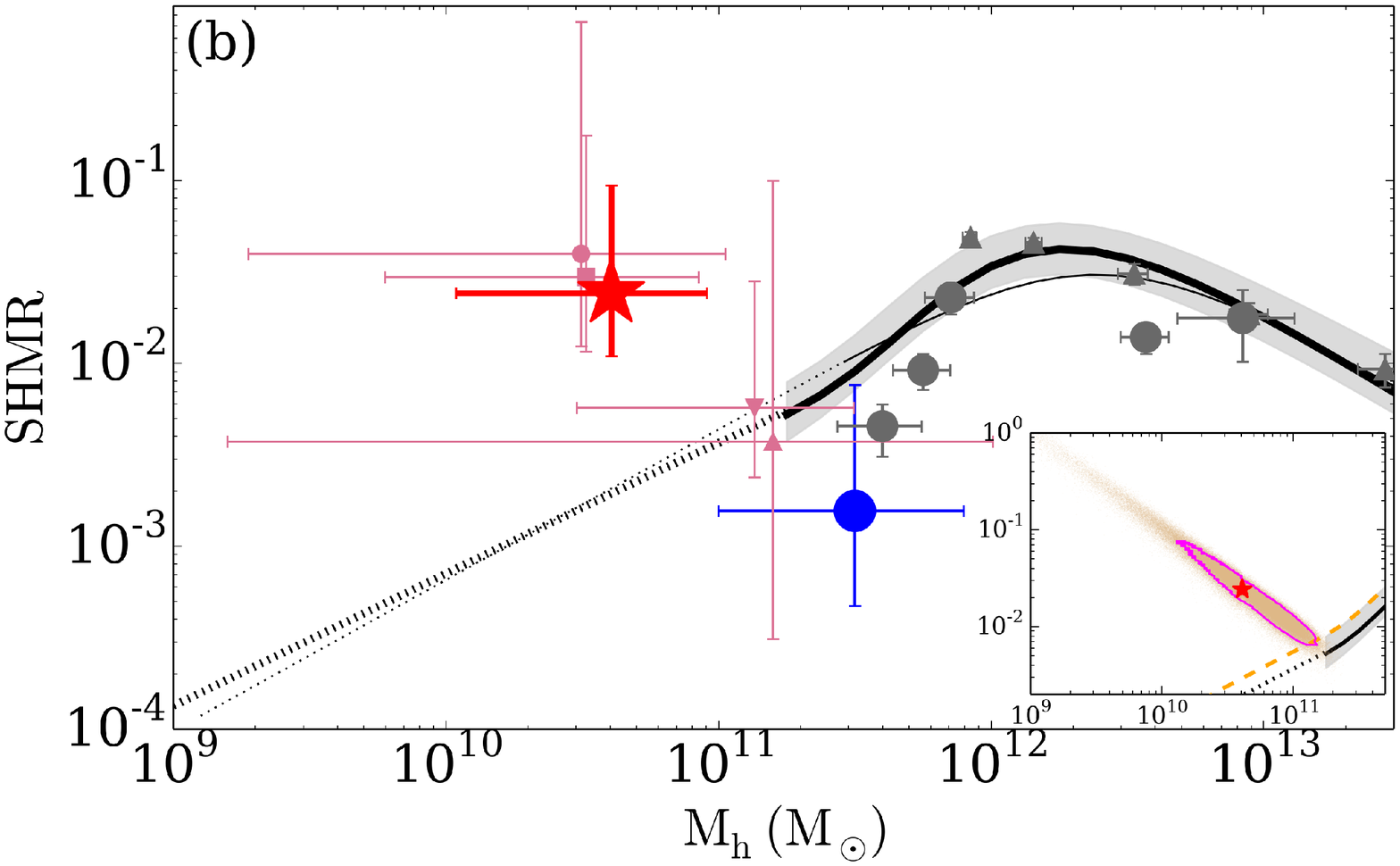}
  \caption{
(a) $M_\star$ vs $M_{\rm h}$ and (b) $SHMR$ vs $M_{\rm h}$.
For each panel, a filled pink square, circle, inverted triangle, and triangle represent average (stacked) LAEs with $NB387_{\rm tot} \le 25.5$ mag in the SXDS, COSMOS, HDFN, and CDFS fields, respectively, and a large red star shows the average over the four fields.  A blue circle indicates median (stacked) LAEs at $z \sim 2$ in \citet{Guaita2011}. Black thick and thin solid lines represent the average relation of galaxies at $z \sim 2$ in \citet{Behroozi2013} and \citet{Moster2013}, respectively;   their extrapolations are shown by dotted black lines. A gray shaded region indicates the $1\,\sigma$ uncertainty in $M_{\star}$ in the relation in \citet{Behroozi2013}. Gray circles and gray triangles denote BzK galaxies in \citet{Lin2012} and gzK galaxies in \citet{Ishikawa2016} and \citet{Ishikawa2017}, respectively. {\color{black} For each data point, the horizontal error bars indicate the $\pm1\sigma$ ($68\%$) range of the $M_h$ measurement, and the vertical error bars the $\pm1\sigma$ ($68\%$) range of the $M_\star$ (panel [a]) and $SHMR$ ([b]) measurement. The inset of the panel (b) shows the two-dimensional probability distribution of our four-field average Mh and SHMR values calculated from a Monte Carlo simulation with 500,000 trials. A magenta contour presents the $68\%$ confidence interval while brown dots indicate randomly selected 150,000 trials for the presentation purpose. An orange dashed line indicates the $+1\sigma$ limit of the average relation.} {\color{black}All data are rescaled to a Salpeter IMF according to footnote \ref{ft:imf}}. See also footnotes \ref{ft:g10b}--\ref{ft:B13extrapolation}. (Color online)}
  \label{fig:mh_ms}
\end{figure*}

{\color{black}Figure \ref{fig:mh_ms}(a) shows $M_{\star}$ and $M_{\rm h}$ of our LAEs in each of the four fields (pink symbols) and those values averaged over the four fields: {\color{black}$M_{\star}=10.2\,\pm\,1.8\times10^{8}\ {\rm M_{\odot}}$} and $M_{\rm h}=4.0_{-2.9}^{+5.1}\times10^{10}\ {\rm M_{\odot}}$ (a red star). Those of LAEs at $z\sim2.1$ \citep[][]{Guaita2010}\footnote{\label{ft:g10b}The SFR and stellar mass in \citet{Guaita2010,Guaita2011}are derived from SED fitting to a median-stacked SED and their halo mass is a median halo mass. We plot them without any correction (see also section \ref{subsec:bias}).}, star forming galaxies based on clustering analysis \citep[][]{Lin2012, Ishikawa2016, Ishikawa2017}\footnote{\label{ft:L12I16}We recalculate halo masses in \citet{Lin2012} from the effective biases given in their table using the same method as ours.}, and the average relation based on abundance matching \citep{Behroozi2013,  Moster2013}\footnote{\label{B13cosmo}The values of cosmological parameters adopted in \citet{Behroozi2013} and \citet{Moster2013} are slightly different from ours, but we have not corrected for those differences in this study. The $M_{\rm h}$ value in \citet{Behroozi2013} becomes $\sim0.15$ dex higher at $M_{\rm h}\le10^{12}\ {\rm M_{\odot}}$ when our values are used (P. Behroozi 2017, private communication). } at $z\sim2$ are shown in figure \ref{fig:mh_ms} (a) and (b) for comparison.} In contrast to Guaita et al.'s result (a blue circle), our LAEs averaged over the four fields {\color{black}{(a red star)}} lie above a simple lower-mass extrapolation (without changing the slope in the log-log space) of the $M_{\star}$-$M_{\rm h}$ relation of star forming galaxies and the average relation. {\color{black}Due to the high stellar mass and low halo mass,} our LAEs have a $SHMR$ of {\color{black}$0.02^{+0.07}_{-0.01}$} as {\color{black}high as galaxies at the pivot mass}, $M_{\rm h} \simeq 2-3 \times 10^{12}\ {\mathrm M_\odot}$. 
{\color{black}Here, the errors in this $SHMR$ value indicate the $\pm1\sigma$ ($68\%)$ range. The inset of figure \ref{fig:mh_ms}(b) shows the two-dimensional probability distribution of our four-field average $M_{\rm h}$ and $SHMR$ values calculated from a Monte Carlo simulation with 500,000 trials. A magenta contour presents the $68\%$ confidence interval, while brown dots indicate randomly selected 150,000 trials. Although the contour touches the $+1\sigma$ limit of the average relation, only $\sim\,2.5$\% of the entire trials reach the $+1\sigma$ limit (an orange dashed line).}

{\color{black} We discuss whether there are any systematic differences in $M_{\star}$ and/or $M_{\rm h}$ between our LAEs and the average relation, which result in the departure of our results fr om the relations.
The average relation by \citet{Moster2013} expresses the mean stellar mass of the central galaxy as a function of halo mass and has a double power-law form, while that by \citet{Behroozi2013} uses the median stellar mass and has five fitting parameters, whose functional form at low halo masses is approximated by a power law\footnote{The \citet{Behroozi2013} relations including extrapolated parts in figures \ref{fig:mh_ms}--\ref{fig:dutton} are taken from the website of P. Behroozi: http://www.peterbehroozi.com/data.html. see also footnote \ref{B13cosmo}.\label{ft:B13extrapolation}
}. Although the definitions of stellar masses of the two relations are different, the relations are similar to one another. Our average stellar mass is a field-average median stellar mass since stellar masses are derived from SED fitting for median-stacked SEDs, which are commonly used to prevent contamination (see section \ref{sec:sed}). The field-average mean stellar mass of our sample is possibly higher than the field-average median. In fact, the mean value of $K$-band flux densities, which is an approximation of stellar mass, is approximately twice as high as the median one in the SXDS field, the field with the deepest $K$ data. We derive effective halo masses of our LAEs from effective biases directly (see section \ref{subsec:Mh}) assuming a one-to-one correspondence between galaxies and dark matter halos with a narrow range of halo mass. Our field-average effective halo mass probably corresponds to the true mean and/or median within the large uncertainty whose $1\,\sigma$ permitted range is $\sim1$ dex. Even though the uncertainty by cosmic variance discussed in section \ref{subsec:cvbias} is added to the total uncertainty in the field-average halo mass, by which the halo mass and $SHMR$ are written as $M_{\rm h, \,cv}=4.0_{-3.5}^{+8.4} \times 10^{10}\ {\mathrm M_\odot}$ and {\color{black}$SHMR=0.02^{+0.18}_{-0.01}$}, respectively, our result is not consistent with the extrapolated average relations within $1\,\sigma$. Therefore, the departure of our field-average LAEs (a red star) from the average relation are not caused by neither a systematic difference of the definition of $M_\star$ nor $1\,\sigma$ cosmic variance on $M_{\rm h}$.}

{\color{black}
On the other hand, if LAEs represent average galaxies, the average $M_{\rm h}$--$SHMR$ relation must have an upturn at $M_{\rm h} \lesssim 10^{11}\ {\mathrm M_\odot}$. This, however, appears to be unphysical because no such upturn is seen at $z \sim 0$, the only epoch at which the average relation below $M_{\rm h} \sim 10^{11}\ {\mathrm M_\odot}$ has been constrained well \citep{Behroozi2013}, unless the low-mass slope of the average relation evolves drastically from $z \sim 2$ to $\sim 0$. Another possibility is that the scatter of the average relation become significantly larger at lower halo masses and the SHMR of our LAEs is within the scatter. }

{\color{black} Note that the $SHMR$s in the HDFN and CDFS are consistent with the average relations although with large uncertainties. We obtain consistent stellar masses between the four fields and it is just the halo masses that are different. The difference in $M_{\rm h}$, and hence in $b_{\rm g,eff}$, among the four fields seen in figure \ref{fig:lyalim_bias} (see also sections \ref{subsec:bias} and \ref{subsec:Mh}) is not due to a difference in the limiting magnitude because all four fields have the same limit, $NB387_{tot} = 25.5$. As shown in figure \ref{fig:mh_ms}, fitting errors and contamination fraction errors possibly drive the offsets of $M_{\rm h}$ in the two fields to the average values.
The difference is also explained by cosmic variance as shown in figure\ref{fig:cv}(b) (see also section \ref{subsec:cvbias}) and averaging over the four fields reduces the effect of cosmic variance.} 

\clearpage

\clearpage

\subsection{Baryon Conversion Efficiency}\label{subsec:sh_bce}
The baryon conversion efficiency ($BCE$), defined as: 
\begin{equation}
BCE = \frac{{\dot{M_{\star}}}}{\dot{M_{\rm b}}}, 
\end{equation}
measures the efficiency of star formation in dark {\color{black}matter} halos at the observed time, where $\dot{M_{\rm b}}$ is the baryon accretion rate ($BAR$). Here we assume that most of the accreting baryons are in {\color{black}a} (cold) gas phase (i.e., the $BAR$ is equal to the inflow rate of cold gas). The average $BAR$ at a fixed halo mass is proportional to the halo mass accretion rate, $\dot{M_{\rm h}}(z, M_{\rm h})$, which is estimated as a function of redshift and halo mass from cosmological simulations \citep{Dekel2009}:  
\begin{eqnarray}
BAR &=& f_{\rm b}\times\dot{M_{\rm h}}(z,\ M_{\rm h})\\
&\sim& 6\times\left( \frac{M_{\rm h}}{10^{12}{\rm M_{\odot}}}\right)^{1.15}\times(1+z)^{2.25} \ {\rm M_{\odot}}\ {\rm yr^{-1}},  
\label{eq:bar}
\end{eqnarray}
where $f_{\rm b}\equiv\Omega_{\rm b}/\Omega_{\rm m}=0.15$. 

\begin{figure*}[ht]
      \includegraphics[width=1.0\linewidth]{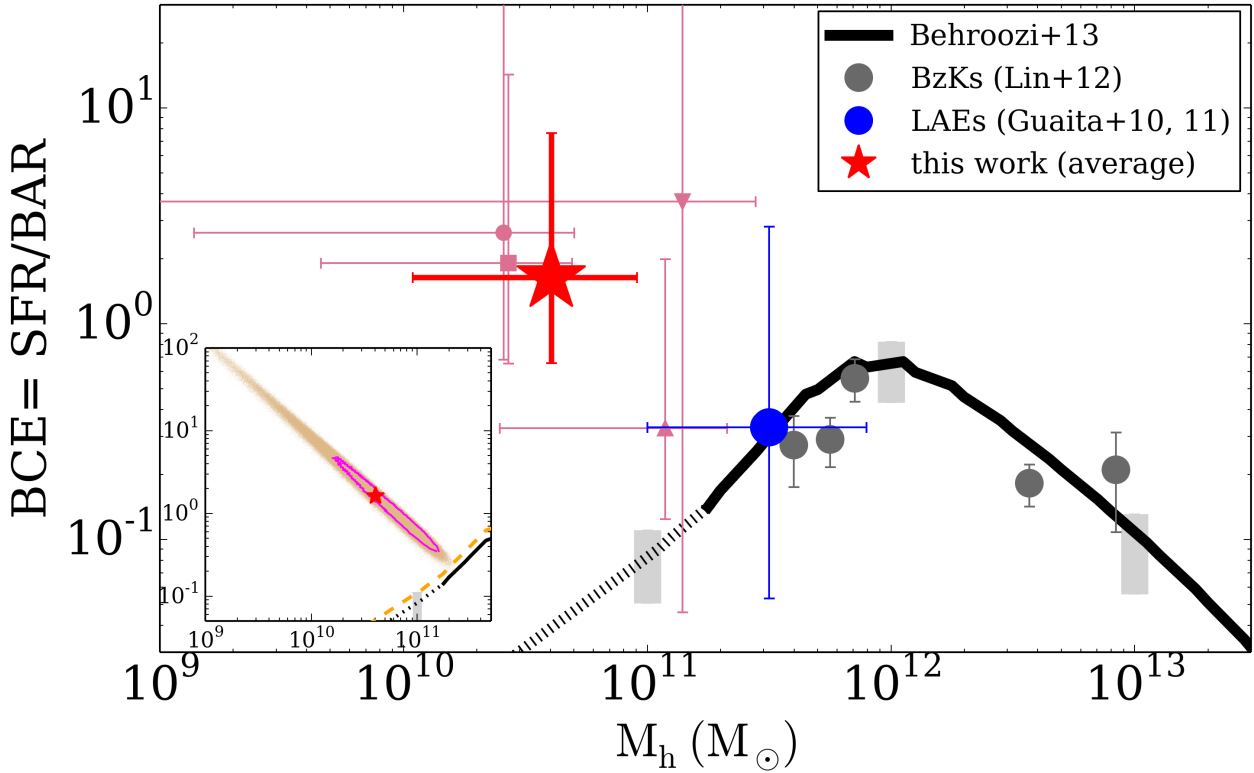}
 \caption{
  Baryon conversion efficiency (BCE) as a function of $M_{\rm h}$. A filled pink square, circle, inverted triangle, and triangle represent average (stacked) LAEs with $NB387_{\rm tot} \le 25.5$ mag in the SXDS, COSMOS, HDFN, and CDFS fields, respectively, and a red star shows the average over the four fields. A blue circle indicates median (stacked) LAEs at $z \sim 2$ in \citet{Guaita2011}. A black thick solid and gray circles show the average relation of galaxies at $z \sim 2$ in \citet{Behroozi2013} and measurements for BzK galaxies in  \citet{Lin2012}, respectively. {\color{black} For each data point, the horizontal (vertical) error bars indicate the $\pm1\sigma$ ($68\%$) range of the $M_h$ ($BCE$) measurement.} Extrapolations and 1$\,\sigma$ {\color{black}scatter} of BCE at fixed $M_{\rm h}$ are shown by a dotted black line and vertical gray bands, respectively. The {\color{black}scatter of BCE is} estimated from {\color{black}the scatter} of $SFRs$ at $M_{\rm h}=1\times10^{11}$, $1\times10^{12}$, and $1\times10^{13}$. {\color{black}
The inset shows the two-dimensional probability distribution of our four-field average Mh and BCE values calculated from a Monte Carlo simulation with 500,000 trials. A magenta contour presents the $68\%$ confidence interval while brown dots indicate the entire trials. An orange dashed line indicates the $+1\sigma$ limit of the average relation.} {\color{black}All data are rescaled to a Salpeter IMF according to footnote \ref{ft:imf}}. See also footnotes \ref{ft:g10b}--\ref{ft:B13extrapolation}  (Color online)  } 
  \label{fig:bce}
\end{figure*}

{\color{black}Figure \ref{fig:bce} shows the $BCE$ against halo mass. our LAEs have {\color{black}$BCE=1.6^{+6.0}_{-1.0}$} and, as shown by a red star, lie above an extrapolation (keeping the slope unchanged) of the average relation by \citet{Behroozi2013} and most of the BzK galaxies in \citet{Lin2012}. {\color{black}Here, the errors in our $BCE$ value indicate the $\pm1\sigma$ ($68\%)$ range.
 The inset of figure \ref{fig:bce} shows the two-dimensional probability distribution of our four-field average $M_{\rm h}$ and $BCE$ values calculated from a Monte Carlo simulation with 500,000 trials. A magenta contour presents the $68\%$ confidence interval, while brown dots indicate the 500,000 trials. Only $\sim\,0.3$\% of the entire trials reach the $+1\sigma$ limit of the average relation (an orange dashed line).} On the other hand, \citet[][]{Guaita2010, Guaita2011}'s LAEs at $z\sim2$ have {\color{black}a moderate $BCE$, although with large uncertainties, which is consistent with} the average relation as shown by a blue circle. The average $SFR$s of both samples are nearly equivalent and it is the clustering measurements that differ and drive our $BCE$ up. 
So the difference in the clustering affects the discrepancy in both axes in figure \ref{fig:bce} making the offset worse.
}

We discuss whether there are any systematic differences in $SFR$ and/or $M_{\rm h}$ between our LAEs and the average relation, which result in the departure of our results from the relations.
The average relation by \citet{Behroozi2013} expresses the mean $SFR$ as a function of halo mass. Our field-average $SFR$ is derived from SED fitting for median-stacked SEDs and probably does not overestimate the true average $SFR$, since the median of $B$-band flux densities, which trace rest-frame UV, is similar to the average $B$-band flux density. Even when we neglect dust attenuation at UV, {\color{black}$A_{1600}=0.6\pm0.1\ {\rm mag}$}, the field-average {\color{black}$SFR$ ($=3.4\pm0.4\ {\rm M_{\odot}}\ {\rm yr^{-1}}$)} decreases only a factor of $\sim2$. Moreover, even when the uncertainty by cosmic variance discussed in section \ref{subsec:cvbias} is added to the measured value, {\color{black}$BCE=1.6^{+6.0}_{-1.0}$}, the $1\,\sigma$ lower limit of the field-average $BCE$ is still larger than {\color{black}$0.4$}.
Thus, it seems difficult for our LAEs to fall on the average relation shown in figure \ref{fig:bce}. 

As described in section \ref{subsec:sh_shmr}, logically we cannot rule out the possibilities that our LAEs lie indeed on or near the average relation which changes the slope and/or scatter below $M_{\rm h} \sim 1 \times 10^{11}\ {\mathrm M_\odot}$ for some reason.

%

\section{Discussion}\label{sec:discussion}
In this section, we {\color{black}interpret our results on LAEs in terms of the general evolution of galaxies} and discuss the physical origin of their high $SHMR$ and $BCE$, as well as predicting their present-day descendants. We assume that the three average relations shown in figures \ref{fig:ms_sfr_smc}, \ref{fig:mh_ms}, and \ref{fig:bce} do  not change either the slope (in log-log plane) or the scatter at low masses.
We also assume that our LAEs are central galaxies. If they are satellite galaxies, their dark matter halo (sub halo) masses will be overestimated and their true $SHMR$ and $BCE$ {\color{black}would be higher than reported in this study.} 

\subsection{Duty Cycle}\label{sec:discussion_dc}
The duty cycle of LAEs, $f_{\rm duty}^{\rm LAEs}$, is defined as the fraction of dark mater halos hosting LAEs. Previous studies find that $f_{\rm duty}^{\rm LAEs}$ at $z\sim3$ is a few tenths to a few percent \citep{Ouchi2010, Chiang2015}. 
We estimate the duty cycle of our LAEs to be: 
\begin{equation}
f_{\rm duty}^{\rm LAEs}= \frac{ND_{\rm LAE}}{ND_{\rm DMH}}\sim 2\%,  
\end{equation}
where $ND_{\rm LAE}$ and $ND_{\rm DMH}$ are the number density of LAEs with $NB_{\rm tot}{\le}25.5\ {\rm mag}$ and that of dark matter halos estimated from the halo mass function at $z\sim2$ using the calculator provided by  \citet{Murray2013}, respectively. {\color{black}For this calculation,} we assume that dark matter halos hosting our LAEs have a one dex range of mass, $10^{10}$--$10^{11}\ {\rm M_{\odot}}$, since the $K$-band magnitudes, an approximation of stellar mass, of our LAEs are distributed with {\color{black}FWHM} of $\sim 3.2$ mag, or $\sim 1.3$ dex. Our result is comparable with those of previous studies.

We also estimate the fraction of galaxies in a given stellar mass range classified as LAEs (LAE fraction), $f_{\rm gals}^{\rm LAEs}$. Assuming that our LAEs have a one dex range of stellar mass, $10^{8.5}$--$10^{9.5}\ {\rm M_{\odot}}$, we obtain: 
\begin{equation}
f_{\rm gals}^{\rm LAEs}= \frac{ND_{\rm LAE}}{ND_{\rm gal}}\sim 10\%,  
\end{equation}
where $ND_{\rm gal}$ is the number density of galaxies estimated by extrapolating \citet{Tomczak2013}'s stellar mass function at $z\sim2$--$2.5$ below $10^9\  {\mathrm M_\odot}$. This result is comparable with those of previous spectroscopic observations of star forming galaxies at $z\sim2$--$2.5$ {\color{black} \citep[$\sim10$\%, ][]{Hathi2016} and BX galaxies at $z\sim1.9$--$2.7$ \citep[$\sim12$\% with $EW_{\rm Ly \alpha}\geqq20\ $\AA; ][]{Reddy2008}.} Note that typical galaxies embedded in dark matter halos with $M_{\rm h}=10^{10}$--$10^{11}\ {\rm M_{\odot}}$ have lower stellar masses than $M_{\star}=10^{8.5}$--$10^{9.5}\ {\rm M_{\odot}}$ because of the high $SHMR$ of our LAEs. The low fractions obtained above imply that {\color{black} only a few percent of galaxies within these mass ranges studied here can evolve into LAEs and/or that galaxies within these mass ranges can experience the LAE phase only for a very short time.}

\subsection{Physical Origin of Ly$\alpha$ Emission}\label{subsec:discussion_lya}
The result that our LAEs have a higher $SHMR$ than average galaxies with the same stellar mass {\color{black}may explain} why they have strong \lya emission. A higher $SHMR$ at a fixed $M_\star$ means a lower $M_{\rm h}$ and hence a lower gas mass ($M_{\rm gas}$), since the $M_{\rm gas}$ of a galaxy is written as $M_{\rm gas} \simeq f_{\rm b} M_{\rm h} - M_\star$. Galaxies with a low $M_{\rm gas}$ likely have a low HI column density, thus making it {\color{black}easier} for \lya photons to escape because of a reduced number of resonant {\color{black}scatterings}. Indeed, \citet{Pardy2014} have found a tentative anticorrelation of HI gas mass {\color{black}with} the \lya escape fraction and the \lya equivalent width using 14 local galaxies \citep[\lya Reference Sample;][]{Hayes2013, Ostlin2014}. 

Furthermore, our LAEs may have high outflow velocities because a high $BCE$ means a high $SFR$ at a fixed $M_{\rm h}$ (recall BAR $\propto M_{\rm h}^{1.15}$) and hence a high kinetic energy from star formation at a fixed gravitational binding energy of {\color{black}dark mater halos}. In high-velocity outflowing HI gas, the probability of the resonant scattering of \lya photons is reduced because of reduced cross sections of HI atoms due to large relative velocities \citep[e.g.,][]{Kunth1998, Verhamme2006, Hashimoto2015}. Note also that our LAEs have absolutely low dust attenuation due probably to a low stellar mass as shown in figure \ref{fig:ms_irx}, which also helps \lya photons survive in galaxies. To summarize, the high $SHMR$, high $BCE$, and moderate $SFR$ obtained for our LAEs are in concord with the strong \lya emission observed.

 \begin{figure*}[t]
  \begin{center}
    \begin{tabular}{c}
      \begin{minipage}{0.35\hsize}
        \begin{center}
     \includegraphics[width=0.9\linewidth]{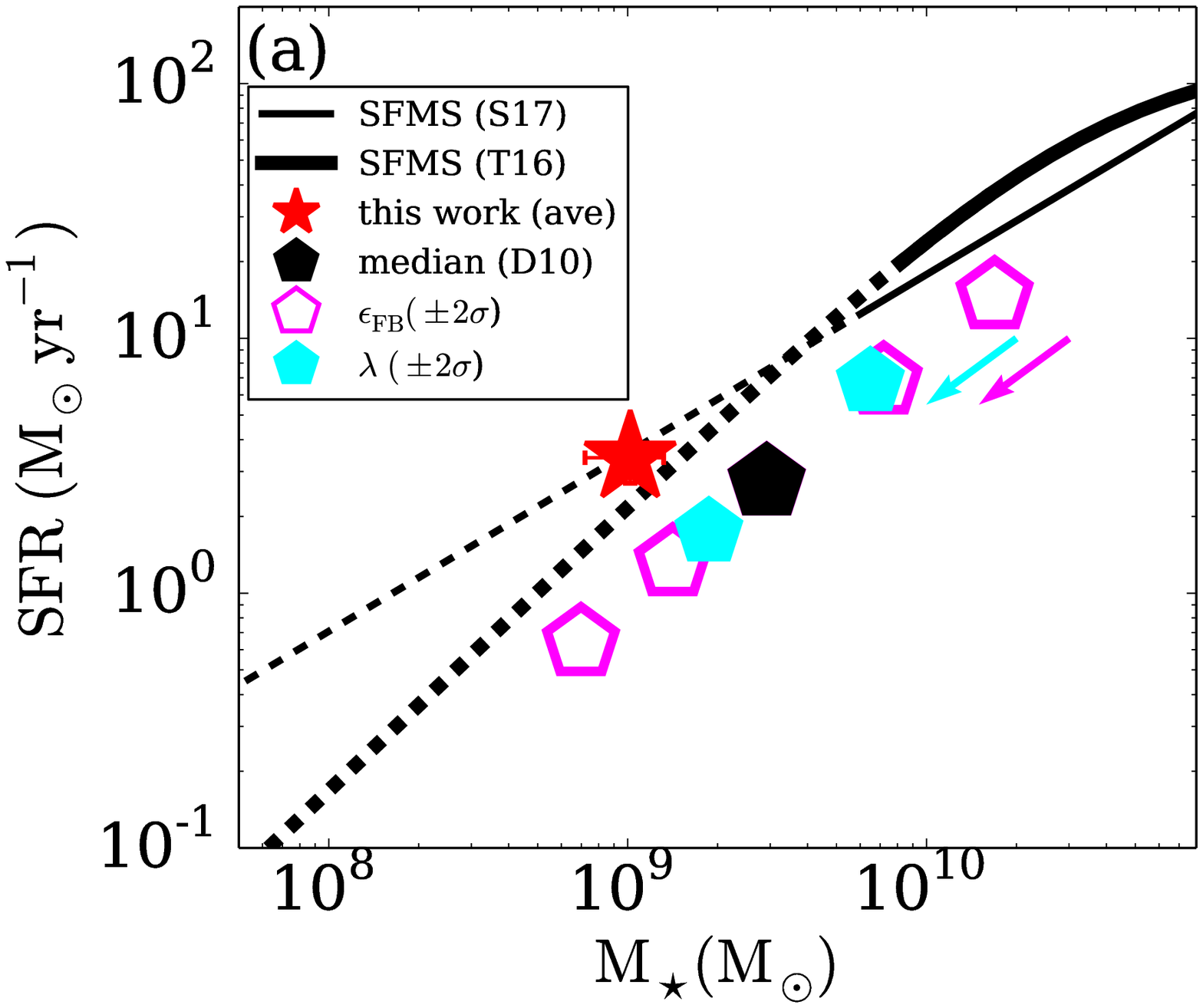}
        \end{center}
      \end{minipage}
      \begin{minipage}{0.35\hsize}
        \begin{center}
      \includegraphics[width=0.9\linewidth]{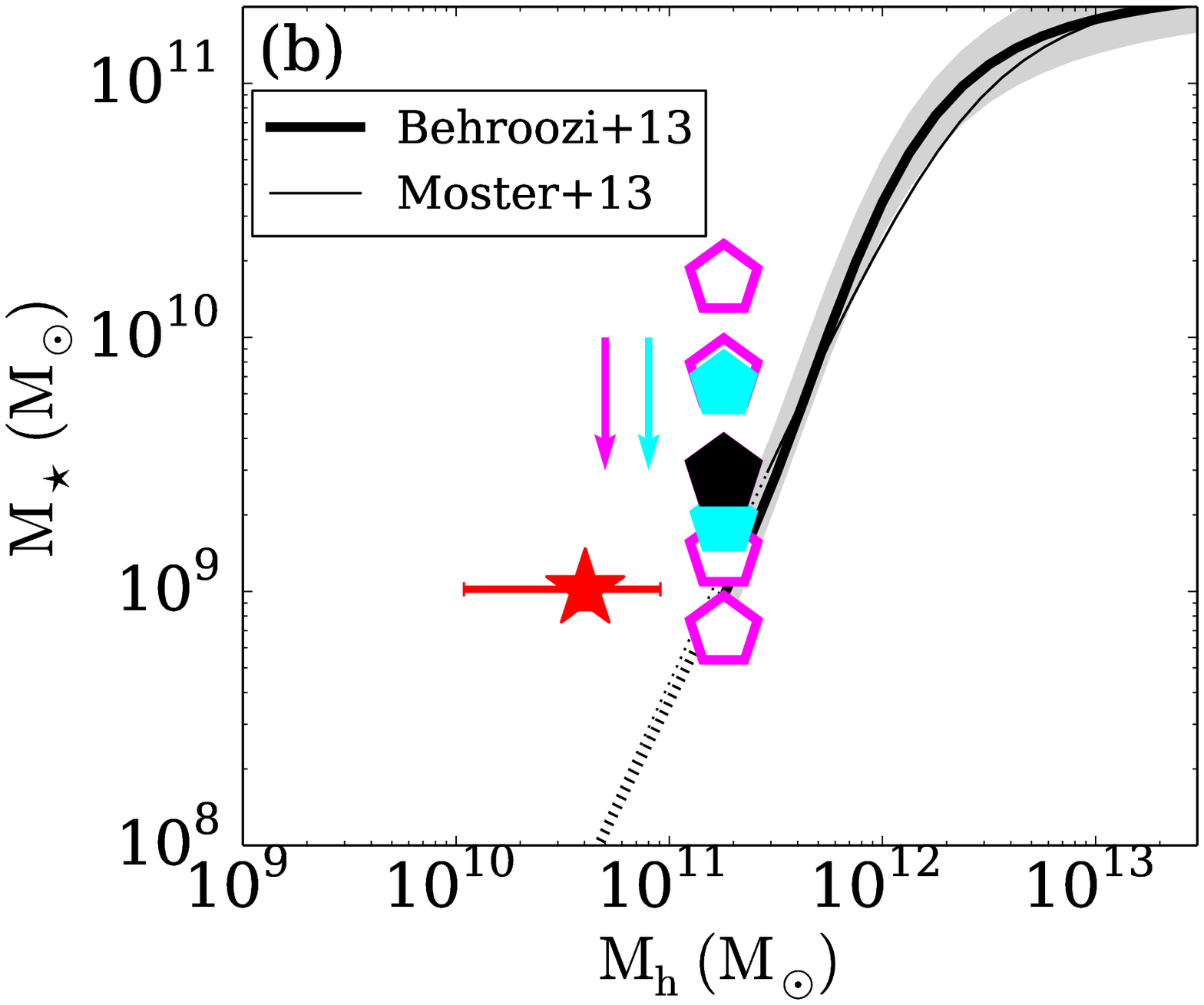}
        \end{center}
      \end{minipage}
       \begin{minipage}{0.35\hsize}
        \begin{center}
     \includegraphics[width=0.9\linewidth]{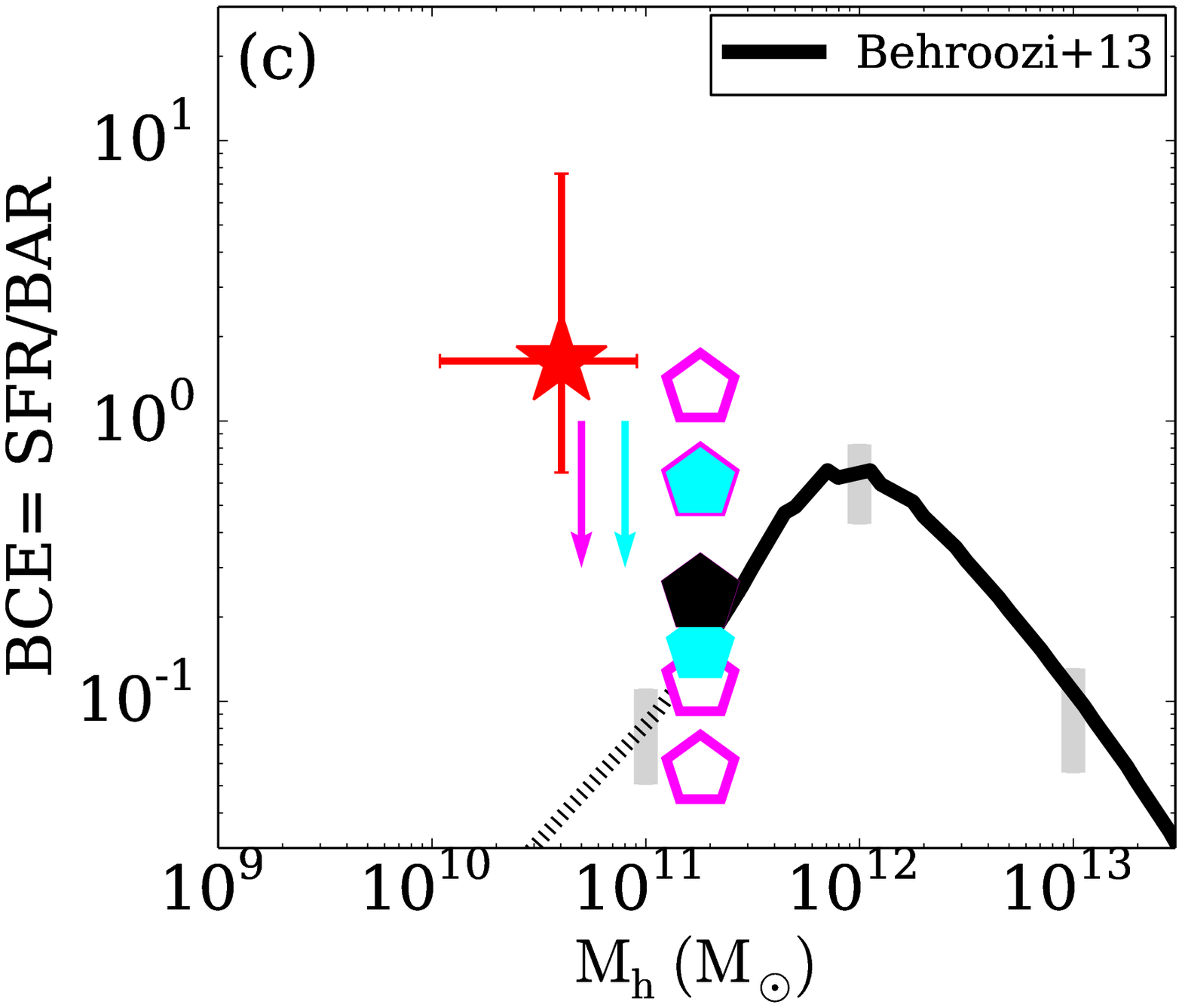}
        \end{center}
      \end{minipage}     
  \end{tabular} 
   \end{center}
  \caption{
  Changes in the position of model galaxies in the $M_\star$-SFR plane (panel (a)), $M_{\rm h}$-$M_\star$ plane ([b]), and $M_{\rm h}$-$BCE$ plane ([c]) due to variations in the halo spin parameter, $\lambda$, and the feedback efficiency, $\epsilon_{\rm FB}$, calculated by \citet[][hereafter D10]{Dutton2010b}. Pentagons show D10’s model galaxies with a fixed halo mass ($M_{\rm h, \ z=0} = 4 \times 10^{11} {\mathrm M_\odot}$, corresponding to $\sim2 \times 10^{11} {\mathrm M_\odot}$ at $z = 2$ according to figures 7 and 8 in \citet{Behroozi2013}), where black, cyan, and magenta colors denote, respectively, positions with median halo parameters, those with $\pm2\sigma$ variation in $\lambda$, and those with $\pm2\sigma$ variation in $\epsilon_{\rm FB}$. All model data of $M_\star$ and $SFR$ are taken from figure 12 in D10 {\color{black}(In D10 four data points are shown as $\pm2\sigma$ variation in $\epsilon_{\rm FB}$)}. The BARs of model galaxies are calculated from equation \ref{eq:bar}. Cyan and magenta arrows indicate the direction in which galaxies move when $\lambda$ and $\epsilon_{\rm FB}$ increase. In all panels, red stars represent the average LAEs with $NB387_{\rm tot} \le 25.5$ mag. In panel (a), several SFMS measurements in previous studies are shown by black lines in the same manner as figure \ref{fig:ms_sfr_smc}. The average relations in \citet{Behroozi2013} and \citet{Moster2013} are plotted by black lines in panels (b) and (c), similar to figures \ref{fig:mh_ms} and \ref{fig:bce}. {\color{black}All data are rescaled to a Salpeter IMF according to footnote \ref{ft:imf}.} See also footnotes \ref{ft:g10b}--\ref{ft:B13extrapolation}. (Color online)
}
  \label{fig:dutton}
\end{figure*}

\subsection{Physical Origin of Moderate Star Formation Mode, High SHMR, and High BCE}\label{subsec:discussion_physical}
Our LAEs have a higher $SHMR$ and a higher $BCE$ than average galaxies but have a moderate $SFR$, being located on the (extrapolated) SFMS defined by average galaxies. Indeed, it is not trivial for galaxy formation models to reproduce these three properties simultaneously.

\citet{Dutton2010b} have used a semi-analytic model to study the evolution of the SFMS and its dependence on several key parameters in the model. As shown in their figure 12 and our figure \ref{fig:dutton}, model galaxies (at $z \sim 2$) at a fixed halo mass move along the SFMS upward when the {\color{black}supernova} (SN) feedback is weakened or the halo's spin parameter is reduced, thus having a higher $SHMR$ and a higher $BCE$ on the SFMS. With a lower feedback efficiency, a {\color{black}larger amount of cold gas can be stored, thus resulting in} a higher $SFR$ and a higher stellar mass. A lower spin {\color{black}causes the gas density to be higher, thereby the SFR per unit gas mass} is elevated. Although these results may not necessarily be applicable to our LAEs whose halo mass is ten times lower, it is interesting to note that there is a relatively simple way to {\color{black}explain} MS galaxies with an elevated SHMR and BCE.

It is beyond our scope to identify the mechanism(s) by which our LAEs acquire a high $SHMR$ and a high $BCE$. If, however, the high $SHMR$ and $BCE$ of our LAEs are due to some systematic differences in one or more parameters controlling the star formation and/or internal structure of halos similar to \citet{Dutton2010b}'s study, then it implies that not all but only a certain fraction of (low-mass) halos at $z \sim 2$ experience the LAE phase.

\subsection{Present-day Descendants of Our LAEs}\label{subsec:discussion_lmc}
\begin{figure}[ht]
     \includegraphics[width=1.0\linewidth]{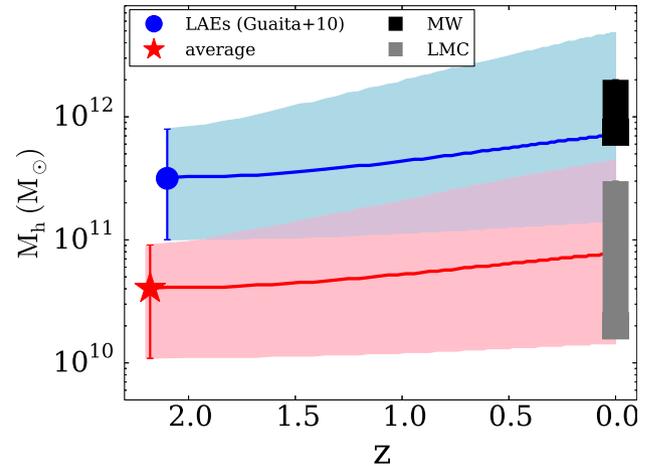}
  \caption{
  Dark matter halo mass evolution as a function of redshift predicted by the EPS formalism. A red (blue) curve indicates the evolution of the mode of the $M_{\rm h}$ distribution starting from the mass of our $z=2.2$ LAEs shown by a red star \citep[]['s $z=2.1$ LAEs shown by a blue circle]{Guaita2010}, with a shaded region indicating the $68\%$ confidence interval of the distribution. Black and gray rectangles represent the measured halo mass ranges of the MW and the LMC, respectively \citep[e.g., ][see also \citealt{Wang2015a}]{Wilkinson1999, Kafle2014, vanderMarel2014, Eadie2015, Penarrubia2016}.   }
  \label{fig:eps}
\end{figure}

LAEs are found to reside in low-mass halos with $M_{\rm h}\sim10^{10}$--$10^{12}\ {\rm M_{\odot}}$ over the wide redshift range $z\sim2$--$7$ as found in section \ref{subsec:Mh} \citep[e.g., ][]{Ouchi2005b, Ouchi2010, Kovac2007, Gawiser2007, Shioya2009, Guaita2010, Bielby2016, Diener2017arXiv, Ouchi2017arXiv}. In other words, the bias value of LAEs tends to decrease with decreasing redshift more rapidly than that of dark matter halos \citep[see figure 7 in][]{Ouchi2017arXiv}. {\color{black} Although this trend may be} biased because faint LAEs in lower-mass halos are missed at high redshifts, it implies that at lower redshifts, only galaxies with relatively lower masses in the halo mass function can be LAEs, which is analogous to and/or maybe related to downsizing \citep{Cowie1996}.

{\color{black}A} roughly constant halo mass {\color{black}with} redshift also implies that local descendants of LAEs vary depending on their redshift. The growth of dark matter halos is statistically predicted by the extended Press-Schehter \citep[EPS:][]{press1974, Bond1991, Bower1991} model. {\color{black}An application of the EPS model to distant galaxies can be found in, e.g., \citet[]{Hamana2006}.} Previous studies suggest that LAEs at $z\sim 4$--$7$ evolve into massive elliptical galaxies at $z=0$ \citep{Ouchi2005b,Kovac2007, Ouchi2010}, while LAEs at $z\sim3$ are expected to be progenitors of present-day $L_{\star}$ galaxies \citep{Gawiser2007, Ouchi2010}. \citet{Guaita2010} show that LAEs at $z\sim2$ could be progenitors of present-day $L_{\star}$ galaxies like the Milky Way (MW) and that they could also be descendants of $z\sim3$ LAEs, depending on star formation and dust formation histories \citep[see also][]{Acquaviva2012}.  

With the EPS model\footnote{We use a publicly released code by T. Hamana: http://th.nao.ac.jp/MEMBER/hamanatk/OPENPRO/index.html.},we find that at $z=0$ our LAEs are embedded in dark matter halos with a median mass similar to the mass of the Large Magellanic Cloud \citep[LMC: $M_{\rm h}\sim0.2$--$3\times10^{11}\ {\rm M_{\odot}}$;][and references therein]{vanderMarel2014, Penarrubia2016}, not in MW-like halos \citep[$M_{\rm h}\sim8\times10^{11}$--$2\times10^{12}\ {\rm M_{\odot}}$; e.g., ][summarized in figure 1 in \citealt{Wang2015a}]{Wilkinson1999, Kafle2014, Eadie2015}, as shown in figure \ref{fig:eps}. This is consistent with the prediction by \citet{Acquaviva2012} from SED fitting that LAEs at $z\sim3$, which are progenitors of present-day $L_{\star}$ galaxies, do not evolve into LAEs at $z\sim2$. Combined with the previous studies, our result {\color{black}imply} that the mass of present-day descendants of halos hosting LAEs depends on the redshift at which they are observed, with higher-z LAEs evolving into more massive halos.

Since the stellar mass of our LAEs, {\color{black}$10.2\pm1.8 \times10^8\ {\rm M_{\odot}}$}, is comparable to that of the LMC within only a factor of $\sim 3$  \citep[$M_{\star}\sim2.9\times10^9\ {\rm M_{\odot}}$:][]{vanderMarel2002}, their star-formation has to be largely suppressed over most of the cosmic time until $z=0$, or even be quenched, if they really become LMC-like galaxies. The star formation history of the LMC has been inferred to have multiple components, i.e., an initial burst and subsequent periods with moderate or quiescent star formation \cite[e.g., ][]{Harris2009}. For example, \citet{Rezaeikh2014} argue that it consists of two components: an initial burst of $\sim10\ {\rm Gyr}$ ago, or at $z\sim2$, with a $SFR\sim2.4\ {\rm M_{\odot}}\ {\rm yr^{-1}}$ assembling $\sim$90\% of the total mass, and a much milder star formation with $SFR\sim0.3\ {\rm M_{\odot}}\ {\rm yr^{-1}}$ after that as shown in their figure 4 {\color{black}(see however \citet{Weisz2013}, who obtained a much lower $SFR$)}. If our LAEs follow such a history with suppressed star formation over $\sim5-10\times10^9\ {\rm Gyr}$, they will grow to be LMC-like galaxies at $z=0$. {\color{black}In this case, if at $z\sim2$ they lie above the average $M_{h}$--$SHMR$ relation, they will evolve into galaxies with an $SHMR$ consistent with the average relation at $z\sim0$ \citep{Behroozi2013,Moster2013}.}

\subsection{Future Survey}\label{subsec:HSC}
In the near future, we will extend this work using new $NB387$ data from $\simeq 25$ deg$^2$ taken with Hyper Suprime-Cam as part of a large imaging survey program \citep{Aihara2017arxiv}. This program uses five broadband and four NB filters, among which the new $NB387$ is included. We call the LAE surveys with the four NB filters SILVERRUSH \citep{Ouchi2017arXiv,Shibuya2017arxiv}. The survey volume for $NB387$ ($z \sim 2$) LAEs is $6\times10^6\ {\rm (h^{-1}_{100}\ Mpc )^3}$ with an expected number of $\sim 9000$ objects. As shown in figures \ref{fig:cv}(a) and \ref{fig:cv}(b), the uncertainty from cosmic variance is expected to be negligibly small, $\sim3$\%, compared with other uncertainties. With the HSC data, we will be able to determine the $SHMR$ and $BCE$ of $z\sim2$ LAEs without suffering from cosmic variance.

%

\section{Conclusions}\label{sec:conclusion}
We have investigated stellar populations and halo masses of LAEs at $z \sim 2$, low-mass galaxies at cosmic noon, using $\sim 1250$ $NB387$-selected LAEs from four separate fields with $\sim1 \ {\rm deg}^2$ in total. In particular, we have derived the average SF mode, $SHMR$, and $BCE$ of objects with $NB387 \le 25.5$ for which measurements for all four fields are available, and discussed star formation activity and its dependence on halo mass. Our main results are as follows.

\begin{enumerate}
\item  The bias parameter of $NB387{\le}25.5$ objects averaged over the four fields is $b_{\rm g,\, eff}^{\rm ave}=1.22^{+0.16}_{-0.18}$, which is lower than that in \citet{Guaita2010} from $0.3$ deg$^2$ with a probability of $96\%$. 
We estimate an external error from cosmic variance which inversely scales with the square root of the survey area. The high bias value obtained by \citet{Guaita2010} becomes consistent with our value if the uncertainties from cosmic variance, $\pm\, 26 \%$ and $\pm\, 51 \%$ for this work and \citet{Guaita2010}, are considered. We have also found that {\color{black}$b_{\rm g,\, eff}$} does not significantly change with limiting NB387 magnitude, or limiting \lya luminosity, which may be partly due to two trends canceling out with each other: galaxies in more massive halos have brighter intrinsic \lya luminosities but lower \lya escape fractions.

\item 
The halo mass corresponding to the above $b_{\rm g,\, eff}^{\rm ave}$ value is $4.0^{+5.1}_{-2.9}\times10^{10}\ {\rm M_{\odot}}$. This value is roughly comparable to previous measurements for $z \sim 3$ -- $7$ LAEs with similar Ly$\alpha$ luminosities, $M_{\rm h} \sim 10^{10}$--$10^{12}\ {\rm M_{\odot}}$ \citep[e.g.,][]{Ouchi2010}, suggesting that the mass of dark halos which can host typical LAEs is roughly unchanged with time.

\item
The mean of each stellar parameter over the four fields is: {\color{black}$M_{\star}=10.2\,\pm\,1.8\times10^{8}\ {\rm M_{\odot}}$, $A_{1600}=0.6\,\pm\,0.1\ {\rm mag}$, Age$=3.8\,\pm\,0.3\,\times10^8\, {\rm yr}$, and SFR$=3.4\,\pm\,0.4\ \ {\rm M_{\odot}}\ {\rm yr^{-1}}$.}
Our LAEs are thus located near an extrapolation of the consensus relation of $IRX$ against stellar mass with an assumption of a SMC-like attenuation curve (see figure \ref{fig:ms_irx}). We have also found that our LAEs are on average placed near a lower-mass extrapolation of the SFMS, confirming the results obtained by  \citet{Kusakabe2015} with a $\sim6$ times larger survey area (shown in figure \ref{fig:ms_sfr_smc}).

\item 
With {\color{black}$SHMR = 0.02^{+0.07}_{-0.01}$}, our LAEs lie above a simple lower-mass extrapolation of the average $M_{\star}$-$M_{\rm h}$ relation (figure \ref{fig:mh_ms}). The higher $SHMR$ than average galaxies with the same $M_\star$ may make it easy for \lya photons to escape since they are expected to have lower gas masses (baryon mass) and thus lower HI column densities. 
Our LAEs also have a high {\color{black}$BCE=1.6^{+6.0}_{-1.0}$}, lying above the average $BCE$-$M_{\rm h}$ relation (figure \ref{fig:bce}). Thus, our LAEs have been converting baryons into stars more efficiently than average galaxies with similar $M_{\rm h}$ both in the past and at the observed epoch but with a moderate SF similar to average galaxies. Galaxies with weak SN feedback and small halo's spin parameters possibly have such properties according to the semi-analytic model by \citet{Dutton2010b}.

\item 
The duty cycle of LAEs (fraction of $M_{\rm h} \sim 3 \times 10^{10}\ {\mathrm M_\odot}$ halos hosting LAEs) is estimated to be $\sim$ 2\%, and the LAE fraction (fraction of $M_\star \sim 1 \times 10^9\ {\mathrm M_\odot}$ galaxies classified as LAEs) is found to be $\sim 10\%$. These low fractions imply either that only a small fraction of all galaxies can evolve into LAEs and/or that even low-mass galaxies can emit \lya only for a very short time.

\item 
We have calculated the halo mass evolution of our LAEs with the EPS model, to find that at $z=0$ our LAEs are embedded in dark matter halos with a median halo mass similar to the mass of the Large Magellanic Cloud (LMC). If their star-formation is largely suppressed after the observed time until $z=0$ similar to the star-formation history of the LMC, they would have a similar $SHMR$ to the present-day LMC. This result, combined with the previous studies, implies that the mass of present-day descendant halos of LAEs depends on the redshift at which the LAEs are observed, with higher-$z$ LAEs evolving into more massive halos.
\end{enumerate}

\section*{Acknowledgments}
{\color{black}We thank the anonymous referee for his/her helpful comments and suggestions. } 
We are grateful to Lihwai Lin and Li-Ting Hsu for kindly providing us with $J, H$ and $Ks$ images of the HDFN field and data in \citet{Lin2012} plotted in figures \ref{fig:ms_sfr_smc}, \ref{fig:mh_ms} and \ref{fig:bce}. We are also grateful to Yoshiaki Ono for giving insightful comments and suggestions on SED fitting. We would like to show our appreciation to Takashi Hamana for helpful comments on cosmic variance and computer programs of the covariance of dark matter angular correlation function and the EPS model. We would like to express our gratitude to David Sobral, Naveen A. Reddy, Giulia Rodighiero, and Shogo Ishikawa for kindly providing their data plotted in figures \ref{fig:cv}(a), \ref{fig:ms_irx}, \ref{fig:ms_sfr_smc}, and \ref{fig:mh_ms}, respectively. We would like to thank Alex Hagen, James E. Rhoads, Jorryt Matthee, and Peter S. Behroozi for useful comments on their results. We also would like to thank Akio K. Inoue, Cai-Na Hao, Hidenobu Yajima, Ikkoh Shimizu, Ken Mawatari, Kotaro Kohno, Kyoung-Soo Lee, Tsutomu T. Takeuchi, Mana Niida and Yuki Yoshiura for insightful discussion. 
We acknowledge Ryota Kawamata, Taku Okamura, and Kazushi Irikura for constructive discussions at weekly meetings. This work is based on observations taken by the Subaru Telescope which is operated by the National Astronomical Observatory of Japan. The authors wish to recognize and acknowledge the very significant cultural role and reverence that the summit of Maunakea has always had within the indigenous Hawaiian community. {\color{black} Based on data products from observations made with ESO Telescopes at the La Silla Paranal Observatory under ESO programme ID 179.A-2005 and on data products produced by TERAPIX and the Cambridge Astronomy Survey Unit on behalf of the UltraVISTA consortium.} This research made use of IRAF, which is distributed by NOAO, which is operated by AURA under a cooperative agreement with the National Science Foundation and of Python packages for Astronomy: Astropy\citep{TheAstropyCollaboration2013}, Colossus, CosmoloPy and PyRAF, which is produced by the Space Telescope Science Institute, which is operated by AURA for NASA. H.K acknowledges support from the JSPS through the JSPS Research Fellowship for Young Scientists. {\color{black}This work is supported in part by KAKENHI (16K05286) Grant-in-Aid for Scientific Research (C) through the JSPS.}

\vspace{2cm}


\begin{appendix}
\section{Result of SED fitting with different assumptions}\label{sec:appendix_sed}
We show the SED fitting results with the Calzetti curve and without nebular emission below. 

\subsection{The Calzetti Curve}\label{sec:appendix_cal}
\begin{figure*}[ht]
  \begin{center}
      \includegraphics[width=1.0\linewidth]{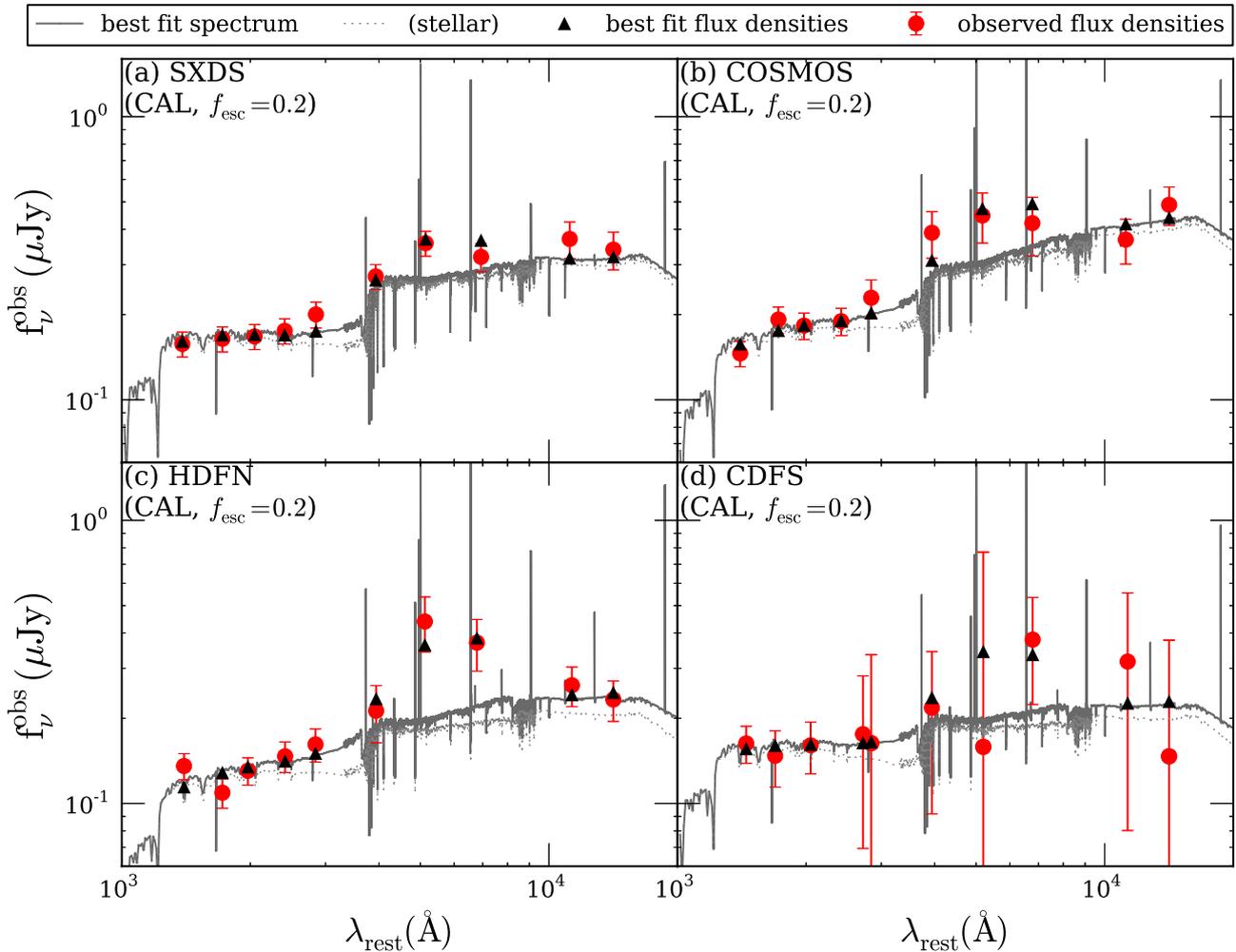} 
  \end{center} 
  \caption{
  Same as figure \ref{fig:sedfit_smc} but with the Calzetti curve. Panels (a) to (d) show results for SXDS, COSMOS, HDFN, and CDFS, respectively. (Color online)}
  \label{fig:sedfit_cal}
\end{figure*}
 We also examine the cases of the Calzetti curve for comparison. The best-fit parameters with a SMC-like curve and the Calzetti curve are listed in table \ref{tbl:sed_para_calsmc}. Figures \ref{fig:sedfit_smc} and \ref{fig:sedfit_cal} show the best-fit SEDs with the observed ones in the case with a SMC-like curve and the Calzetti curve, respectively. We compare the best-fit parameters in subsection \ref{subsec:sed_result}.

\subsection{Without nebular emission}\label{sec:appendix_fesc}

\begin{table*}
\tbl{Results of SED fitting with a SMC-like curve and the Calzetti curve ($f_{\rm esc}^{\rm ion}=0.2$).  }{
\begin{tabular}{lccccc}
\hline
 attenuation & $M_{\star}$ &$E(B-V)_{\star}\ [A_{1600}]$ &Age &$SFR$ &$\chi^2_r$   \\
curve &($10^8 {\rm M_{\odot}}$) &(mag) &($10^8\,$yr) &(M$_{\odot}$yr$^{-1}$) &    \\
& (1)& (2) & (3) & (4) & (5)\\
\hline SXDS \\ \hline 
SMC 
& $9.7^{+3.6}_{-1.7}$
& $0.05^{+0.01}_{-0.02}\ [0.6^{+0.1}_{-0.2}]$ 
& $3.6^{+2.8}_{-1.1}$
& $ 3.3^{+ 0.5}_{- 0.7}$
& $0.604$ \\ 
 Calzetti 
& $7.8^{+3.4}_{-1.9}$
& $0.11^{+0.02}_{-0.05}\ [1.1^{+0.2}_{-0.5}]$ 
& $1.6^{+2.4}_{-0.7}$
& $ 5.7^{+ 1.7}_{- 2.3}$
& $0.665$ \\ 
\hline COSMOS \\ \hline 
 SMC 
& $14.0^{+3.4}_{-3.6}$
& $0.07^{+0.02}_{-0.02}\ [0.8^{+0.2}_{-0.2}]$ 
& $4.1^{+2.4}_{-1.8}$
& $ 4.2^{+ 1.2}_{- 0.8}$
& $0.473$ \\  
 Calzetti 
& $7.9^{+5.1}_{-2.5}$
& $0.18^{+0.03}_{-0.05}\ [1.8^{+0.3}_{-0.5}]$ 
& $0.7^{+1.6}_{-0.4}$
& $12.3^{+ 6.4}_{- 5.5}$
& $0.648$ \\ 
\hline HDFN \\ \hline 
SMC 
& $7.6^{+4.0}_{-1.9}$
& $0.06^{+0.02}_{-0.03}\ [0.7^{+0.2}_{-0.4}]$ 
& $3.2^{+4.0}_{-1.4}$
& $ 2.9^{+ 0.8}_{- 0.8}$
& $1.298$ \\  
 Calzetti 
& $3.2^{+0.6}_{-0.8}$
& $0.20^{+0.02}_{-0.03}\ [2.0^{+0.2}_{-0.3}]$ 
& $0.3^{+0.2}_{-0.1}$
& $13.3^{+ 5.1}_{- 3.9}$
& $0.866$ \\ 
\hline CDFS \\ \hline 
SMC 
& $10.3^{+11.1}_{-9.7}$
& $0.02^{+0.07}_{-0.01}\ [0.2^{+0.8}_{-0.1}]$ 
& $5.7^{+8.6}_{-5.7}$
& $ 2.2^{+534}_{- 0.4}$
& $0.120$ \\ 
 Calzetti 
& $3.1^{+17.5}_{-2.4}$
& $0.14^{+0.05}_{-0.13}\ [1.4^{+0.5}_{-1.3}]$ 
& $0.4^{+14.0}_{-0.3}$
& $ 9.0^{+23.4}_{- 7.1}$
& $0.101$ \\ 
\hline Average \\ \hline 
SMC 
& $10.2\pm{\color{black}1.8}$ 
& $0.06\pm{\color{black}0.01}\ [0.6\pm{\color{black}0.1}]$ 
& $3.8\pm{\color{black}0.3}$
& $ 3.4\pm{\color{black}0.4}$ 
& \\ 
 Calzetti 
& $3.4\pm{\color{black}0.4}$  
& $0.19\pm{\color{black}0.01}\ [1.9\pm{\color{black}0.1}]$ 
& $0.3\pm{\color{black}0.04}$
& $ 12.7\pm{\color{black}0.6}$
& \\ 
\hline
\end{tabular}}\label{tbl:sed_para_calsmc}
 \tabnote{
 Note. (1) The best fit stellar mass; (2) the best-fit color excess [UV attenuation]; (3) the best fit age;  (4) the best fit SFR; (5) reduced chi-squared value. The UV attenuation is derived from the attenuation curve listed in the first column. Metallicity, redshift, and $f_{\rm esc}^{\rm ion}$ are fixed to $0.2Z_{\odot}$, 2.18, and 0.2, respectively. }
 \end{table*}

\begin{figure*}[ht]
  \begin{center}
      \includegraphics[width=0.8\linewidth]{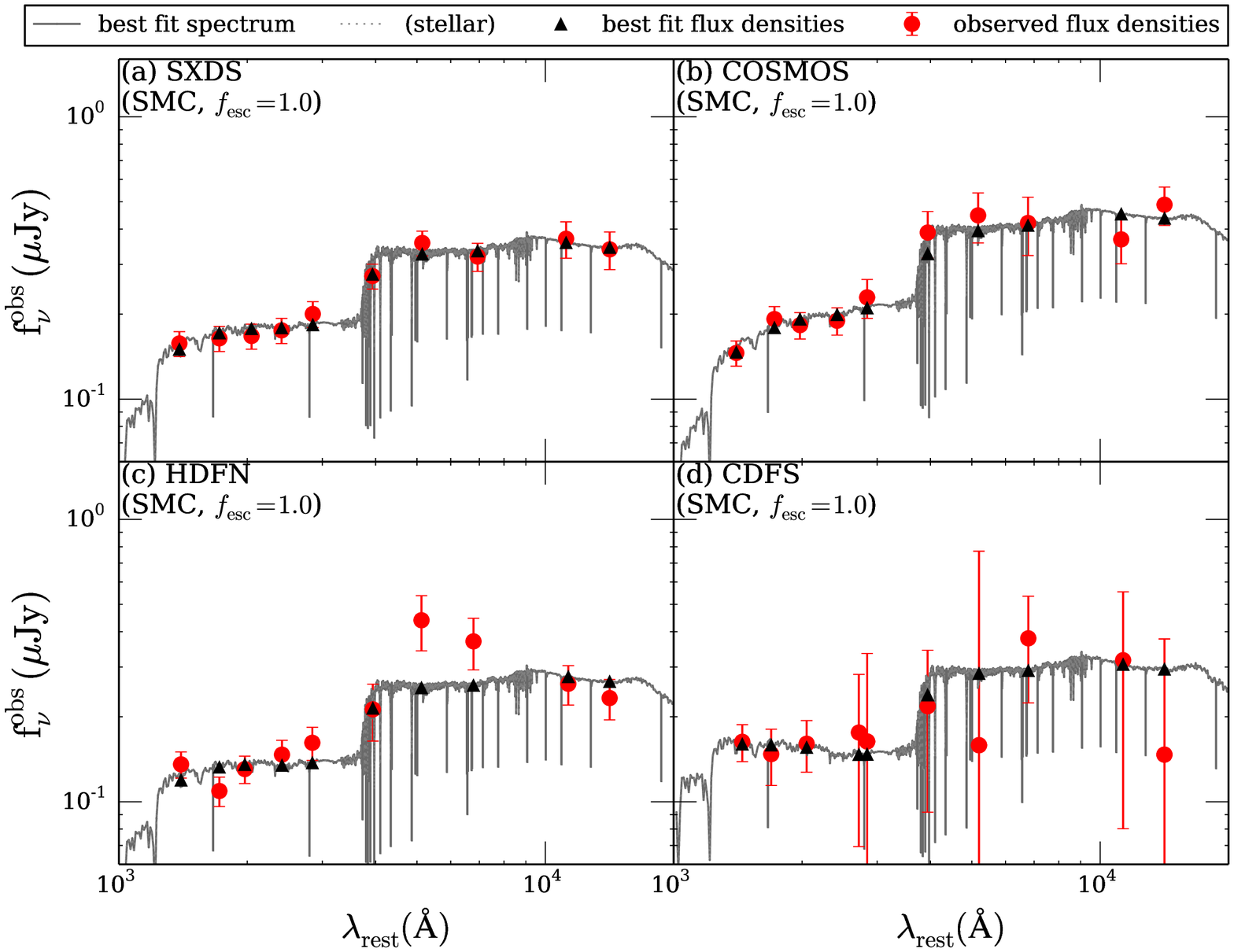}
       \includegraphics[width=0.8\linewidth]{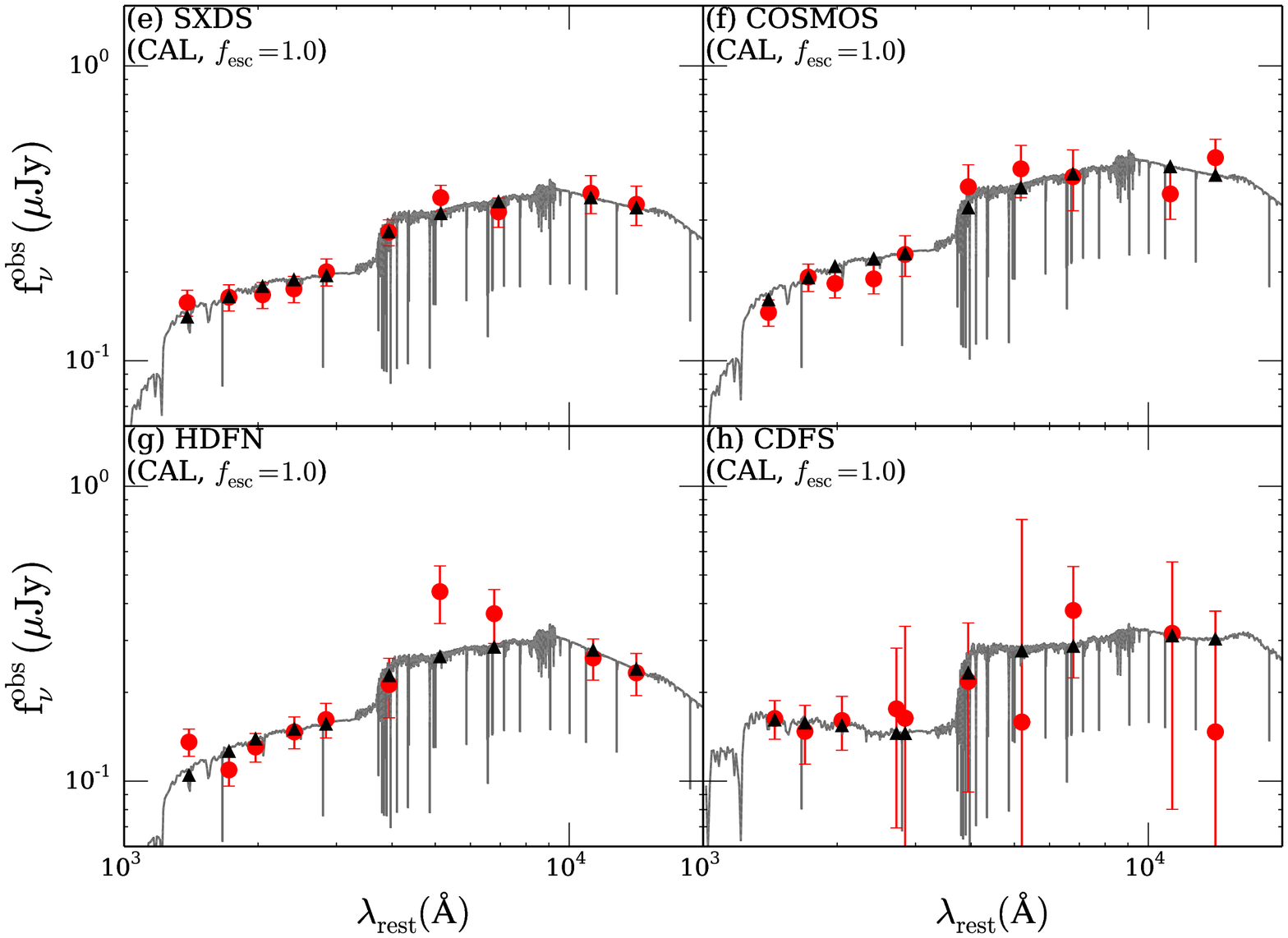}
  \end{center} 
  \caption{
 Same as figure \ref{fig:sedfit_smc} but without nebular emission, $f^{\rm ion}_{\rm esc}=1$. Panels (a) to (d) show results with a SMC-like curve for SXDS, COSMOS, HDFN, and CDFS, respectively. Panels (e) to (h) show results with the Calzetti curve for SXDS, COSMOS, HDFN, and CDFS, respectively. (Color online)}
  \label{fig:sedfit_smcfec1}
\end{figure*}

\begin{table*}[ht]
\tbl{Results of SED fitting without nebular emission, $f_{\rm esc}^{\rm ion}=1$. }{
\begin{tabular}{lccccc}
\hline
 attenuation & $M_{\star}$ &$E(B-V)_{\star}\ [A_{1600}]$ &Age &$SFR$ &$\chi^2_r$   \\
curve &($10^8 {\rm M_{\odot}}$) &(mag) &($10^8\,$yr) &(M$_{\odot}$yr$^{-1}$) &    \\
& (1)& (2) & (3) & (4) & (5)\\
\hline SXDS \\ \hline 
SMC 
& $11.4^{+2.7}_{-1.3}$
& $0.06^{+0.02}_{-0.02}\ [0.7^{+0.2}_{-0.2}]$ 
& $3.6^{+2.1}_{-1.1}$
& $ 3.9^{+ 0.9}_{- 0.8}$
& $0.350$ \\ 
 Calzetti 
& $5.1^{+7.1}_{-0.4}$
& $0.27^{+0.02}_{-0.16}\ [2.7^{+0.2}_{-1.6}]$ 
& $0.1^{+2.4}_{-0.0}$
& $45.3^{+12.0}_{-40.0}$
& $0.586$ \\ 
\hline COSMOS \\ \hline 
SMC 
& $14.6^{+5.2}_{-2.7}$
& $0.08^{+0.02}_{-0.02}\ [1.0^{+0.2}_{-0.2}]$ 
& $3.6^{+2.8}_{-1.3}$
& $ 4.9^{+ 1.3}_{- 1.1}$
& $0.611$ \\ 
 Calzetti 
& $6.6^{+1.5}_{-0.7}$
& $0.29^{+0.01}_{-0.04}\ [2.9^{+0.1}_{-0.4}]$ 
& $0.1^{+0.2}_{-0.0}$
& $56.2^{+12.8}_{-26.4}$
& $0.821$ \\ 
\hline HDFN \\ \hline 
SMC 
& $9.8^{+2.4}_{-2.5}$
& $0.05^{+0.03}_{-0.02}\ [0.6^{+0.4}_{-0.2}]$ 
& $4.5^{+2.7}_{-2.0}$
& $ 2.7^{+ 1.0}_{- 0.5}$
& $1.865$ \\ 
 Calzetti 
& $4.4^{+0.0}_{-0.8}$
& $0.30^{+0.00}_{-0.04}\ [3.0^{+0.0}_{-0.4}]$ 
& $ 0.09^{+ 0.03}_{-0.01}$
& $51.8^{+ 5.9}_{-18.9}$
& $1.653$ \\ 
\hline CDFS \\ \hline 
SMC 
& $13.1^{+10.9}_{-8.9}$
& $0.02^{+0.06}_{-0.01}\ [0.2^{+0.7}_{-0.1}]$ 
& $7.1^{+8.9}_{-6.2}$
& $ 2.3^{+ 3.0}_{- 0.3}$
& $0.148$ \\ 
 Calzetti 
& $12.1^{+12.7}_{-10.0}$
& $0.05^{+0.25}_{-0.04}\ [0.5^{+2.5}_{-0.4}]$ 
& $5.1^{+11.9}_{-5.1}$
& $ 2.9^{+135.8}_{- 1.0}$
& $0.157$ \\ 
\hline Average \\ \hline 
SMC 
& $11.2\pm{\color{black}1.2}$ 
& $0.06\pm{\color{black}0.01}\ [0.6\pm{\color{black}0.1}]$  
& $4.1\pm{\color{black}0.5}$
& $ 3.2\pm{\color{black}0.6}$ 
& \\ 
 Calzetti 
& $4.7\pm{\color{black}0.4}$  
& $0.29\pm{\color{black}0.02}\ [2.9\pm{\color{black}0.2}]$ 
& $0.09\pm{\color{black}0.01}$
& $51.8\pm{\color{black}4.5}$ 
& \\ 
\hline
\end{tabular}}\label{tbl:sed_para_fesc1}
\tabnote{Note. (1) The best fit stellar mass; (2) the best-fit color excess [UV attenuation]; (3) the best fit age;  (4) the best fit SFR; (5) reduced chi-squared value. The UV attenuation is derived from the attenuation curve listed in the first column.  Metallicity, redshift, and $f_{\rm esc}^{\rm ion}$ are fixed to $0.2Z_{\odot}$, $2.18$, and $1$, respectively. }
 \end{table*}

It is well known that considering nebular emission generally leads to a lower stellar mass \citep[e.g., ][]{DeBarros2014}. To obtain  upper limits of stellar mass and determine the star formation mode of our LAEs, we also examine the case without nebular emission, $f^{\rm ion}_{\rm esc}=1$. The best-fit parameters with a SMC-like curve and the Calzetti curve are listed in table \ref{tbl:sed_para_fesc1}. {\color{black}Figure \ref{fig:sedfit_smcfec1}} shows the best-fit SEDs with the observed ones in the case with a SMC-like curve and the Calzetti curve.  

When we assume a SMC-like curve, the average stellar mass and SFR without nebular emission, {\color{black}$M_{\star}=11.2\,\pm\,1.2\times10^{8}\ {\rm M_{\odot}}$ and $SFR=3.2\,\pm\,0.6\ {\rm M_{\odot}}\ {\rm yr^{-1}}$}, are consistent with those with nebular emission, {\color{black}$M_{\star}=10.2\,\pm\,1.8\times10^{8}\ {\rm M_{\odot}}$ and $SFR=3.4\,\pm\,0.4\ {\rm M_{\odot}}\ {\rm yr^{-1}}$}. This means that the average stellar mass and star formation mode of our LAEs are insensitive to $f^{\rm ion}_{\rm esc}$ when a SMC-like curve is used. 
On the other hand, if we assume the Calzetti curve, the average SFR without nebular emission, {\color{black}$SFR=51.8\,\pm\,4.5\ {\rm M_{\odot}}\ {\rm yr^{-1}}$}, is {\color{black} about four times} higher than that with nebular emission, {\color{black}$SFR=12.7\,\pm\,1.0\ {\rm M_{\odot}}\ {\rm yr^{-1}}$}. 
{\color{black}Their average stellar mass without nebular emission, $M_{\star}=4.7\,\pm\,0.7\times10^{8}\ {\rm M_{\odot}}$ is slightly higher than that with nebular emission,  $M_{\star}=3.4\,\pm\,0.8\times10^{8}\ {\rm M_{\odot}}$}. 
With this high SFR, our LAEs lie above the SFMS at $z\sim2$. However, this case seems unrealistic because our LAEs have \lya emission, one of nebular emission lines. Indeed, the reduced $\chi$ square values in the case without nebular emission are larger than those with nebular emission in all the fields except SXDS. In addition, results with $f^{\rm ion}_{\rm esc}=1$ give a high UV attenuation of {\color{black}$A_{1600}=2.9\pm0.2$} mag and hence a high {\color{black}$IRX$ ($=22^{+5}_{-4}$)}, which is significantly higher than predicted by the consensus relation (see figure \ref{fig:ms_irx}).

\begin{figure*}[ht]
  \begin{center}
    \begin{tabular}{c}
      \begin{minipage}{0.5\hsize}
        \begin{center}
     \includegraphics[width=1.0\linewidth]{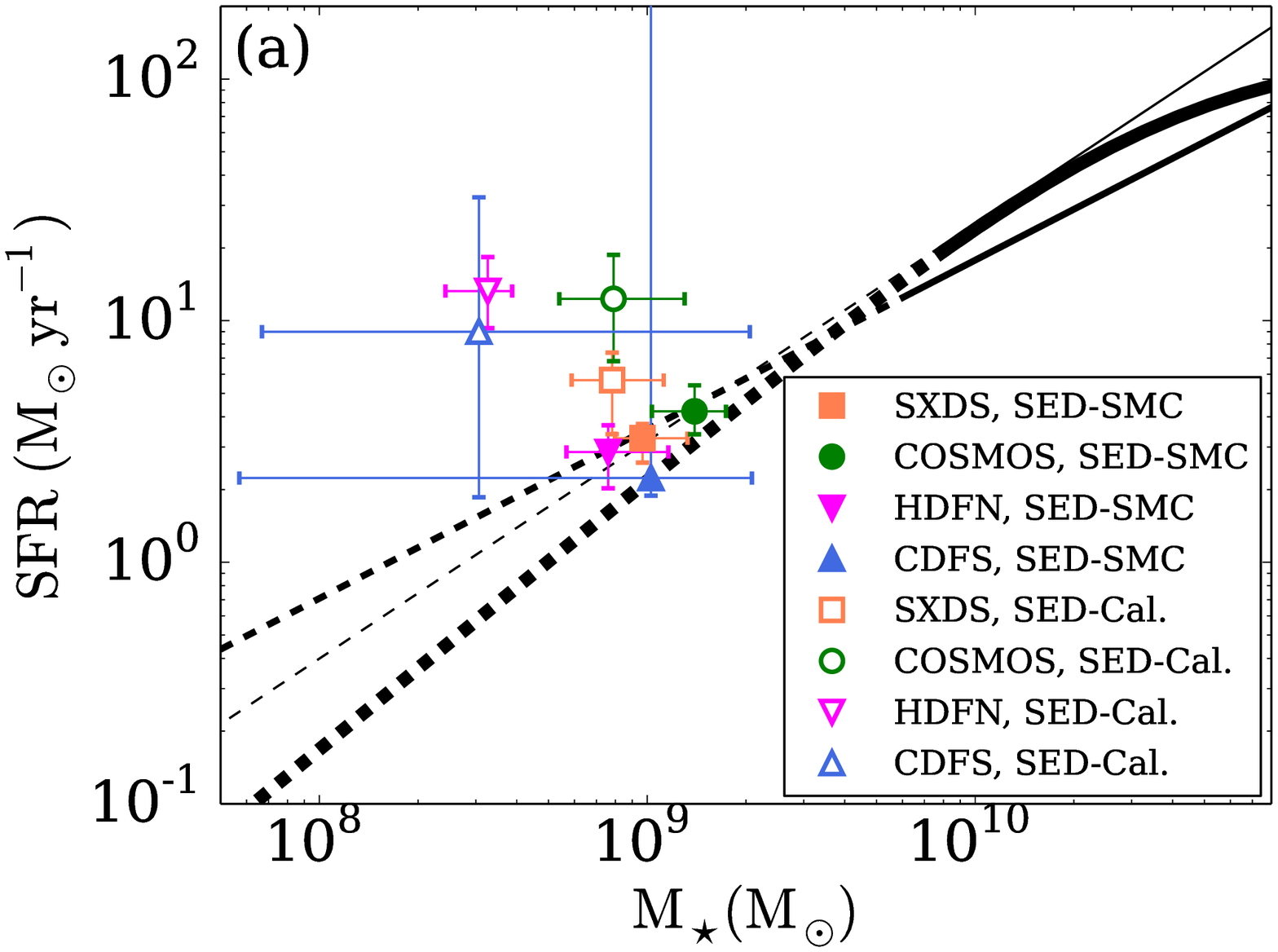}
        \end{center}
      \end{minipage}
      \begin{minipage}{0.5\hsize}
        \begin{center}
      \includegraphics[width=1.0\linewidth]{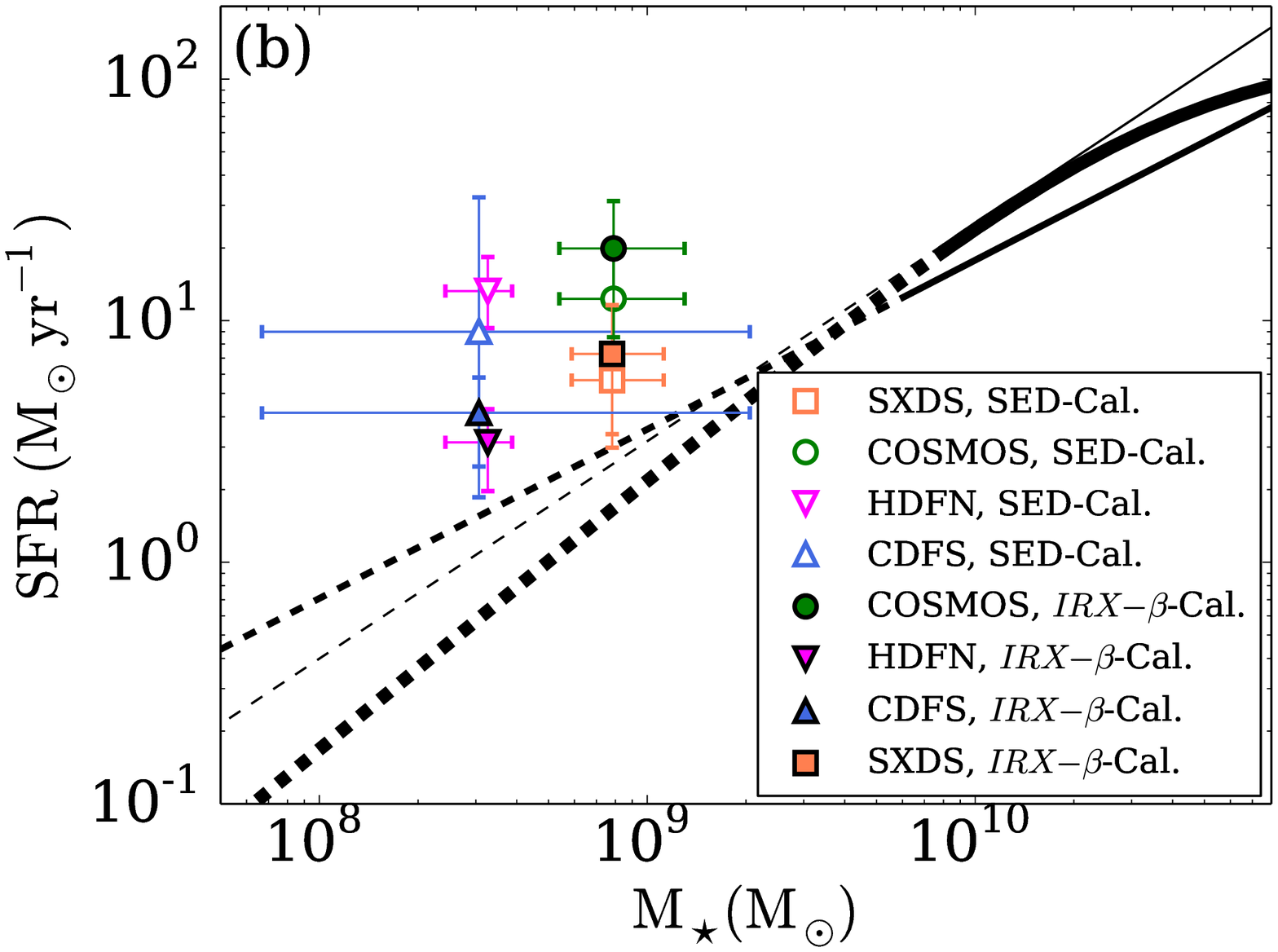}
        \end{center}
      \end{minipage}\\
      \begin{minipage}{0.5\hsize}
        \begin{center}
      \includegraphics[width=1.0\linewidth]{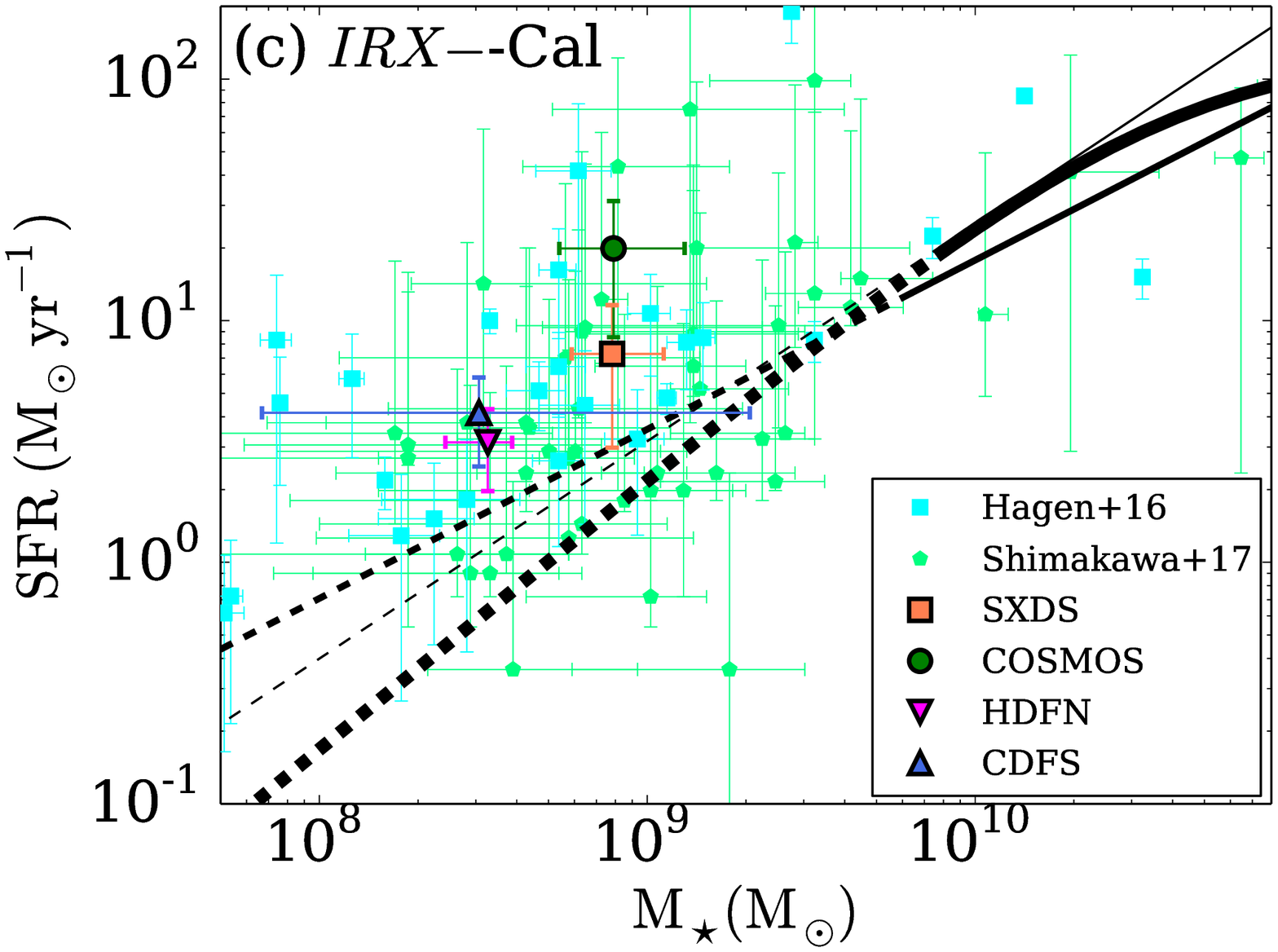}
        \end{center}
      \end{minipage}\\
      
  \end{tabular} 
  \end{center} 
  \caption{
 
  SFR plotted against $M_\star$. Panels (a) and (b) compare different SFR calculation methods for our LAEs; in panel (a) $SFRs$ calculated from SED fitting with two different attenuation curves are compared; in panel (b) $SFRs$ from SED fitting are compared with those from the $IRX-\beta$ relation, where the Calzetti curve is used in both calculations. Panel (c) uses the $IRX-\beta$ relation with the Calzetti curve and compares our LAEs with \citet{Hagen2016}'s and \citet{Shimakawa2017a}'s. In panel (a), orange squares, green circles, magenta inverted triangles, and blue triangles represent stacked LAEs with $NB387_{\rm tot} \le 25.5$ mag in the SXDS, COSMOS, HDFN, and CDFS fields, respectively; filled and open symbols are for a SMC-like curve and the Calzetti curve, respectively. In panel (b),  encircled symbols indicate that $SFRs$ are derived from the $IRX-\beta$ relation with the Calzetti curve \citep{Meurer1999}. In panel (c), cyan squares and light green pentagons show individual LAEs at $z \sim 2$ in \citet{Hagen2016} and \citet{Shimakawa2017a}, respectively; in both studies, $SFRs$ are derived from the $IRX-\beta$ relation with the Calzetti curve \citep{Meurer1999}. Our results based on the $IRX-\beta$ relation with the Calzetti curve are also plotted (encircled symbols). In all panels, several SFMS measurements in previous studies are shown by black lines in the same manner as figure \ref{fig:ms_sfr_smc}. {\color{black}All data are rescaled to a Salpeter IMF according to footnote \ref{ft:imf}.} (Color online)}
\label{fig:ms_sfr_cal}
\end{figure*}

\section{SFMS based on the IRX-$\beta$ relation with the Calzetti curve}\label{sec:appendix_sfms_cal}
In the discussion of the star formation mode of LAEs at $z\sim2$ in section \ref{subsubsec:sed_Ms-SFR}, we derive the average SFR of our LAEs using SED fitting with a SMC-like curve, while \citet{Hagen2016} and \citet{Shimakawa2017a} derive SFRs using the $IRX-\beta$ relation with the Calzetti curve. For a fair comparison, figure \ref{fig:ms_sfr_cal}(c) shows our results with the $IRX-\beta$ relation with the Calzetti curve \citep{Meurer1999}. We find our LAEs to have higher sSFRs similar to LAEs in \citet{Hagen2016}. Note that the selections of these three samples are different as described in section \ref{subsubsec:sed_Ms-SFR}. We also compare our results by the three different methods discussed in appendix \ref{sec:appendix_cal} and in this section (see figures \ref{fig:ms_sfr_cal}(a) and (b)). 

\clearpage
\end{appendix}

\clearpage
\bibliographystyle{aasjournal}

\end{document}